\documentclass[fleqn,usenatbib]{mnras}

\usepackage{newtxtext,newtxmath}
\usepackage[T1]{fontenc}

\DeclareRobustCommand{\VAN}[3]{#2}
\let\VANthebibliography\thebibliography
\def\thebibliography{\DeclareRobustCommand{\VAN}[3]{##3}\VANthebibliography}

\usepackage{graphicx}	% Including figure files
\usepackage{amsmath}	% Advanced maths commands
\usepackage{dcolumn}% Align table columns on decimal point
\usepackage{bm}% bold math
\usepackage{hyperref}% add hypertext capabilities
\usepackage{algpseudocode}
\usepackage{float}
\usepackage{fontawesome}
\usepackage[normalem]{ulem}
\usepackage{color}
\usepackage[table, dvipsnames]{xcolor}% http://ctan.org/pkg/xcolor
\usepackage{tabularx}

\usepackage{appendix}

\usepackage{booktabs}

\newcommand{\react}{\texttt{ReACT}}
\newcommand{\halofit}{\texttt{halofit}}
\newcommand{\hmcode}{\texttt{HMcode}}
\newcommand{\bacon}{\texttt{BaCoN-II}}
\newcommand{\lcdm}{$\Lambda$CDM}

\title[Cosmology with Neural Networks]{Classifying Modified Gravity and Dark Energy Theories with Bayesian Neural Networks: Massive Neutrinos, Baryonic Feedback, \\ and the Theoretical Error} %Beyond-\lcdm{} Cosmologies

\author[L. Thummel et al.]{
Linus Thummel,$^{1,2}$\thanks{E-mail: linus.thummel@ed.ac.uk}
Benjamin Bose,$^{1,3}$
Alkistis Pourtsidou,$^{1,2}$
Lucas Lombriser$^{4}$
\\
$^{1}$Institute for Astronomy, University of Edinburgh, Royal Observatory, Blackford Hill, Edinburgh, EH9 3HJ, UK\\
$^{2}$Higgs Centre for Theoretical Physics, University of Edinburgh, James Clerk Maxwell Building, Edinburgh, EH9 3FD, UK\\
$^{3}$Basic Research Community for Physics e.V., Mariannenstraße 89, Leipzig, Germany\\
$^{4}$D\'epartement de Physique Th\'eorique, Universit\'e de Gen\`eve, 1211~Gen\`eve~4, Switzerland\\
}

\date{Accepted XXX. Received YYY; in original form ZZZ}

\pubyear{2024}

\begin{document}
\label{firstpage}
\pagerange{\pageref{firstpage}--\pageref{lastpage}}
\maketitle

% Abstract of the paper
\begin{abstract}
We study the capacity of Bayesian Neural Networks (BNNs) to detect new physics in the dark matter power spectrum. As in previous studies, the Bayesian Cosmological Network (\texttt{BaCoN}) classifies spectra into one of 5 classes: \lcdm{}, $f(R)$, $w$CDM, Dvali-Gabadaze-Porrati (DGP) gravity and a `random' class, with this work extending it to include the effects of massive neutrinos and baryonic feedback. We further develop the treatment of theoretical errors in \bacon{}, investigating several approaches and identifying the one that best allows the trained network to generalise to other power spectrum modelling prescriptions. In particular, we compare power spectra data produced by \texttt{EuclidEmulator2}, \hmcode{}  and \halofit{}, all supplemented with the halo model reaction to model beyond-\lcdm{} physics. 
We investigate BNN-classifiers trained on these sets of spectra, adding in Stage-IV survey noise and various theoretical error models. Using our optimal theoretical error model, our fiducial classifier achieves a total classification accuracy of $\sim$ 95\% when it is trained on \texttt{EuclidEmulator2}-based spectra with modification parameters drawn from a Gaussian distribution centred around \lcdm{} ($f(R)$:~$\sigma_{fR0} = 10^{-5.5}$, DGP:~$\sigma_{r\mathrm{c}} = 0.173$, $w$CDM:~$\sigma_{w0} = 0.097$, $\sigma_{wa}=0.32$).  This strengthens the promise of this method to glean the maximal amount of unbiased gravitational and cosmological information from forthcoming Stage-IV galaxy surveys. 
\end{abstract}

% Select between one and six entries from the list of approved keywords.
% Don't make up new ones.

%%%%%%%%%%%%%%%%%%%%%%%%%%%%%%%%%%%%%%%%%%%%%%%%%%

%%%%%%%%%%%%%%%%% BODY OF PAPER %%%%%%%%%%%%%%%%%%

\section{Introduction}

\subsection{Cosmology}
The concordance model of cosmology (\lcdm{}) remains largely consistent with the vast amount of cosmological data produced during the last three decades \citep[see][for example]{Huterer:2017buf}. The model is largely characterised by the two `dark' cosmological components that dominate our present Universe:  dark energy in the form of a cosmological constant ($\Lambda$) and cold dark matter (CDM). These cosmological energy densities are needed for the theory to match a wide range of data sets such as the cosmic microwave background (CMB) measurements \citep{Aghanim:2018eyx} and cosmic large scale structure (LSS) measurements \citep{Anderson:2012sa,Song:2015oza,Beutler:2016arn,Hildebrandt:2016iqg,2021A&A...646A.140H,Abbott:2020knk,DES:2021lsy}. Dark matter is also needed to match smaller astrophysical and cluster scale data \citep{Clowe:2006eq,Corbelli:1999af,2011ARA&A..49..409A}. The nature of these components, as well as their links to other fundamental physical problems, such as the cosmological constant problem \citep[see][for a recent review]{Bernardo:2022cck}, have been the focus of intense investigation over the past 25 years. 

In this direction, many different theoretical ideas have been imagined and tested \citep[for example][]{Dvali:2000hr,Burgess:2004ib,Brax:2016kin}. Often these can be represented as a scalar tensor theory of gravity, or `modified gravity' more generally \citep{cliftonModifiedGravityCosmology2012}, or some non-trivial dark energy component \citep{Copeland:2006wr}. Many of these model's degrees of freedom have been well constrained by data or have exhibited theoretical or conceptual issues \citep[see][for a review]{Koyama:2015vza}. Nevertheless, such characterisations of departures from \lcdm{} and General Relativity (GR) should be tested by new data sets in order to guide future theoretical development.

 An important anchor in such tests is that any new physics should be unobserved in our local part of the Universe. This instruction comes from the exquisite tests of GR \citep[see][for a review]{Will:2014kxa} and various table-top experiments of modified gravity \citep[see][for example]{Burrage:2019yle}. Thus, we must shut-off the effects of any additional degrees of freedom in the local Universe. Such theoretical mechanisms have been dubbed `screening mechanisms' and are well studied in the context of modified gravity \citep{Brax:2021wcv}. 

Other noteworthy observational constraints come from the large-scale data sets, including CMB and large scale structure  measurements. These constraints have been placed on a wide range of possible departures from \lcdm{}, particularly in the context of the Effective field theory of Dark Energy (EFTofDE) \citep{Gubitosi:2012hu,Salvatelli:2016mgy,Perenon:2019dpc,Noller:2020afd}, with statistical preference still generally given to \lcdm{}. One is also instructed by the measurement of incredibly consistent electromagnetic radiation and gravitational wave speeds \citep{Lombriser:2015sxa,Monitor:2017mdv,Lombriser:2016yzn,Creminelli:2017sry,Ezquiaga:2017ekz,Baker:2017hug,Sakstein:2017xjx,Battye:2018ssx,deRham:2018red,Creminelli:2018xsv}. Despite this, we must keep in mind the energy scales that these constraints probe which may have no bearing on the large scale gravitational or cosmological theory \citep[see][for example]{deRham:2018red}. 

One major probe of new physics still left to be tapped is the smaller (nonlinear) scales of the large scale structure, which will be exquisitely measured by ongoing and forthcoming galaxy surveys such as Euclid \citep{2011arXiv1110.3193L,Amendola:2016saw,Blanchard:2019oqi}, DESI \citep{Levi:2019ggs}, the Nancy Grace Roman Space Telescope  \citep{2019arXiv190205569A} and the Vera Rubin Observatory  \citep{Abate:2012za}. At smaller scales we are benefiting from improved galaxy statistics and so measurement errors become much smaller. The extension of data analyses to the nonlinear regime can be achieved with accurate nonlinear prescriptions such as emulators of key quantities  \citep{Angulo:2020vky,Euclid:2020rfv,Nishimichi:2018etk,Donald-McCann:2021nxc,Kobayashi:2020zsw}. This is generally only the case for consistency tests of \lcdm{}, or particular extensions thereof \citep{Winther:2019mus,Arnold:2021xtm,Ramachandra:2020lue}. 

In this direction, a theoretically general and accurate approach to the nonlinear regime for beyond-\lcdm{} models was proposed in \cite{Cataneo:2018cic,Cataneo:2019fjp,Bose:2021mkz}, named the {\it halo model reaction}, with a general parametrisation for this framework further developed in \cite{Bose:2022vwi}. The associated \react{} code~\citep[\href{https://github.com/nebblu/ACTio-ReACTio}{\faicon{github}}]{Bose:2020wch,Bose:2022vwi}, allows fast computations of beyond-\lcdm{} corrections of cosmological matter density field correlations. Coupled with \lcdm{} prescriptions for these correlations, one has a means of predicting these beyond-\lcdm{} correlations in the form of the {\it matter power spectrum}. This has allowed its use in real and mock data analyses for galaxy surveys to obtain and forecast constraints on deviations from \lcdm{} \citep{KiDS:2020ghu,SpurioMancini:2023mpt,Euclid:2023rjj, Carrion:2024itc}.

Modelling beyond-\lcdm{} physics is not the only major challenge. In the nonlinear regime many physical phenomena start to affect matter clustering in a significant way. In particular, the effects of massive neutrinos \citep{2018MNRAS.481.1486B,2012MNRAS.420.2551B,2014JCAP...11..039B,2016MNRAS.459.1468M,2017ApJ...847...50L,Tram:2018znz,Massara:2014kba,2020arXiv200406245A} and baryons \citep[for example][]{vanDaalen2011,Mummery2017,Springel2018,vanDaalen:2019pst,Schneider:2018pfw,Chisari:2019tus} have been shown to be extremely important. Neglecting any of these effects will lead to biased estimates of cosmology and gravity \citep{Semboloni:2011fe,Schneider:2019snl,Schneider:2019xpf,Euclid:2020tff}. Fortunately, baryonic effects have been well studied and can be efficiently modelled with the addition of a few degrees of freedom \citep{Arico:2020lhq,Mead:2020vgs,Giri:2021qin}. It also appears they are loosely coupled to modifications to gravity \citep[see][for example]{Mead:2016zqy,Arnold:2019zup}. On the other hand, massive neutrinos are known to be highly degenerate with particular modifications to gravity and do couple in non-trivial ways to the gravitational degrees of freedom \citep[see for example][]{Mead:2016zqy,Wright:2017dkw,Wright:2019qhf}. Models for these effects have been tested in tandem with the halo model reaction against various beyond-\lcdm{} numerical and hydrodynamical simulations, with very promising agreement ($\sim 3\%$ for $k \leq 1~h/{\rm Mpc}$) shown \citep{Bose:2021mkz,Parimbelli:2022pmr} at the level of the  power spectrum.

\subsection{Machine Learning}

The aforementioned modelling methods can be implemented in statistical data analyses to derive cosmological parameter constraints, marginalised over a suite of nuisance parameters. Performing  statistical analyses such as Markov Chain Monte Carlo (MCMC) on very high dimensional models is not straightforward. Computational expense and complex parameter degeneracies are among the major problems of such analyses, and the efficient sampling of the posterior distribution can be challenging. In addition to the huge number of extra parameters, one deals with a vast model space of cosmological theories.
This is where machine learning can help. A trained neural network can be more efficient and preselect models that are worth testing in a MCMC.

A deep neural network (DNN) provides a highly nonlinear map between the input (in this case a cosmological quantity) and the output (in this case a classification of theoretical model). DNNs have shown to be successful in many classification problems, predominantly outside physics \citep{2019PhR...810....1M, Lin2017, LeCun2015}. Bayesian Neural Networks (BNNs) try to model the uncertainty by replacing each weight in the network by a distribution initialised to a prior \citep{MacKay, Neal, Blundell2015,  gal2016dropout, Charnock2020, jospin2020handson}. In a previous paper \citep{Mancarella:2020jyu}, the authors suggested the employment of BNNs to supplement cosmological MCMC analyses. This works by classifying an input power spectrum using the BNN, and then using that classification to select various degrees of freedom or apply stronger priors on model parameters, both of which can be applied to heighten efficiency of subsequent MCMC analyses.  

Using the \react{} code, the authors of \citet{Mancarella:2020jyu} built a very large data set of power spectra for various gravitational and cosmological models with 21,000 realisations each. This was then used to train and test a BNN called the \textbf{Ba}yesian \textbf{Co}smological \textbf{N}etwork, ({\tt BaCoN}, \href{https://github.com/Mik3M4n/BaCoN}{\faicon{github}}). The authors also implemented a constant systematic error in training which aimed at accounting for theoretical inaccuracies in the beyond-\lcdm{} and nonlinear theoretical modelling. This BNN was shown to be remarkably accurate at classifying power spectra into various modified gravity or dark energy models, with a test set accuracy of $\sim 95\%$. Post training, this classification could be done almost instantly given an input spectrum. This suggested that a well calibrated and appropriately trained network could be invaluable in guiding more intensive cosmological analyses.

One major advantage of using a Bayesian instead of a Deterministic Neural Network is the treatment of classification uncertainties.
The uncertainty in the classification by a neural network can be divided into the intrinsic uncertainty of the data and the uncertainty due to the machine learning model itself. These components are known as aleatoric and epistemic uncertainties, respectively
\citep{Kiureghian-2009-AleatoryEpistemicDoes}. 
Our approach to deal with uncertainties is threefold. First, we introduce a fifth `random' class that accounts for out of distribution samples. It represents cosmologies that are not represented by the cosmological models chosen for the other training data sets. It flags previously unseen cosmological classes and might also get activated when the cosmological parameters of the test samples are far outside of the prior range of the training data. We investigate the effectiveness of this `random' class in \autoref{sec:conclusions}.  
Second, we marginalise over the weight distribution in the Bayesian layers. This is achieved by averaging over multiple passes through the network, each drawing different Monte Carlo (MC) samples from the weight distributions. At the same time, we can use the distribution of output probabilities to construct an estimate for the uncertainty of the classification result. The computation of the distribution of MC averages is presented in \autoref{sec:uncertainty F}.

However, this does not erase all of the epistemic uncertainty of the network. The training data is not a perfect representation of real observations. We try to compensate this with the third part of our uncertainty treatment, by introducing a theoretical error in the training process. A detailed discussion can be found in \autoref{sec:theoryerr}. While this can account for the biggest part of the systematic error of the training data, some of the code-specific features of the training data will remain and bias the neural network. We investigate the error induced by different power spectrum prescriptions in \autoref{sec:powerSpectrumPrescription}. 
In the next section we will have a look at other implementations of neural networks for the analysis of modified gravity and dark energy models.

\subsection{Neural Networks for Beyond-\lcdm{} models}

When testing the application of machine learning techniques to modified gravity scenarios,  the well studied $f(R)$ class is particularly popular. It's Hu-Sawicki parameterisation~\citep{Hu:2007nk} has been implemented in many simulation suites.

\citet{Garcia-Farieta-2024-BayesianDeepLearning} use real-space density fields and power spectra of $f(R)$ DM-only simulations as training data for their network. It is constructed from different Bayesian modifications of the ResNet18 architecture -- a convolutional neural network with 18 layers.  The main goal of their work is to constrain cosmological parameters from $f(R)$ simulations. The Bayesian networks perform well with exception for the estimation of the MG parameter itself. They adopt multiplicative normalising flows for the approximate posteriors of the BNN parameters. This method of BNN training was tested in their previous work \citep{Hortua-2023-ConstrainingCosmologicalParameters} for the estimation of \lcdm{} parameters from N-body simulations.
An interesting combination of a BNN and a Recurrent Neural Network are used in \citet{Escamilla-Rivera-2020-DeepLearningApproach} to investigate dark energy models. This study is not based on LSS data but uses the distance modulus from supernovae data. They implement a Bayesian network to obtain uncertainty contours.

Moving away from BNNs, \citet{Ocampo-2024-EnhancingCosmologicalModel} use a deterministic neural network to classify $f(R)$ spectra based on the evolution of $f \sigma_8$ values. Their focus is on the interpretability of the chosen network architecture, \textit{e.g.} studying feature importance. They follow a similar approach in \citet{Ocampo-2024-NeuralNetworksCosmological} but this time using Cosmic Microwave Background data.
Another test of a $f(R)$-vs-\lcdm{} classification has been performed in \citet{Peel-2019-DistinguishingStandardModified}. They use a deterministic convolutional neural network to classify simulated weak-lensing convergence maps and specifically test the influence of massive neutrinos. They find that adding noise significantly lowers the network's performance.

Apart from neural networks, many other machine learning approaches have been used to study alternatives to \lcdm{}. \citet{Gangopadhyay-2023-PhantomDarkEnergy} use a genetic algorithm to construct a more model-independent test of dark energy theories. A similar approach has been taken in \citet{Arjona-2020-WhatCanMachine} to search for deviations from \lcdm{} in the background expansion. A genetic algorithm can be employed to optimise the design of neural network architectures, as it was shown in \citet{Gomez-Vargas-2023-NeuralNetworksOptimized} for the equation of state of a quintessence model.

It seems that to our best knowledge not many multi-class MG classifiers have been constructed. Most likely this is the case because the risk of degeneracies increases when more MG models are considered at the same time. \bacon{} tries to circumvent this problem by using LSS data from highly nonlinear scales. This requires accurate models for physics affecting nonlinear scales like massive neutrinos and baryonic feedback.

\subsection{Overview}

In this work, we extend the training of {\tt BaCoN} to include massive neutrinos and baryonic effects using the baryonic feedback model described in \cite{Mead:2020vgs} and the massive neutrino implementation of \react{} \citep{Bose:2021mkz}. We also investigate the most effective way to model the theoretical (systematic) error, going beyond the constant assumption of the previous work. Finally, we also adopt a far more accurate prescription for the power spectrum training data by making use of the state-of-the-art Euclid Emulator \citep{Euclid:2020rfv}. 

%%%%%%%%% CONTINUE %%%%%%%%%%%%% 

This paper is laid out as follows. 
\autoref{sec:setup} presents our theoretical and numerical setup, along with a discussion of the BNN's architecture.  
\autoref{sec:training} then describes the data preparation, the training and testing process as well as an analysis of the calibration and uncertainty of the trained network.
\autoref{sec:results} presents our results and discusses the performance of the BNN. 
\autoref{sec:conclusions} presents our conclusions and outlines future plans.

%%%%%%%%%%%%%%%%%%%%%% Theory ###################

\section{Theoretical Setup}
\label{sec:setup}

\subsection{Bayesian and Convolutional Neural Networks}

A Bayesian neural network combines a neural network with Bayesian statistics. The general idea is that instead of optimising fixed weights for every node the network uses marginal distributions.
The two main advantages of BNNs are avoiding overfitting and providing a measure of uncertainty for the predictions. If we pass a sample repeatedly through the network we obtain a distribution instead of a single prediction for the classification result.

\bacon{} is based on the architecture of convolutional neural networks (CNNs) so it combines Bayesian inference with image analysis in broad terms. A CNN does not fully connect all the nodes of adjacent layers. A convolutional layer convolves the image layer with a smaller set of filters and the recognised features are saved in activation maps.

For \bacon{} we use a matter power spectrum, $P(k,z)$, at four redshifts as the input image. The network performs its classification based on visual features in the spectrum. The patterns recognised in the convolutional layers are in our case shapes in the power spectrum. This makes the substructure of the graph more important than the overall amplitude. Hence, our choice of network architecture influences what the network is most sensitive to, which will differ from an established cosmological MCMC analyses. This also alters our approach to the noise model, which we will discuss in \autoref{sec:theoryerr}. 
In general, it had been expected that CNNs do not keep track of the position of a structure in the input image, but it was shown recently by \cite{islamHOWMUCHPOSITION2020} that the convolutional layers may implicitly learn absolute position information. This would explain the high accuracy that we obtain when analysing spectral data. Furthermore, \citet{zhouLearningDeepFeatures2016} have developed Class Activation Maps (CAMs) that are able to visualise the areas of the image that were most influential on the classification decision of a CNN. \citet{zhongExplainableMachineLearning2022} have shown that they can be deployed successfully for spectra retrieving position information of class-specific features in the input.

Our network architecture ends with a final fully connected layer to be able to output probabilities for 5 classes. The actual classification is decided by a probability threshold. If one class has a probability above $50 \%$ the spectrum counts as being classified. 
The final output for a test of a trained \bacon{} network is a {\it confusion matrix} which shows the percentages of all correctly, wrongly, and not classified spectra. This is ideally a fully diagonal matrix and we can use the off-diagonal entries to interpret degeneracies between classes. The contribution to true positives, false positives, true negatives and false negatives are broken down by the specific classes.  We evaluate all our networks using the full confusion matrix for the test data but we will sometimes only state the total test accuracy (true positive rate over all classes) for conciseness. 

The architecture and machine learning techniques used here are described in \autoref{sec:architecture}, they are adopted from our previous work in \citet{Mancarella:2020jyu}.

%.....................
\subsection{Halo model reaction} \label{sec:nl-hmrapproach}
%.....................

To create our training data we use the approach of \cite{Cataneo:2018cic}, which models the nonlinear power spectrum as
\begin{equation}
 P_{\rm NL}(k,z) = \mathcal{R}(k,z) \, P_{\rm pseudo}(k,z)\,.
 \label{eq:nonlinpk}
\end{equation}
Here, $\mathcal{R}$ is the halo model reaction function and quantifies corrections to the pseudo spectrum coming from the nonlinear, non-standard physics in the beyond-\lcdm{} universe. We refer the reader to \cite{Bose:2021mkz} for the analytic formulae on which this function is based. It can be efficiently computed using the publicly available \react{} code ~\citep[\href{https://github.com/nebblu/ACTio-ReACTio}{\faicon{github}}]{Bose:2020wch,Bose:2022vwi}. The accuracy of this function was found to be at the $1\%$-level in \cite{Cataneo:2018cic} for $k \leq 1~h/{\rm Mpc}$ for both modified gravity and dynamical dark energy models. 

$P_{\rm pseudo}(k,z)$ is called the pseudo power spectrum and is defined as a nonlinear spectrum in a \lcdm{} universe but whose initial conditions are tuned so that the linear clustering matches the beyond-\lcdm{} universe at a given redshift, $z$. The purpose of using the pseudo power spectrum is to ensure the halo mass functions in both beyond-\lcdm{} and pseudo universes are similar, which gives a better transition between linear and nonlinear regimes. This quantity can be modelled using nonlinear formulas such as \hmcode{} \citep{Mead:2015yca,Mead:2016zqy,Mead:2020vgs} or \halofit{} \citep{Takahashi:2012em}, which require the specification of a linear power spectrum, allowing the user to provide the modified linear clustering while keeping the nonlinear clustering based on \lcdm{} physics. The drawback is that these fitting formulae introduce a significant inaccuracy in the calculation. These inaccuracies have been qualified at $5\%$ in both cases for $k\leq 1~h/{\rm Mpc}$, but with \hmcode{} typically achieving a higher $\sim 2.5\%$ accuracy for most cosmologies \citep{Mead:2020vgs,Takahashi:2012em}. These inaccuracies dominate the error budget of \autoref{eq:nonlinpk}. 

In \cite{Mancarella:2020jyu}, the authors used the \halofit{} formula to generate the training set, and account for theoretical uncertainties using a constant systematic. This was highly underestimated as we highlight in \autoref{sec:theoryerr}. In addition to improving the error prescription, we also improve upon the `old' training data by  using two, improved prescriptions for our predictions. 

{\it The first} employs the improved \hmcode{} formula of \cite{Mead:2020vgs} 
\begin{equation}
     P^{\rm HMcode2020}_{\rm NL}(k,z) = \mathcal{R}(k,z) \, P_{\rm pseudo}^{\rm HMcode2020}(k,z)\,.\label{eq:nlpkhmc}
\end{equation}
We expect this prediction to be $\sim 6\%$ accurate for $k \leq 1~h/{\rm Mpc}$ as we consider a wide range of cosmologies. In particular, the pseudo cosmologies may have a fairly large amplitude of linear clustering due to enhancements from modified gravity forces. This will affect the accuracy of the pseudo power spectrum as noted in \cite{luis2024,maria2024}. This accuracy increases to $\geq 9\%$ for $k\leq 3~h/{\rm Mpc}$ as estimated from \cite{Cataneo:2018cic}, which did not include massive neutrinos. 

{\it The second} also employs \hmcode{}, but as a ratio of modified-to-\lcdm{} predictions, i.e., a boost. This effectively factors out some of the inaccuracy inherent in the \hmcode{} prediction. To get $P_{\rm NL}$ we then multiply the \hmcode{}-based boost with a highly accurate prediction for $P_{\rm NL}^{\rm \Lambda CDM}$. This is available via the sophisticated \lcdm{} power spectrum emulator, \texttt{EuclidEmulator2}~\citep{Euclid:2020rfv} (EE2), which efficiently emulates $N$-body predictions for \lcdm{} cosmologies. This `optimal' accuracy version for the spectra predictions is then given as 
\begin{equation}
     P^{\rm EE2}_{\rm NL}(k,z) = B^{\rm HMcode2020}(k,z) \times P_{\rm \Lambda CDM}^{\rm EE2} \, , \label{eq:nlnpkee2} 
\end{equation}
where 
\begin{equation}
    B^{\rm HMcode2020}(k,z) \equiv  \frac{ \mathcal{R}(k,z) \, P_{\rm pseudo}^{\rm HMcode2020}(k,z)}{P_{\rm \Lambda CDM}^{\rm HMcode2020}(k,z)}\,.
\end{equation}
EE2 is $\sim 1\%$ accurate for $k\leq 10~h/{\rm Mpc}$ for \lcdm{} cosmologies. The \hmcode{} boost was found to have an accuracy of $\sim 2\%$ for a range of gravitational and dark energy theories including massive neutrinos, for $k\leq 1~h/{\rm Mpc}$ \citep{Bose:2021mkz}. This gives us an estimated accuracy of $3-4\%$ for $k\leq 1 h/{\rm Mpc}$. This degrades to $\sim 6\%$ for $k\leq 3~h/{\rm Mpc}$ \citep{Bose:2021mkz}. From these references we estimate an accuracy of $5\%$ for  $k\leq 2.5~h/{\rm Mpc}$, which will be used in \autoref{sec:results}. 

In this work we consider both \autoref{eq:nlpkhmc} and \autoref{eq:nlnpkee2} to train our network, and importantly use both of these predictions to calibrate the theoretical error assumed in the network's predictions. The effects of massive neutrinos and beyond-\lcdm{} physics are primarily included in $\mathcal{R}$, but also in the linear spectrum that goes into  $P_{\rm pseudo}$. We also look to include the effects of baryonic physics, which is known to greatly affect the matter power spectrum at nonlinear scales \citep{Schneider:2018pfw,Schneider:2019xpf,Schneider:2019snl}. This can now easily be included via $P_{\rm pseudo}^{\rm HMcode2020}$ through the single-parameter baryonic feedback modelling available within \hmcode{} \citep{Mead:2020vgs}. This is a less comprehensive modelling of feedback processes than the more sophisticated emulators such as those of \cite{Arico:2020lhq,Giri:2021qin}, but will serve as a first test of our network's capacity to distinguish between baryonic physics, massive neutrinos and non-standard physics. We leave more sophisticated feedback modelling to a future work.

\subsection{Data Sets}\label{sec:data}

We will consider 5 classes of cosmological and gravitational models, following \cite{Mancarella:2020jyu}:
\begin{enumerate}
    \item 
    \lcdm{}, which assumes general relativity as the underlying gravitational model. 
    \item 
    The Hu-Sawicki $f(R)$ gravity model \citep{Hu:2007nk}, parametrised by the value of the additional scalar field today, $f_{\rm R0}$. This model exhibits the Chameleon screening mechanism \citep{Khoury:2003rn}. We assume a \lcdm{} background for this model. 
    \item 
    The normal branch of the Dvali-Gabadadze-Porrati (DGP) brane-world model \citep{Dvali:2000hr}, parametrised by $\Omega_{\rm rc} = 1/(4 r_c^2 H_0^2)$, where $H_0$ is the Hubble constant and $r_c$ is a scale governing where gravitational interactions begin to dilute into the 5-dimensional bulk. This model exhibits the Vainshtein screening mechanism \citep{Vainshtein:1972sx}. We assume a \lcdm{} background for this model.
    \item 
    An evolving dark energy model as parameterised in \cite{Chevallier:2000qy,Linder:2002et} ($w$CDM), with the parameter pair $\{ w_0, w_a\}$ giving the value of the dark energy equation of state today and its time evolution respectively, where $w(a) = w_0 + (1-a)\,w_a$, $a$ being the scale factor of the FLRW metric. Here the background is distinct from \lcdm{}. 
    \item 
    A random class as considered in \cite{Mancarella:2020jyu}, but with slightly different settings as described in Sec. \ref{app:randoms}. This class aims to capture any unknown/unconsidered models of gravity or dark energy whose phenomenology is largely distinct from the other classes. The new features are correlated in their redshifts and scales instead of being completely random. Hence these spectra can be interpreted as the `other' class that represents feasible other models with new physics.
\end{enumerate}
For each of these scenarios we consider different sets of baseline \lcdm{} cosmological parameters, $\{\Omega_{\rm m}, \Omega_{\rm b}, H_0, n_s, A_s \}$ - the total matter fraction, the total baryonic matter fraction, the Hubble constant, the spectral index and the primordial amplitude of perturbations, respectively.  In addition to these, we also include the effects of massive neutrinos, parametrised by the sum of the neutrino masses $\sum m_\nu$ as well as baryonic feedback effects, parametrised by the $T_{\rm AGN}$ parameter of \hmcode{}.

We sample these parameters to generate large sets of power spectra for each scenario. The parameters are sampled from Gaussian distributions with the following means and standard deviations: 
\begin{description}
\item 
$\{\Omega_{\rm m}, \Omega_{\rm b}, H_0, n_s, A_s \}$  are sampled using the Planck 2018 \citep{Planck:2018vyg} best fits as a mean and we take the standard deviation forecasted for the combination of weak lensing and spectroscopic clustering probes of the Euclid mission \citep{Blanchard:2019oqi}. When using \autoref{eq:nlnpkee2} we also impose the hard EE2 priors, which particularly restrict $\Omega_{\rm b} \in [0.04,0.06]$.  
\item 
$\{ \Omega_{\rm rc}, f_{\rm R0}\}$ are sampled with their \lcdm{}-limits as a mean ($\Omega_{\rm rc} = f_{\rm R0} = 0$, though the mean itself is not part of the data set) and a standard deviation taken to be the $3\sigma$ cosmic shear constraint forecasted for an LSST-like survey in \cite{Bose:2020wch} using only linear scales. 
\item 
$\{w_0,w_a \}$ are sampled using their \lcdm{}-limits ($\{w_0,w_a \} = \{-1,0\}$) and the standard deviation is taken to be the value forecasted for the combination of weak lensing and spectroscopic clustering probes of the Euclid mission \citep{Blanchard:2019oqi}. We also impose the following hard limits to ensure stability of the \react{} code, $w_0 \in [-1.3,-0.7]$ and $w_a \in [-1.5,0.3] $.
\item 
The $\sum m_\nu$ parameter is taken to have the same standard deviation and mean as the fiducial value assumed in \cite{Blanchard:2019oqi}, a lower bound estimate based on observations from neutrino oscillation experiments \citep{Esteban:2020cvm}. In parameter files we quote the massive neutrino energy density fraction today, $\Omega_\nu$, with $\sum m_\nu = \Omega_\nu h^2 
 \, 93.14 ~{\rm eV}$.
\item 
The baryonic parameter $\log_{10}[T_{\rm AGN}]$ is taken to have the default mean value of \hmcode{}, 7.8, and a standard deviation covering the fits to the BAHAMAS simulations \citep{McCarthy:2016mry} given in \cite{Mead:2020vgs}.
\end{description}
These choices aim to give an estimate of what the BNN can achieve given the data from forthcoming galaxy surveys such as Euclid and VRO/LSST, while remaining consistent with the Planck CMB observations. The parameter ranges are summarised in \autoref{tab:bac2data}. We demonstrate the effects of dynamical dark energy, modified gravity, massive neutrinos and baryonic feedback on the matter power spectrum in  \autoref{app:priors}. \autoref{fig:mg_params} shows the characteristic changes of $P_\mathrm{NL}$ when the model parameter is varied, keeping the baseline cosmological parameters constant.

We generate power spectra for all these scenarios using the following codes (see \autoref{eq:nlpkhmc} and \autoref{eq:nlnpkee2}):
\begin{enumerate}
    \item 
    $\mathcal{R}$ is computed using \react{} \href{https://github.com/nebblu/ACTio-ReACTio}{\faicon{github}}. 
    \item 
    $P_{\rm pseudo}^{\rm HMcode2020}$, $B^{\rm HMcode2020}$ and the baryonic boost are computed using \hmcode{} \href{https://github.com/alexander-mead/HMcode}{\faicon{github}}.
    \item 
    $P_{\rm \Lambda CDM}^{\rm EE2}$ is computed using EE2 \href{https://github.com/miknab/EuclidEmulator2}{\faicon{github}}. 
    \item 
    The modified linear spectra with massive neutrinos are computed using \texttt{MGCAMB} \href{https://github.com/sfu-cosmo/MGCAMB}{\faicon{github}} for the \lcdm{}, $f(R)$ and $w$CDM scenarios. 
    \item 
    The DGP linear spectra were generated using a private, modified version of \texttt{CLASS} \citep{Lesgourgues:2011re,Blas:2011rf} which was also employed in \cite{Euclid:2023rjj}. 
\end{enumerate}
On the data repository \href{https://zenodo.org/records/10688282}{here} we include the pipelines used to generate spectra for all scenarios and using both \autoref{eq:nlpkhmc} and \autoref{eq:nlnpkee2}.  

Each power spectrum data file is generated using a parameter set sampled from the ranges detailed in \autoref{tab:bac2data}. These data files, used for both training and testing of the network, have a shape of $5$ columns $\times$  $500$ rows. The first column is the values of the Fourier mode sampled in $h/{\rm Mpc}$, where we sample logarithmically in the range $[0.01,10]~h/{\rm Mpc}$. The 2nd to 5th columns are the values of the power spectrum at those Fourier modes, with each column corresponding to the following redshifts $z\in\{1.5, 0.785, 0.478, 0.1\}$. These redshifts are chosen to roughly sample the range of tomographic bins Euclid's weak lensing survey will be making measurements at \citep{2011arXiv1110.3193L,Euclid:2019clj}, with an omission of the highest bins which can go beyond $z=2$. We do not expect strong beyond-\lcdm{} effects at high redshift, but aim to consider this in a more comprehensive future iteration of the network where we train directly on the observables and not the matter power spectrum.  

\vspace{\baselineskip}

The complete list of data sets available \href{https://zenodo.org/records/10688282}{here} is summarised in \autoref{tab:datasets}.

\begin{table*}
\centering
\caption{Parameter ranges for \bacon{} data. These are sampled assuming a Gaussian distribution with means and standard deviations given below. The fiducial (mean) cosmology (without baryons but with massive neutrinos) gives a $\sigma_8(z=0)=0.812$, which is the Planck 2018 best fit. We normalise with a \lcdm{} spectrum with $\sum m_\nu =0$, where we use $A_s=2.025 \times 10^{-9}$ to get $\sigma_8 = 0.812$. Hard limits are placed on $w_0$ and $w_a$ as described in the main text to ensure stability of \react{}. When using EE2, $\Omega_{\rm b}$ is also restricted to the emulator range of $[0.04,0.06]$.}
\begin{tabularx}{\textwidth}{| c | c | c | X |} \hline 
 {\bf Parameter} & {\bf Mean} & {\bf Std. Dev.} & {\bf Reference \& Notes} \\ \hline 
  $\Omega_{\rm m}$ & 0.3158 & 0.009& Table.~1 of \cite{Planck:2018vyg} (Plik, best fit) \& Table.~9  of \cite{Euclid:2019clj} (${\rm GC}_s+{\rm WL}$, pessimistic) \\ 
 $\Omega_{\rm b}$ & 0.0494 & 0.016 &  \textbar \,  We use $\sigma_{\rm Planck}/2$ as large $\Omega_{\rm b}$ leads to computational issues in \react{}. \\ 
 $H_0$ [km/s/Mpc] & 67.32 & 0.41 &  \textbar \\ 
 $n_{ s}$ & 0.966 & 0.007 &  \textbar  \\ 
   $A_{ s}$ & $2.199\times 10^{-9}$ & $2.199\times 10^{-11}$ & \textbar \, Mean cosmology corresponds to $\sigma_8=0.812$. We convert $\%$ error on $\sigma_8$ to $A_s$. \\ \hline 
 $|f_{\rm R0}|$ & $10^{-10}$ & $10^{-5.5}$ & \lcdm{}-limit mean and $3\sigma$, $\ell_{\rm max}=500$ from Table.~2  of \cite{Bose:2020wch} \\
 $\Omega_{\rm rc}$ & $10^{-10}$ & 0.173 & \lcdm{}-limit mean and  $3\sigma$, $\ell_{\rm max}=500$ taken from chains of  \cite{Bose:2020wch} \\
   $\{w_0,w_a\}$ & $\{-1,0\}$ & $\{ 0.097 , 0.32 \}$ &  \lcdm{}-limit mean and \& Table.~11 of \cite{Euclid:2019clj} (${\rm GC}_s+{\rm WL}$, pessimistic) \\  \hline 
     $\sum m_\nu$ [eV] & 0.06 & 0.06& Fiducial of \cite{Euclid:2019clj} and take same for Std. Dev. \\ 
    $\log_{10}[T_{\rm AGN}]$ & 7.8 & 0.2  & Range of BAHAMAS simulation from Table.~4 of \cite{Mead:2020vgs}.  \\ \hline
\end{tabularx}
\label{tab:bac2data}
\end{table*}

\begin{table*}
\centering
\caption{Data sets available (\href{https://zenodo.org/records/10688282}{here}) and used in this work. }
\begin{tabular}{ | c | c | c | c |  } \hline 
 {\bf P(k) prescription} & {\bf Classes} & {\bf Set size} & {\bf Notes} \\ \hline 
\hmcode-based & All & 20,000 & Main training set \\ 
\hmcode-based  & All & 1,000 & Main test set  \\ \hline 
EE2-based & All & 20,000 & Main training set \\ 
EE2-based  & All & 1,000 & Main test set  \\  
EE2-based  & \lcdm{} & 20,000 & No baryons or massive neutrinos  \\ 
EE2-based  & All & 1,000 & No baryons or massive neutrinos  \\ \hline
halofit-based & All & 19,900 & No baryons or massive neutrinos, used in \cite{Mancarella:2020jyu}  \\ \hline

\end{tabular}
\label{tab:datasets}
\end{table*}

\subsection{Network Architecture and Training Methods}
\label{sec:architecture}
The key question behind \texttt{BaCoN} is whether we can design a neural network that can recognise features of different cosmological models in the matter power spectrum and distinguish them from \lcdm{}. This requires 3 main ingredients: layers that perform the feature analysis of the data, components that regularise the training and ensure a well-calibrated model, and a mathematical model for the weight distributions and their optimisation in the training process.

\begin{figure}
    \centering
        \includegraphics[width=0.8\columnwidth]{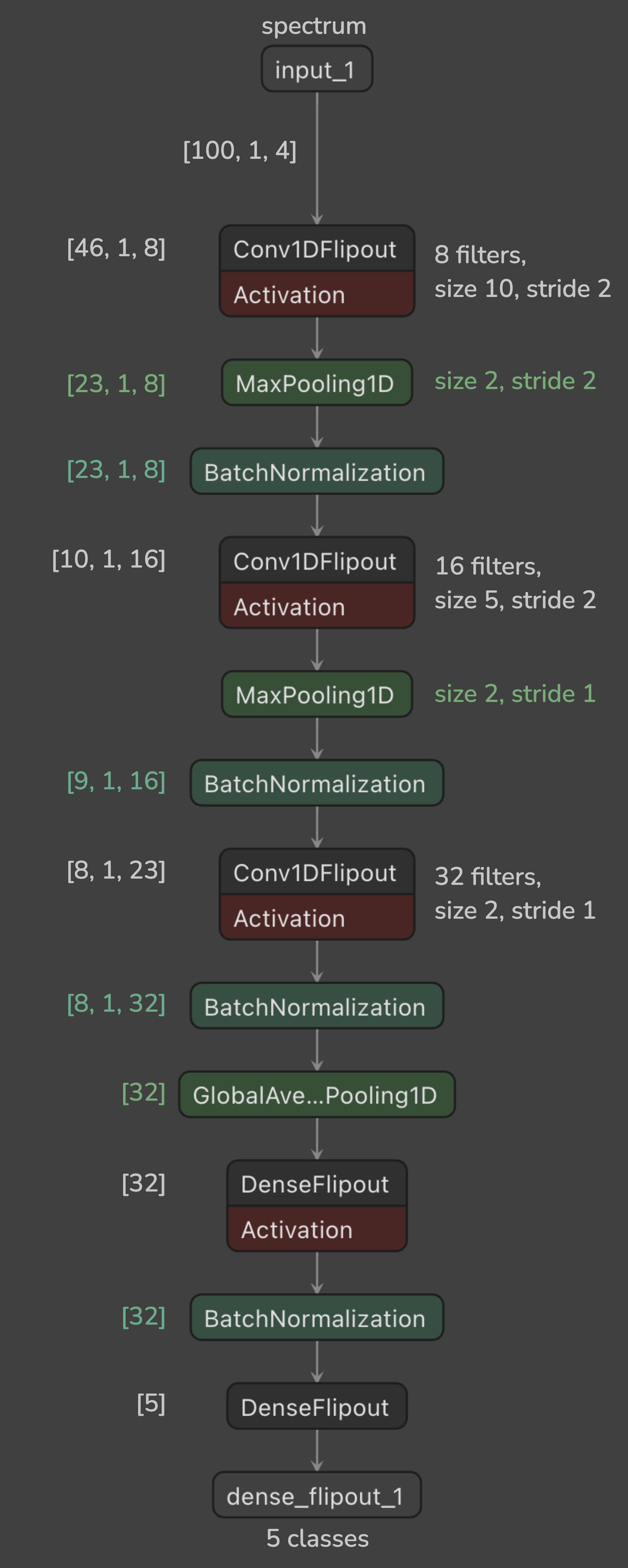}
    \caption{\textbf{Architecture of \bacon{}:} Schematic representation of network layers for the classification of a power spectrum with 100 $k$-bins and 4 $z$-bins into 5 classes. The output dimension for each layer is noted on the left side. Additional details for the 1D-Convolutional layers (filter number, kernel size and stride) and for the 1D-MaxPooling layers (pooling size and stride) are shown on the right side. Schematic created with \href{https://netron.app/}{Netron}.}
    \label{fig:architecture}
\end{figure}
We start with the first point, the core of \bacon{}: 1D-convolutional layers. They extract features from the spectrum by convolving it with a set of smaller matrices, called convolutional filters or kernels. We choose one-dimensional layers to treat the four redshift bins separately and then slide the filters along the direction of the 100 $k$ bins. The activation of these kernels builds feature maps in the next layer of the network. The dimensions of these layers depend on the number and size of the filters. We show the output dimensions of every layer in \autoref{fig:architecture} next to  the architecture of \bacon{}. We are building onto the code by \cite{Mancarella:2020jyu} using \textit{TensorFlow-2.10.0} \citep{tensorflow2015-whitepaper}. The full code is available at \href{https://github.com/cosmicLinux/BaCoN-II}{\faicon{github}}.

 Apart from our image-analysing layer we need methods that reduce the dimensionality and renormalise the layer's output. We use 1D max pooling layers and batch normalisation layers for this, respectively. A global average pooling layer is used to flatten the feature maps into a dense layer with 32 nodes. Finally, a softmax layer with 5 nodes produces the classification output.

 In total, the architecture of \bacon{}, shown in \autoref{fig:architecture}, consists of 6605 trainable parameters. They are tuned by passing training data through the network and  minimising a loss function with an optimisation algorithm. A backpropagation algorithm finds the minimum of the derivative of the loss to update the parameters and match the network's output closer to the training labels. This method is called gradient descent. Here, we are using the \textit{adam} optimisation algorithm \citep{kingma2017adam}. We also have to set a learning rate (lr) that dictates the size of the steps in the direction of the gradient. It changes after every full pass of all training samples through the network, called an epoch $e$. For $N_\mathrm{T}$ samples in the training set we get the learning rate 
 \begin{equation}
    \text{lr}(e) = 0.01 \times 0.95^{(e/N_\mathrm{T})} \, .
     \label{eq:learning rate}
 \end{equation}
In a BNN, the single values for weights in a traditional deterministic layer are replaced by parameterised distributions \citep{MacKay, Neal}. The goal of training the network is to find the posterior distributions $p(w | \mathcal{D})$  of the network's weights $w$ given the training data $\mathcal{D}$ starting from a prior distribution of the weights $p(w)$.
A common way to find an approximation for the posterior weight distributions is variational inference
\citep{ Jordan1999, NIPS2011_4329, Charnock-2020-BayesianNeuralNetworks}. 
It uses a family of distributions $q_{\theta}(w)$ parameterised by the parameter $\theta$ that are closely matched to the posterior $p(w | \mathcal{D})$ in the training process. One way to achieve this is by minimising the Kullback-Leibler divergence term (KL) \citep{kullback1951}. 
Our loss function is constructed from a negative log-likelihood term in combination with a KL-divergence between the variational distribution and the prior that acts as a regularisation term.

Unfortunately, computing the gradients for variational inference is difficult. A reparameterisation trick \citep{Kingma2013, Kingma2015} can be used to perform backpropagation but then the gradients of a batch become correlated. \citet{wen2018flipout} have developed the Flipout technique to decorrelate the weights in a mini-batch. Flipout has become widely popular for the use in BNNs and has been implemented in TensorFlow Probability \citep{dillon2017tensorflow}. We are using the implementation of the Flipout technique in \textit{TensorFlow-Probability-0.14.0} in all convolutional and dense layers pictured in \autoref{fig:architecture}.  
In comparison with other established BNN training methods, Flipout has been shown to be one of the higher performing variational inference methods \citep{Hortua-2020-ParameterEstimationCosmic}. 
Other works have implemented alternatives to the Gaussian variational posterior $q_{\theta}(w)$ for a further improved calibration  \citep{Hortua-2023-ConstrainingCosmologicalParameters}. 
Therefore, it is important to test whether the network architecture in combination with the training data leads to a well-calibrated network. We will present the results of the calibration test for \bacon{} in \autoref{sec:calibration}.

To evaluate the training progress we are using a dedicated validation set. After every epoch we compute the classification accuracy for the spectra in the validation set to see how the training progresses. The training is finished when the validation accuracy converges to a stable configuration and the loss stabilises at a minimum.
The next section will explain the details of the practical procedures for training and testing the network.

%%%%%%%%%%%%%%%%%%%%%% Training ###################

\section{Training the Network}
\label{sec:training}

\begin{figure*}
\centering
\includegraphics[width=0.78\textwidth]{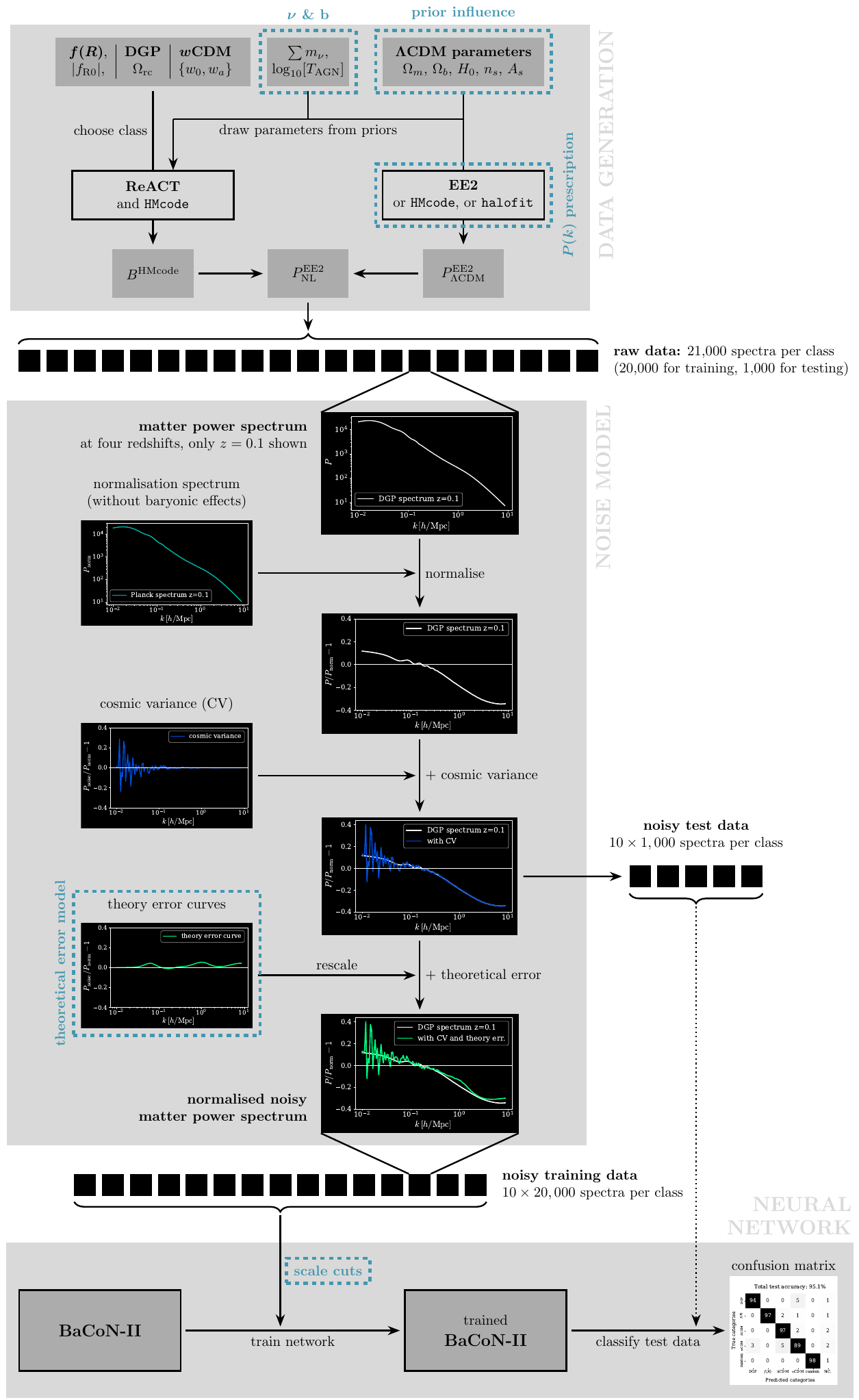}
\caption{Flowchart of the full pipeline from data generation to the final classification result. The noise model is demonstrated on a DGP spectrum. See  \autoref{sec:training} for a detailed explanation.}
\label{fig:data-flowchart-tikz}
\end{figure*}

The whole process from data generation to training and testing the classification network is displayed as a scheme in \autoref{fig:data-flowchart-tikz}. We will describe all the involved stages in this section. Some steps are marked with blue boxes in the graphic and are discussed in the following subsections. We will go through the specific influences of these marked parts in \autoref{sec:results}. 

\subsection{Data Generation}

The theoretical background of the data generation has been described in  \autoref{sec:nl-hmrapproach}. Here, we will only address the practical choices. We start by drawing cosmological parameters from the priors for \lcdm{}. This set of parameters is then fed into a code to produce a nonlinear \lcdm{} matter power spectrum (either EE2, or \hmcode{}, or \halofit{}). Then we choose the cosmological class that we want to generate matter power spectra for. This can be $f(R)$, DGP, $w$CDM or \lcdm{} and draw from the class-specific parameter distribution. If we want to include effects of massive neutrinos and baryonic feedback, we include their respective parameters as well. All parameters are then passed on to \react{} and \hmcode{} to calculate the boost. Finally, this is combined with the \lcdm{} spectrum to get a nonlinear matter power spectrum for the selected class.

\subsection{Noise Model}

We normalise the power spectra data with a generic matter power spectrum to reduce the dynamic range of the data presented to the \bacon{} network.
Throughout this work, we use the same \lcdm{} normalisation spectrum. We have tested the influence of different normalisation spectra including a linear matter power spectrum without Baryonic Acoustic Oscillations (BAOs). There is no noticeable effect as long as the same normalisation is used for the training and the testing phases. We set the default normalisation spectrum as an EE2 \lcdm{} spectrum with the Planck cosmology \citep{Planck:2018vyg} without baryonic feedback or massive neutrinos.

\begin{figure}
    \centering
        \includegraphics[width=\columnwidth]{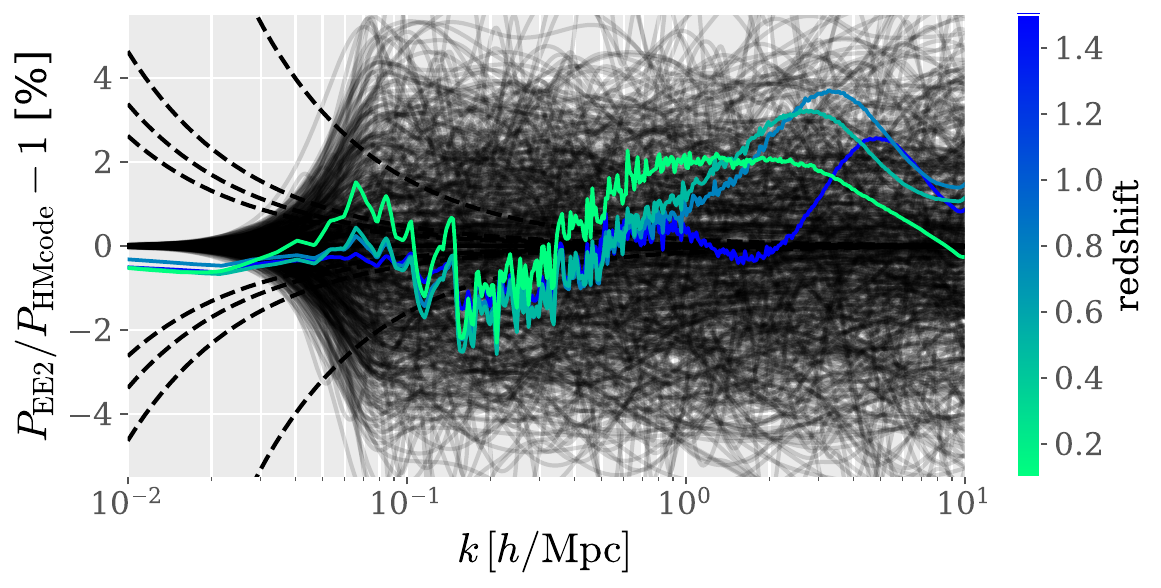}
    \caption{\textbf{Effects of the power spectrum prescription and the theoretical error model:} Ratio of EE2 and \hmcode{} \lcdm{} spectra at different redshifts for the same cosmological parameters shown as coloured lines. The dashed lines show the $1\, \sigma$ level of cosmic variance (see \autoref{eq:Euclid_error}) at the 4 redshifts considered in this work and the black lines are 600 possible theory error curves scaled to $5\%$. Note the noise in the EE2-\hmcode{} ratio is noise inherent in the EE2 spectra emulation.}
    \label{fig:ratio_EEHM_noise}
\end{figure}

For every normalised spectrum we produce 10 realisations sampled from our {\it noise model}. This model has two components:

\begin{description}
    \item 
The first component represents noise coming from a Stage IV-like survey's cosmic variance. This component is given by
\begin{equation}
    \sigma_p(k) = \sqrt{\frac{4 \pi^2}{k^2 \Delta k V(z)}} \times P(k)   \, ,
    \label{eq:Euclid_error}
\end{equation}
where $V(z)$ is the volume probed at a given redshift and $\Delta k$ is the bin-width. The Stage IV-like volumes we adopt are: $V(1.5) =10.43~{\rm Gpc^3} / h^{3} $, $V(0.785) = 6.27~{\rm Gpc^3}/ h^{3} $, $V(0.478)=3.34~{\rm Gpc^3}/ h^{3} $ and $V(0.1) = 0.283~{\rm Gpc^3} / h^{3} $ \citep{2011arXiv1110.3193L,Euclid:2019clj}.
We do not consider shot noise as we are using dark matter spectra, whose associated particle density is very large. 

\item 
The second component is used to model the theoretical uncertainty in our power spectra predictions.
As stated in \autoref{sec:nl-hmrapproach}, all terms used in the power spectra model have associated inaccuracies that contribute to the the overall theoretical error of our nonlinear matter power spectra predictions. This becomes very apparent when we look at the difference of power spectra predictions based on EE2 and \hmcode{} as displayed in \autoref{fig:ratio_EEHM_noise}. Both codes produce characteristic fingerprints in the spectrum and deviate from each other by up to 4\%.  To avoid a fitting of the neural network to these prescription-specific errors, we generate curves with similar features to model such theoretical errors in the training process. The development of our theoretical error models will be discussed in detail in \autoref{sec:theoryerr}. These theory error curves are rescaled with a factor which we treat as a parameter to vary, and which accounts for potential errors of different amplitudes.
\end{description}

Note that while we add cosmic variance to the test data before passing it through the network, we do not add the theoretical error component. This is because the test data should be treated the same way as observational data that will be passed through the network for classification. 

\subsection{Training and Testing the Neural Network}

We train the BNN with the noisy training data. The shape of the data can be adapted at this stage, \textit{e.g.} with scale cuts.  We employ a training data set of 20,000 spectra per class (of which 15\% are used as a validation set) to train our most accurate models. 
The architecture of the BNN, hyperparameters, loss function etc. remain unchanged from the initial \texttt{BaCoN} by \cite{Mancarella:2020jyu} and are described in detail in \autoref{sec:architecture}. We train the network for 50 epochs or until the training accuracy has become stationary. This is the most time consuming part of the machine learning process. The list of data sets used for training and testing is displayed in \autoref{tab:datasets} and has been made publicly available \href{https://zenodo.org/records/10688282}{here}.

Once we have trained  the network, we  test it with the noisy test data. As in \cite{Mancarella:2020jyu}, every example requires a classification probability of 50\% or above to be considered as classified. Otherwise we count it as not classified (N.C.). 
In total we use 1000 test data spectra and pass them through the network with 10 noise realisations each. To make full use of the stochastic features of \bacon{}, we take the average of 100 MC samples per every noisy sample for the confusion matrix. So we technically pass every noisy spectrum 100 times through the network and sample from the weight distributions. The final classification result for every test spectrum is an average of the classification outputs from all MC samples. Hence, the confusion matrix is a generalised representation of the performance of the trained network marginalised over the weights. 
When rerunning the test code for the same trained model and test data, the average over the accuracy of 100 MC samples changes overall by less than 1\%. Therefore, we do not state errors in the confusion matrices.

We conclude this overview with a list of all changes in our methodology in comparison to \citet{Mancarella:2020jyu}:
\begin{enumerate}
    \item Improved physics in the data generation (massive neutrinos and baryonic effects).
    \item Codes with higher accuracy for the \lcdm{} spectra computation (EE2 and \hmcode{} instead of \halofit{}).
    \item Improved random spectra (see \autoref{app:randoms}).
    \item Different model for the theoretical error and no shot noise term is included in training.
    \item We only add the theoretical error to the training data and not to the test data.
\end{enumerate}

We will present the theoretical error model, results for our baseline model and the influence of the blue marked components in \autoref{fig:data-flowchart-tikz} in \autoref{sec:results}.

\subsection{Calibration}
\label{sec:calibration}

\begin{figure*}
    \centering
    \includegraphics[width=0.95\textwidth]{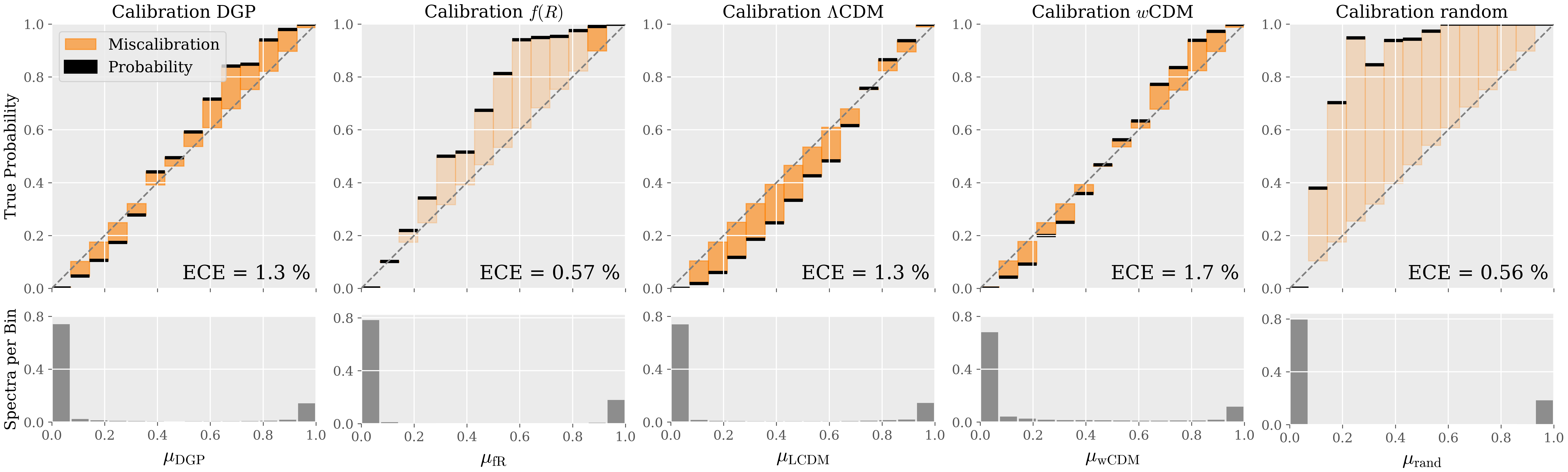}
    \caption{\textbf{Calibration of the baseline model:} The bins represent the difference between the predicted probability $\mu_i$ of a class $i$ compared to the true probability given by the True-False number ratio of test spectra for this class in this $\mu$ bin.The Expected Calibration Error (ECE) defined in \autoref{eq:SCE} is displayed for every class individually.
    The lower row shows the relative bin count (fraction of test spectra per bin to the total test size).
    If the bin contains less than 0.5 \% of the total number of test spectra it is marked with a fainter colour in the calibration diagram as it barely influences the overall calibration. 
    The classifications are made with the baseline model and baseline test data set (see \autoref{fig:cm-EE2-5perc-kmax25} for the full test result).
    }
    \label{fig:reliability-curve}
\end{figure*}

The final output of the softmax layer of the network sums to one and can be interpreted as probabilities. For this to be an reasonable interpretation, we need to ensure that the network is well calibrated. Especially since research on DNNs has suggested that modern networks tend to have miscalibration issues \citep{pmlr-v70-guo17a}.
We construct a calibration diagram of our baseline model to measure the reliability of the network's output. 
First, we run a test of our baseline model with the same test data and testing procedure as in our main result shown in \autoref{fig:cm-EE2-5perc-kmax25}. For consistency, we add 10 noise realisations to every test spectrum and the $\mu$ values are generated from 100 MC samples. 
Then, we bin the predictions $\mu_i$ for every class $i$ and compute the average $\hat{\mu}(b, i)$ in every bin $b$. We compare this to the true probability $P_\mathrm{true}(b,i)$ which is the frequency of the positive label for this class $i$ for the spectra in bin $b$. 
The calibration of every individual class is displayed in \autoref{fig:reliability-curve} following the design by \cite{Hollemans-2024-Reliabilitydiagrams}. A perfectly calibrated network would produce a diagonal line in the calibration diagram. We can see that most classes look well calibrated with the exception of $f(R)$ and the random class that show multiple bins that are strongly underconfident. This is actually not a calibration effect but a result of the low number of spectra in these bins caused by the high confidence for this classes and the lack of spectra with $\mu$ far from 0 or 1. Therefore, we've added a histogram with the bin counts for every class and mark bins with a fainter colour if they contain less than 0.5 \% of the total number of test spectra.

The population of bins is automatically accounted for in the Static Calibration Error (SCE), a measure for the overall miscalibration of a multiclass network. The SCE of BNNs is computed in the same way as for deterministic networks, with the only difference that we average $\mu$ over 100 MC samples for every spectra in the Bayesian network. We can calculate it as an averaged sum over the Expected Calibration Errors (ECE) of the individual classes following \cite{nixon2019measuring} as
\begin{align}\label{eq:SCE}
    \text{SCE} &=\frac{1}{N}\sum_{i=1}^{N} \sum_{b=1}^{B_i} \frac{n_{bi}}{N_{\text{tot}}} \left| P_\mathrm{true}(b,i)- \hat{\mu}(b, i)\right| \\ \nonumber
    &= \frac{1}{N}\sum_{i=1}^{N} \mathrm{ECE}_i\, ,
\end{align}
with the total number of classes $N$, the number of bins per class $B_i$ and the total number of test samples $N_\mathrm{tot}$.
The ECEs for every class are shown in \autoref{fig:reliability-curve}. The individual ECEs are all below 2 \% and combined give in a total SCE for \bacon{} of 1.1 \%. This shows that the network is overall well calibrated. Hence, we can reasonably use the output $\mu$ as an estimate for the probability. We will have a more detailed look at the properties and uncertainties of this output in the next section.

\subsection{Uncertainty Estimates}
\label{sec:uncertainty F}
We now present a test for the uncertainty of the \bacon{} classification. This provides an insight into the internal workings of the Bayesian network and reflects the advantages of using a Bayesian instead of a deterministic neural network.
First, we need to understand the statistics of the classification in a Bayesian network.  The output of the final softmax layer is a vector with one component $p_i$ for each class $i$ of a total of $N$ classes.\footnote{In this work we train \bacon{} for 5 classes, so $N=5$, but this can be adapted easily.} 
As they add to 1, the values of $p_i$ can be interpreted as the probabilities for each class, when the network is calibrated correctly (see \autoref{fig:reliability-curve}).
In a deterministic neural network we only get a single prediction for $p_i$, as the weights are fixed and repeating the analysis will result in the same output vector. Hence, we have no way to tell whether the classification result is reliable. A BNN on the other hand provides an intrinsic method to compute the uncertainty of the classification. When passing a spectrum through the trained network we Monte-Carlo sample from the distribution of the weights $w$. Multiple passes will result in different output probabilities $p_{i,\alpha}$, where we use the greek index $\alpha$ to indicate the MC sample. We can marginalise over the weights of the network by averaging over a number of $N_\mathrm{MC}$ MC samples. This gives an approximate value for the average output probability of the softmax layer
\begin{equation}
    \mu_{i} \approx  \frac{1}{N_\mathrm{MC}} \sum_{\alpha=1}^{N_\mathrm{MC}}  \, p_{i, \alpha}  \, .
    \label{eq:mu}
\end{equation}
We classify the example spectrum by assigning the label of the class with the highest MC average  $\mu_i$, if it exceeds a set threshold probability $p_\mathrm{th}$.
To get a confidence estimate, we need the probability distribution of these MC averages $\mu$, which we denote as $\mathcal{F}$. We construct this function by modelling a distribution for a random variable $x_i$ (representing the softmax output of the network) with mean $\mu_i$ and the full covariance of the classification $\Sigma$. The covariance of a trained network is computed following \cite{uncertainty-cov} and \cite{Mancarella:2020jyu} as
\begin{equation}
    \Sigma = \frac{1}{N_\mathrm{MC}}\sum_{\alpha=1}^{N_\mathrm{MC}} \, \left( \text{diag}(p_\alpha) -   p_\alpha^{\otimes 2}  \right) + \frac{1}{N_\mathrm{MC}}\sum_{\alpha=1}^{N_\mathrm{MC}} (p_\alpha-\mu)^{\otimes 2} \, , 
    \label{eq:covariance-sigma}
\end{equation}
with $y^{\otimes 2} = y y^T$.
The first term of the covariance represents the aleatoric uncertainty and the second term models the epistemic uncertainty.
The resulting output probability distribution is then given by
\begin{align}\label{eq:F-distr}
    \mathcal{F}(x; \mu, \Sigma) & = \,  \delta\Big( 1-\sum_{j=1}^N x_j \Big)  \times \sqrt{N} \\ \nonumber
    & \times \prod_{i=1}^{N-1} {\mathcal{\tilde{N}}} \Big( \left[{B}^{-1} (x-\mu)\right]_i; 0, \left[{B}^{-1} \Sigma{B}\right]_{ii} \Big) \, ,
\end{align}
where $\mathcal{\tilde{N}}$ is a multivariate Gaussian truncated between 0 and 1 and $B$ is the matrix that diagonalises $\Sigma$. The derivation of this form of $\mathcal{F}$ and how to construct it from the distribution of MC-sample probabilities is described in the appendix of \cite{Mancarella:2020jyu}.
Finally, to obtain a classification statement and label of the spectrum we need to impose a threshold probability $p_\mathrm{th}$, in our case $50\,\%$. The probability $P_i$ of labelling the spectrum as a class $i$ is determined from the fraction of samples that fulfil the criterion $x_i > p_\mathrm{th}$. This gives an estimate of the uncertainty for the classification result.

\begin{figure*}
    \centering
    \includegraphics[width=0.9\textwidth]{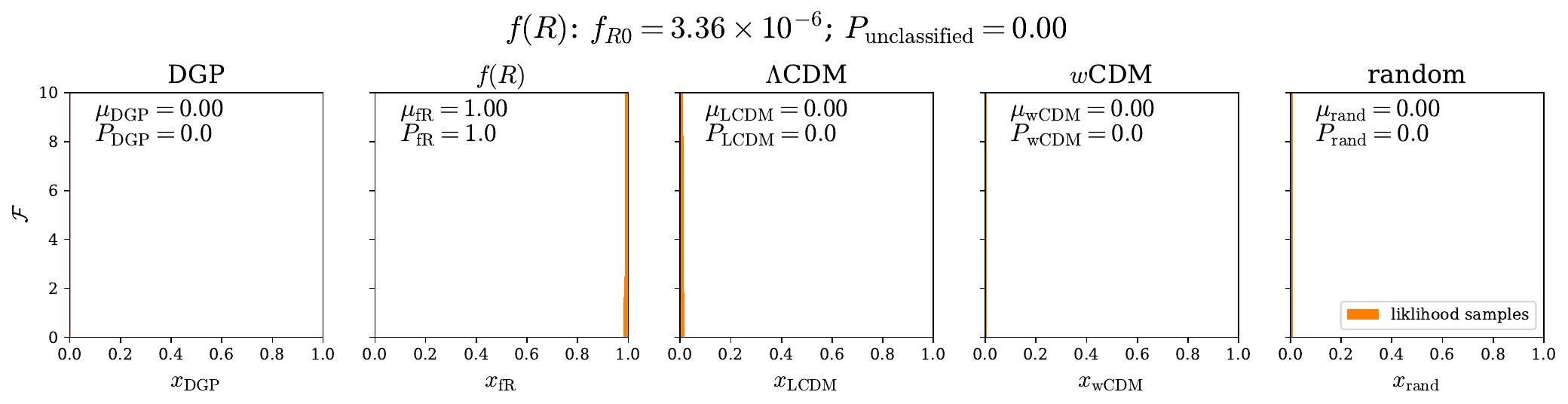}
    \includegraphics[width=0.9\textwidth]{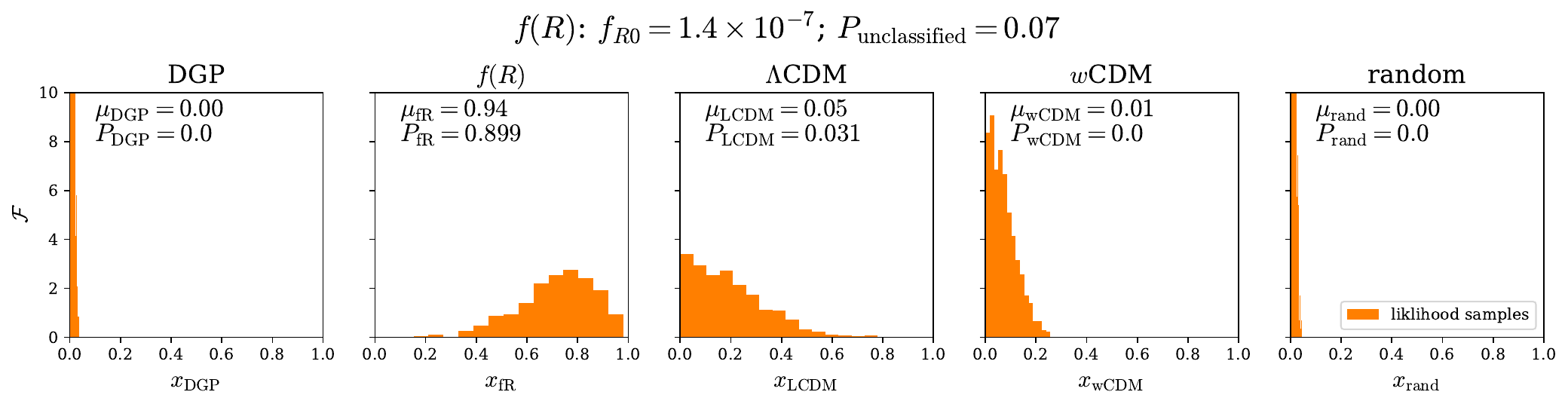}
    \includegraphics[width=0.9\textwidth]{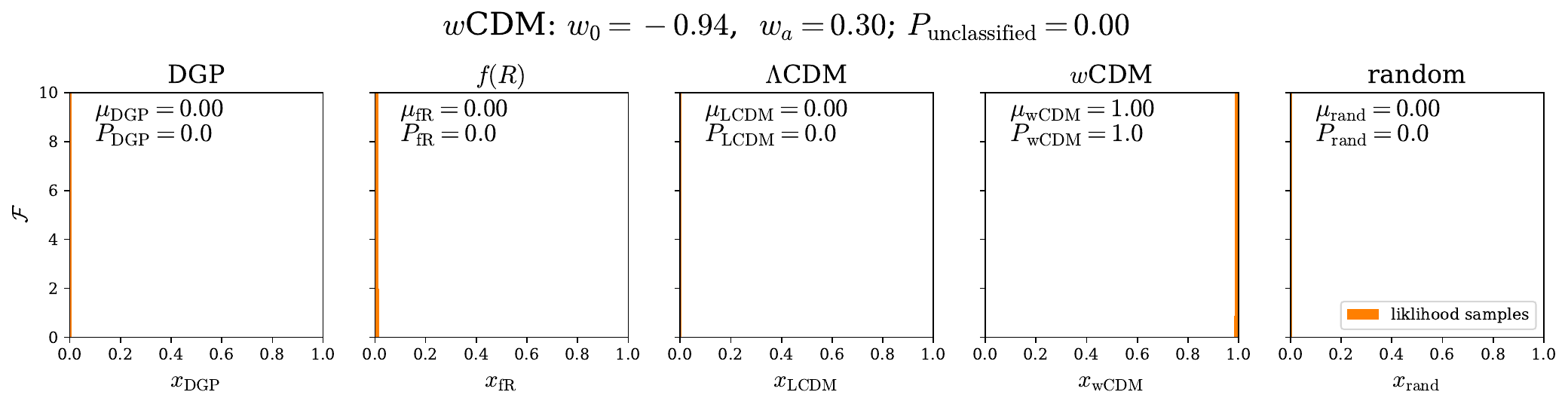}
    \includegraphics[width=0.9\textwidth]{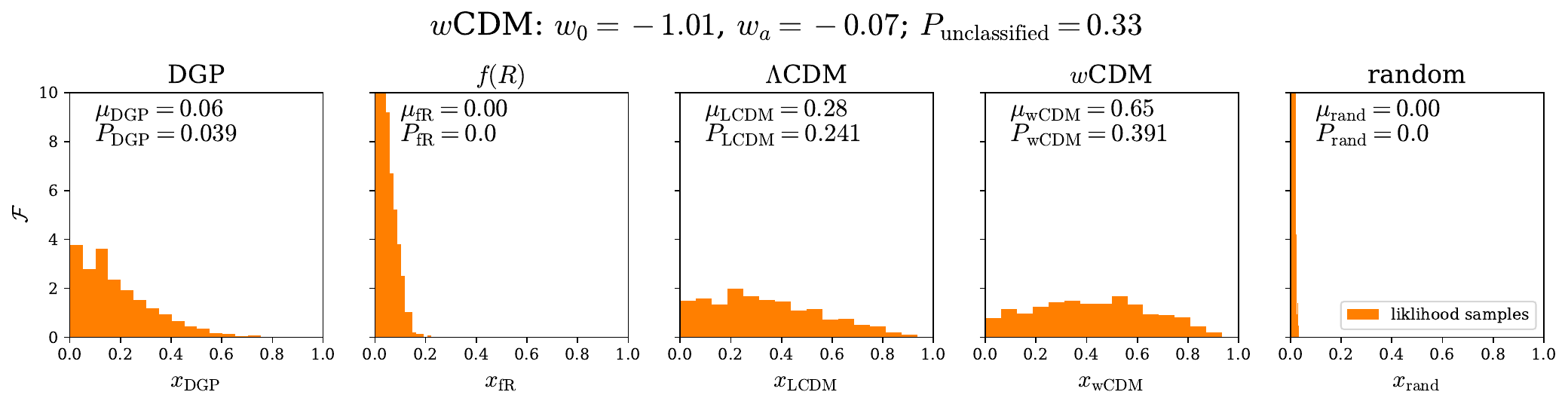}
    \caption{\textbf{Output probability distribution $\mathcal{F}$ for some example spectra:} $x_i$ represents the output of the network, that can be interpreted as the probability for the respective class $i$. $\mathcal{F}$ is the probability distribution of the MC averages for one specific power spectrum with a fixed noise realisation (cosmic variance only). $\mathcal{F}$ is computed with \autoref{eq:F-distr} based on the average $\mu_i$ and variance $\Sigma$ expected for 100 MC samples. We draw 1000 samples from $\mathcal{F}$ to display the distribution here.
    The classifications are made with the baseline EE2-based model (see \autoref{fig:cm-EE2-5perc-kmax25}) for two $f(R)$ and two $w$CDM spectra with less and more extreme class-specific parameters.
    }
    \label{fig:F-hist-uncertainty}
\end{figure*}

\autoref{fig:F-hist-uncertainty} displays the probability distribution $\mathcal{F}$ for a specific spectrum with one fixed noise realisation. We are adding the same noise that we are adding to test data, \textit{i.e.} cosmic variance at large scales.
The spectrum is passed through the network 100 times. We use these 100 MC samples to compute $\mu$ and $\Sigma$ and finally construct $\mathcal{F}$. The plots are created by drawing 1000 samples from the final distribution. We show the label probabilities $P_i$ for each class as well as $P_\mathrm{unclassified}$, when the output does not exceed $p_\mathrm{th}=0.5$ for any of the classes.

\autoref{fig:F-hist-uncertainty} shows $\mathcal{F}$ for four different spectra -- two are $f(R)$ and two are $w$CDM examples. We choose these two classes because, as we will later see in \autoref{fig:cm-EE2-5perc-kmax25}, $w$CDM shows a higher degeneracy with other classes while $f(R)$ gets classified very confidentially.
The two spectra have class parameters that produce a strong deviation from \lcdm{}. They are classified correctly $100\, \%$ of the time with an output probability of $\mu_i=1$ for the correct class throughout. This changes drastically when we look at spectra that have only a small modification compared to \lcdm{}. The $\mu_i$ distributions widen and we see higher chances of unclassified and wrongly classified results. This trend shows how a difficult detection due to small features leads to a low confidence of the classification that can be captured by plotting $\mathcal{F}$ and the label probabilities $P_i$. It is a measure for the uncertainty of the classification while marginalising over the weight distributions of the network. The $w$CDM spectrum in particular shows a high degeneracy with DGP, which we will discuss in more detail in \autoref{sec: classes results}. 
The distribution of $\mathcal{F}$ as plotted in \autoref{fig:F-hist-uncertainty} is a powerful tool to estimate the uncertainty for the classification of one specific spectrum. This is relevant for observational data where we only have a single spectrum with one noise realisation. An estimate of confidence in the classification of observational data is an essential component for the method to be useful. Further, if the spectrum is determined to be 'Not Classified', $\mathcal{F}$ will indicate if there are any classes that can be excluded safely and which classes to focus further analysis on.

%%%%%%%%%%%%%%%%%%%%%% Results ###################

\section{Results}
\label{sec:results}

 In this section, we will investigate how the classification result is affected by the various influences outlined in \autoref{sec:training} and \autoref{fig:data-flowchart-tikz}. We evaluate the network performance using the full confusion matrix, but for clarity we will often only state the total classification accuracy (true positive rate) in the figures. A note on reproducibility of the numbers in this section: The scatter of classification accuracies is very low. When retraining and retesting the network with data produced using the same power spectrum prescription, the total accuracy varies by less than 1\%. The variability of test results can go up to 4\% when testing on data produced using a different spectrum prescription than used in training. We expect this variability to decrease through the refinement of our theoretical error model, which should suppress the influence of the specific power spectrum prescription.

\subsection{Theoretical Error Model} \label{sec:theoryerr}

\begin{figure}
    \centering
        \includegraphics[width=\columnwidth]{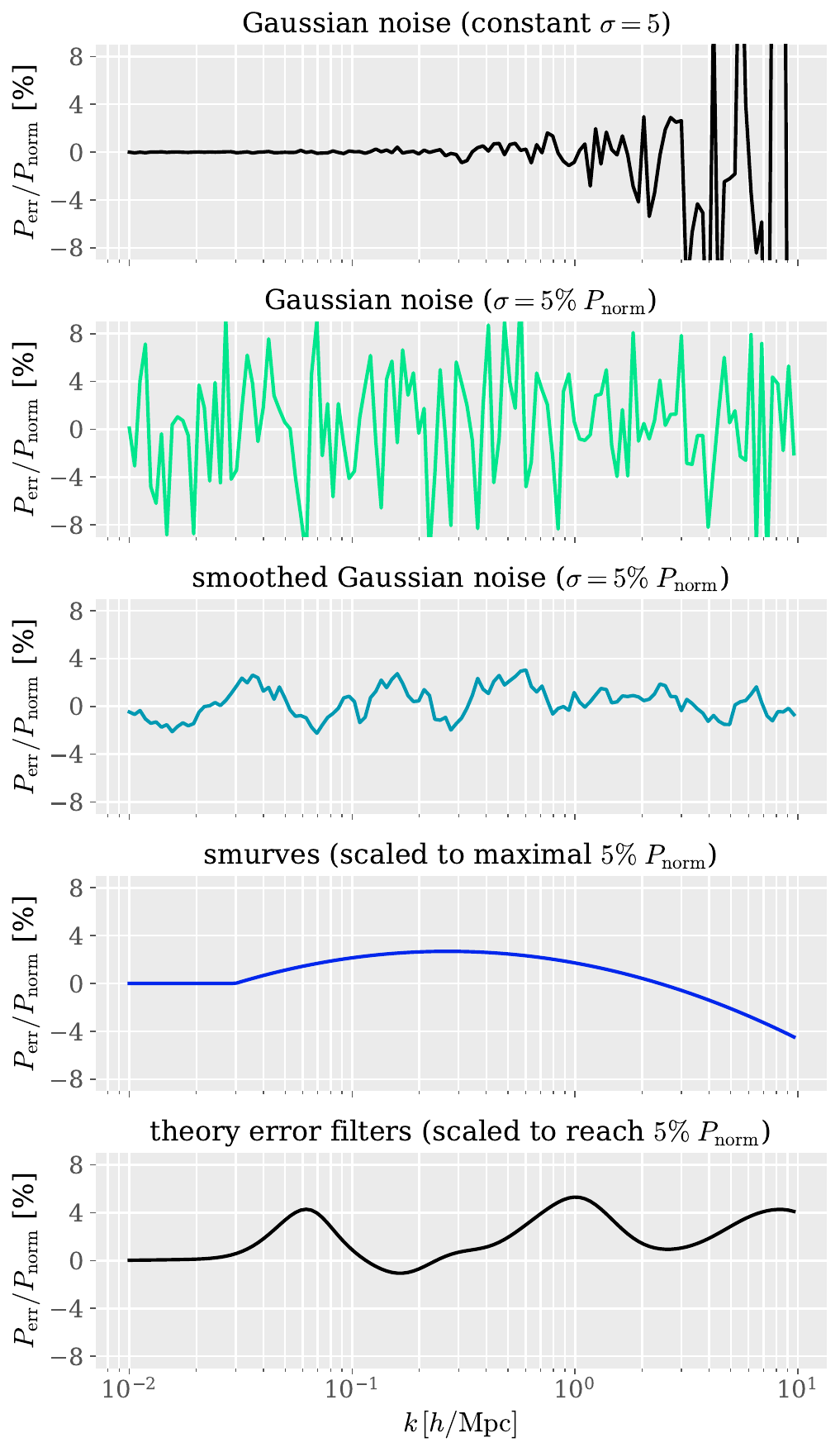}
    \caption{\textbf{Theoretical error models:} Development stages of curves added to the training data that account for the theoretical uncertainty in our training data power spectra predictions. All models are scaled to an equivalent of 5\% of a \lcdm{} Planck normalisation spectrum and are plotted after normalisation.}
    \label{fig:noise-sys-evolution}
\end{figure}

 Before we move to the main classification results for our baseline model, we will explain the shape of the curves that we use to model the theoretical error component in our noise model. This has gone through several stages of development and we show all of them as they provide insights into the nature of this error. \autoref{fig:noise-sys-evolution} shows the different shapes of the models we have considered, which are discussed shortly. Curves generated using these models are added to the noisy spectra before training as highlighted in \autoref{fig:data-flowchart-tikz}. We once again emphasise that they will only be added to the training data and not to the test data, as the latter is meant to represent actual observations.

To test the effectiveness of our theoretical error model, we trained networks on a reduced EE2 training set (1,000 spectra per class) and added the curves for the theoretical error scaled to different percentages. These networks were then tested with EE2 and \hmcode{} test data. As shown in  \autoref{fig:ratio_EEHM_noise}, there is a considerable systematic difference between EE2 and \hmcode{} matter power spectra - for $k_\mathrm{max} = 2.5 \;h/\mathrm{Mpc}$ this is estimated to be roughly 5\%. This difference is representative of the potential error when comparing our spectra prescription to the real Universe. If we model the theoretical error well enough,  the accuracies for the two test data sets should converge when the amplitude of the error in training is increased, without reducing the overall accuracy significantly.

In \citet{Mancarella:2020jyu} the authors employed a Gaussian noise similar to our treatment of cosmic variance, with a constant component associated with the theoretical uncertainty. The top panel in \autoref{fig:noise-sys-evolution} shows that after normalisation this leads to an underestimate of the error on BAO scales and an extremely large error at very small scales. We first attempted to correct this by scaling the constant contribution by the normalisation matter power spectrum. This gives us a Gaussian error of the same amplitude at all scales, which is shown in the second panel. 

\begin{figure}
    \centering
        \includegraphics[width=\columnwidth]{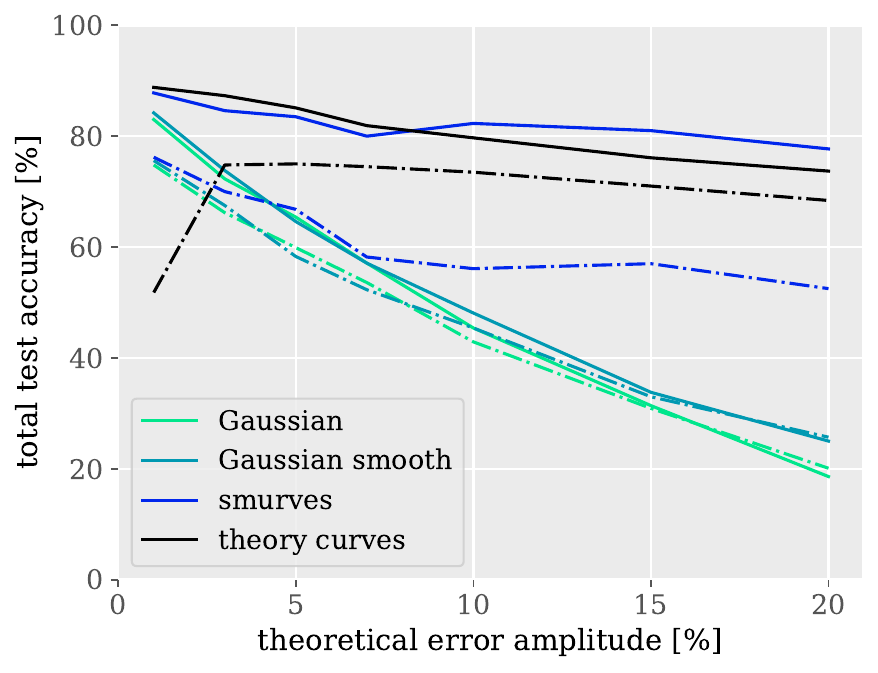}
    \caption{\textbf{Model accuracy under various noise models for the theoretical error:} Dependence of total test accuracy on the size of the theoretical error for different theory error  models. Representatives of the curves corresponding to each model are shown in \autoref{fig:noise-sys-evolution} with the same colour coding. All models have been trained with a subset of the main EE2 training data (1000 training spectra per class). The testing was conducted without adding a theoretical error, using EE2 (solid line) or \hmcode{} test data (dashed-dotted line). }
    \label{fig:sys-acc-noisemodels}
\end{figure}

In \autoref{fig:sys-acc-noisemodels} we show the test accuracies for this Gaussian error model as a function of the overall error amplitude (in this case the standard deviation of the Gaussian). We note the accuracy falls off drastically when we increase the standard deviation. We have found that this decrease does not happen when we  add the same noise to the test spectra. This suggests the dropping accuracies are resulting from the missing noise in the test data compared to the training data. The network has over-trained on the structure of the Gaussian noise and is not able to generalise to noise-free spectra. 

To try to mitigate this, we then smoothed the Guassian noise as shown in the middle panel of \autoref{fig:noise-sys-evolution}. However, the testing results plotted in \autoref{fig:sys-acc-noisemodels} are unaffected by the smoothing. The shapes introduced by this kind of noise still seem to be too distracting for the network. Different smoothing scales had no effect on this trend either.

To overcome this problem, we opted for a completely smooth error curve displayed in the second lowest panel of \autoref{fig:noise-sys-evolution}. These curves have been produced with a modification of the \texttt{smurves} package that was developed by \citet{moews2019stress} (\href{https://github.com/moews/smurves}{\faicon{github}}). We then scale these curves by an overall amplitude factor which we vary, as in the Gaussian error case. These parabola-like shapes have no substructure that the network could become over-fit to. The test accuracy for EE2 data remains high even for large error amplitudes. However, the \hmcode{} tests plotted in  \autoref{fig:sys-acc-noisemodels} have a large decrease in accuracy and are nowhere near convergence with the EE2 accuracies. We deduce that the smurve curves do not distract the network but on the other hand do not add enough features to capture the systematic error between EE2 and \hmcode{} spectra.

Finally, we test a combination of both approaches and produce smooth  curves with features to represent real theoretical uncertainties in the prescription of the matter power spectrum. These are generated using a slightly modified algorithm as used to generate the random class (see \autoref{app:randoms}). These theory error curves are shown in the bottom panel of \autoref{fig:noise-sys-evolution}. The EE2 test results in \autoref{fig:sys-acc-noisemodels} are  of similar high accuracy to the smurve model. There is a small decrease towards very high amplitudes as we would expect. The \hmcode{}-results now, for the first time show converging accuracy to the EE2 test set,  when adding more systematic noise. This indicates that the theory error curves are a step in the right direction, towards mimicking generalised theoretical uncertainties (as represented in this case by the difference between EE2 and \hmcode{} spectra).

In fact, we have plotted a number of  `theory error curves' in the background of the EE2-\hmcode{} ratio in \autoref{fig:ratio_EEHM_noise} for comparison. While the long-wavelength features in the EE2-\hmcode{} ratio itself are well represented, we can see that there are some short-wavelength, fine structure in the error that we have not captured. This is likely responsible for the remaining accuracy discrepancy of $\sim$ 5\% between the EE2 and \hmcode{} test results, present even when we add theory error curves that are scaled up to 20\% of the normalisation spectrum. However, the overall desired effect of the theoretical error component  has been achieved. 

\vspace{\baselineskip}

All the tests that we will discuss in the following sections have been made with networks trained with theory error curves as the theoretical error component to our noise model, added in to the training spectra.

\begin{figure}
    \centering
        \includegraphics[width=0.9\columnwidth]{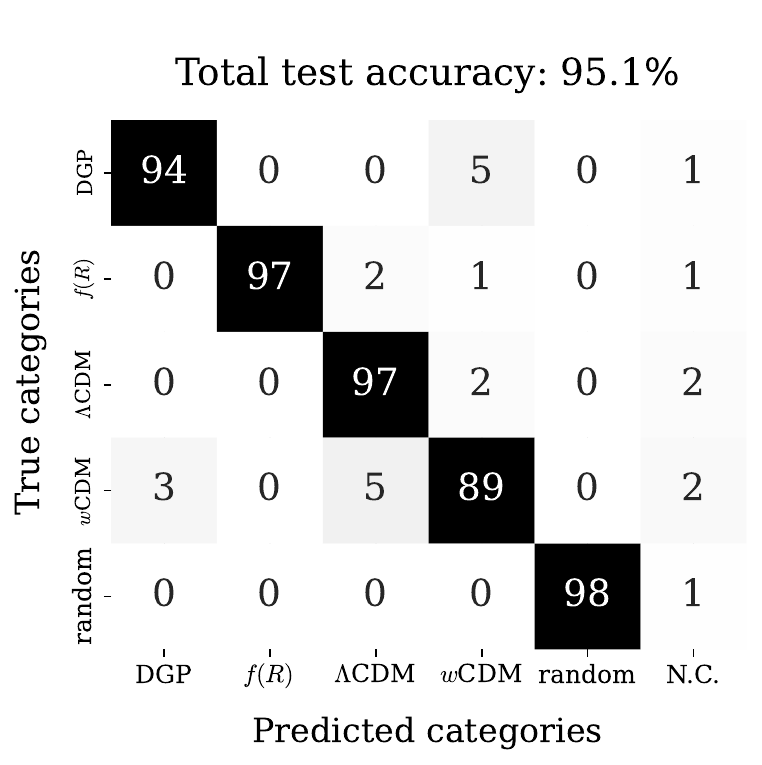}
    \caption{Confusion matrix of the baseline EE2-based model (see \autoref{eq:nlnpkee2}) trained with 5\% theoretical error component and tested with the main EE2 test data. We use data up to  $k_\mathrm{max} = 2.5 \;h/\mathrm{Mpc}$ for training. The class accuracies are shown in percent. (N.C. stands for `Not Classified').}
    \label{fig:cm-EE2-5perc-kmax25}
\end{figure}

\subsection{Cosmological Model Classes}
\label{sec: classes results}

 The main test result of our EE2 baseline model is shown in \autoref{fig:cm-EE2-5perc-kmax25}. The model has been trained with a 5\% theoretical error based on the theory error curves and we have used a scale cut at $k_\mathrm{max} = 2.5 \, h/\mathrm{Mpc}$. The total test accuracy is 95.1\% and the confusion matrix is mostly diagonal. Overall, the trained \bacon{} network can successfully classify a very high proportion of matter power spectra from all classes. As the strength of modifications in the training data are drawn from a Gaussian around \lcdm{} (see \autoref{tab:bac2data} for the parameter distributions), this result shows a high capability of correctly identifying new physics beyond \lcdm{} even for small deviations. 
 
 To understand the few misclassifications, we can study the characteristic effects of each beyond-standard-model class on the matter power spectrum that is displayed in \autoref{fig:mg_params}. Dynamical dark energy can produce a similar amplitude shift to DGP at large scales so the two models are slightly degenerate. Additionally, the $w_0$ parameter allows for values very close to a cosmological constant so that 5\% of $w$CDM spectra get misclassified as \lcdm{}. Both effects reduce the true positive rate of the $w$CDM class. Hence, $w$CDM ends up with the lowest class accuracy. Nevertheless, the overall classification result is remarkable. Given that all cosmological parameters are varied at the same time as the class specific parameters, there are fewer degeneracies than one might expect. This shows that the visual analysis of a matter power spectrum that is performed by the convolutional neural network contains enough class specific information that allows for a robust distinction of the cosmological models.

\begin{figure*}
\centering
\includegraphics[width=\textwidth]{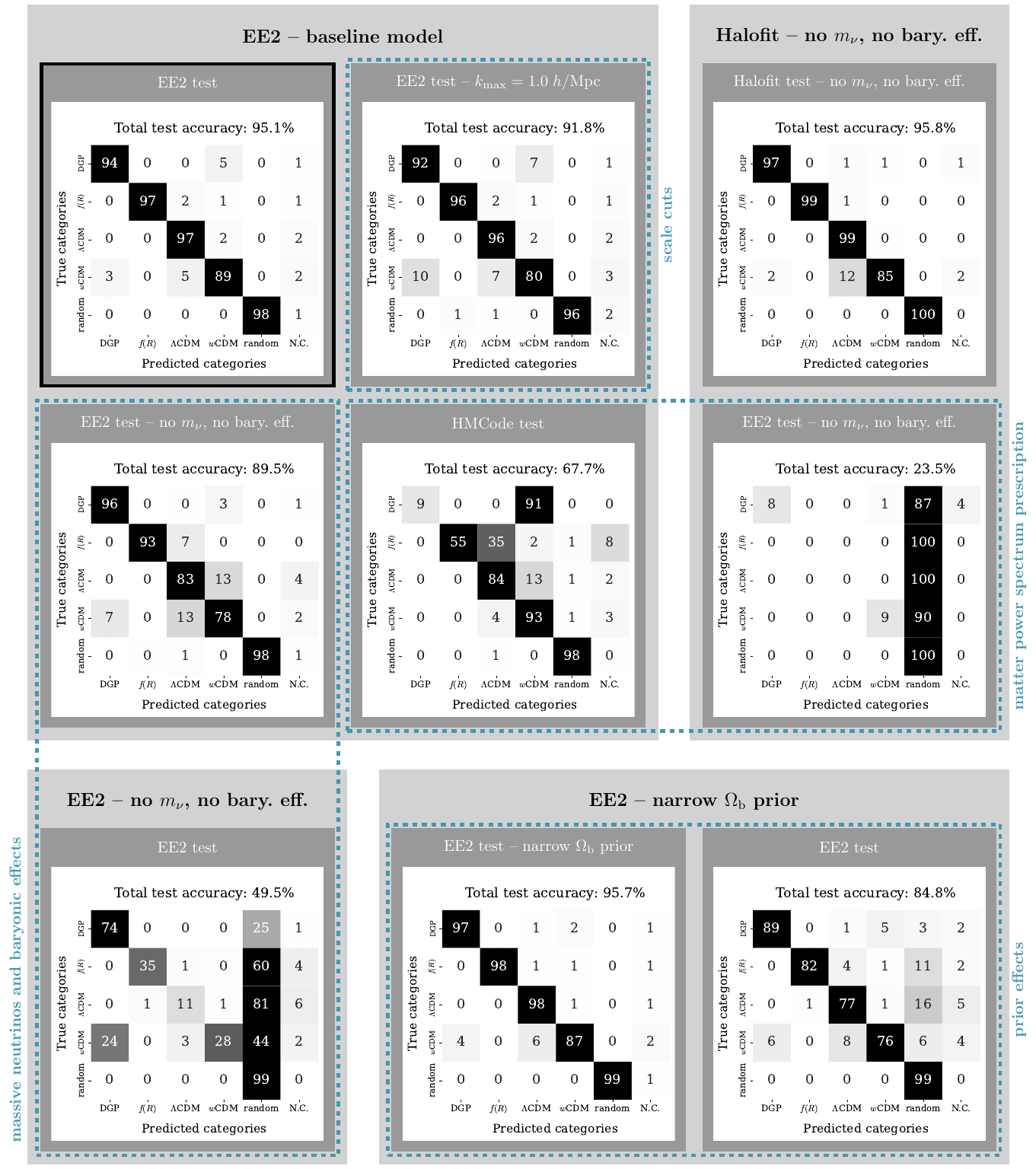}
\caption{Overview of confusion matrices from testing various models with different training and testing data. All the shown models include  theory curves scaled to  $5\%$ as a theoretical error component in the noise model added during training, and the new random class based on EE2 \lcdm{} spectra with baryons and neutrinos. Light gray boxes indicate the training data, dark gray boxes the test data. The blue dashed frames show which trends are investigated with these models/tests. The main baseline test result is shown in the upper left corner. Class accuracies are shown in percent. (N.C. is `Not Classified').  The scale cut is at $k_\mathrm{max} = 2.5 \;h/\mathrm{Mpc}$ if not specified otherwise. }
\label{fig:cm-overview-tikz}
\end{figure*}

We use this main test of the baseline model as a reference for the various effects that we will investigate next. An overview of all the confusion matrices that will be discussed in the coming sections is given in \autoref{fig:cm-overview-tikz}.  Our main test result (\autoref{fig:cm-EE2-5perc-kmax25}) is shown in the upper left corner for comparison. All the displayed models are trained with theory error curves scaled to 5\%.

\begin{figure}
    \centering
        \includegraphics[width=\columnwidth]{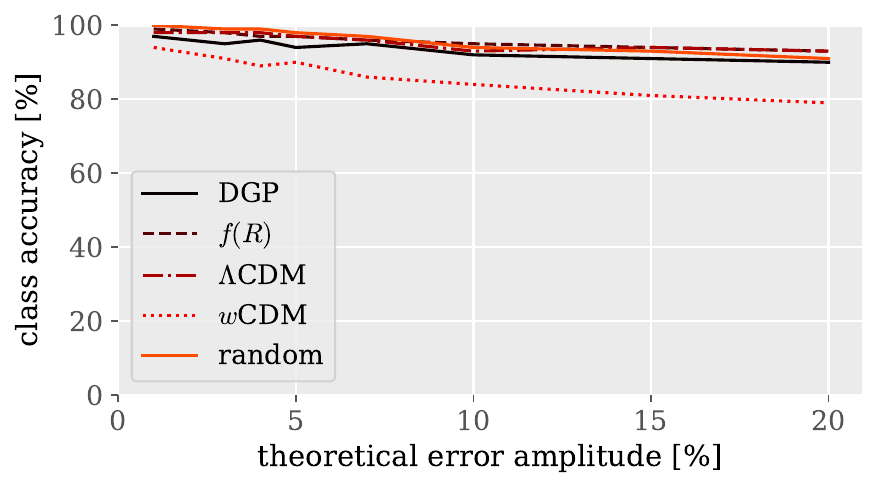}
    \caption{\textbf{Influence of theoretical error amplitude on class accuracies:} EE2 test accuracies for correct class classifications of an EE2 model with added theory curves in training (20,000 training spectra per class) with a scale cut at $k_\mathrm{max} = 2.5 \;h/\mathrm{Mpc}$. The EE2 test and training data have the same cosmological parameter distributions. }
    \label{fig:sys-class-ee}
\end{figure}

\begin{figure}
    \centering
        \includegraphics[width=\columnwidth]{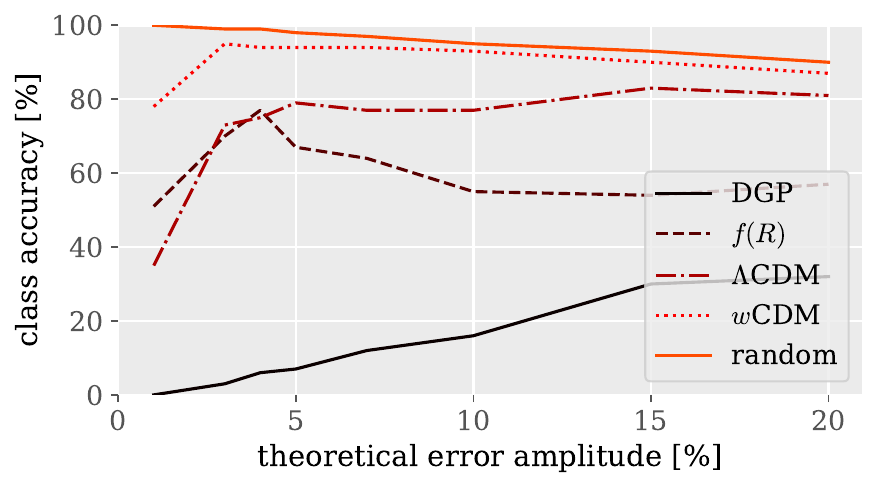}
    \caption{\textbf{Influence of theoretical error amplitude on class accuracies:} \hmcode{} test accuracies for correct class classifications of an EE2 model with added theory curves in training (20,000 training spectra per class) with a scale cut at $k_\mathrm{max} = 2.5 \;h/\mathrm{Mpc}$. The \hmcode{} test and EE2 training data have the same cosmological parameter distributions. }
    \label{fig:sys-class-hm}
\end{figure}

Before we move on we have a final look at the class-specific accuracies in the baseline EE2 model. We have plotted the dependency of the class accuracies on the amplitude of the theoretical error for the EE2 test data in \autoref{fig:sys-class-ee} and for \hmcode{} test data in \autoref{fig:sys-class-hm}. When tested with EE2, the accuracies barely drop with a larger theoretical error amplitude. Only $w$CDM has an accuracy below 90\% for the reasons discussed above. When tested with \hmcode{}, we see a stronger class dependency. DGP has a very low accuracy that rises significantly with an increase of the error amplitude, while the $f(R)$ accuracy is reduced at high amplitudes. The latter class has mostly an effect on the power spectrum on smaller scales so it gets proportionally more affected by the amplitude level. The strong differences between the classes in the \hmcode{} case show that the theoretical error influences the fingerprints of the different cosmological models on the matter power spectrum to a very different extent.

\subsection{Massive Neutrinos and Baryonic Feedback}

One of the main motivations for this updated version of \bacon{} was the inclusion of effects from massive neutrinos and baryonic feedback in our description of the matter power spectrum. The total test accuracy of $\sim$ 95\% for the updated spectra shows that the addition of these effects generates no significant loss in accuracy.

If we test a network trained with massive neutrinos and baryonic feedback but without these effects in the test data, there is only a slight 5\% decrease of the accuracy as can be seen in the confusion matrix in \autoref{fig:cm-overview-tikz}.
On the other hand, for a network trained on spectra without the effects of massive neutrinos and baryonic feedback but tested on data with these effects we see a drop of the test accuracy to only 50\%.
Most classes get predominantly classified as random. This can be explained by the construction of the random spectra based on \lcdm{} spectra that include the neutrino mass and baryon modelling parameters. Hence, we see that the characteristic mark of neutrinos and baryons in the matter power spectrum has a stronger impact on the classification than any class specific features. As a consequence, the inclusion of all known physical effects in the data modelling is important to be able to distinguish between beyond-\lcdm{} models. However, the effect of massive neutrinos and baryons can be interpreted as an unknown `new physics' from the perspective of a network that is trained without them. Hence, this result shows the capability of the random class to flag the presence of new physics that is missing in the other selected classes. 

Also, we have to make sure that the random class is constructed from accurate spectra including all the relevant effects.
The only exception in this specific case shown in the test result in \autoref{fig:cm-overview-tikz} is DGP.  It still has a high true positive rate as it is mostly affecting the large scales while massive neutrinos, and especially baryonic feedback, impact the smaller scales. This can be seen in \autoref{fig:mg_params}.

\begin{figure}
    \centering
    \includegraphics[width=0.48\columnwidth]{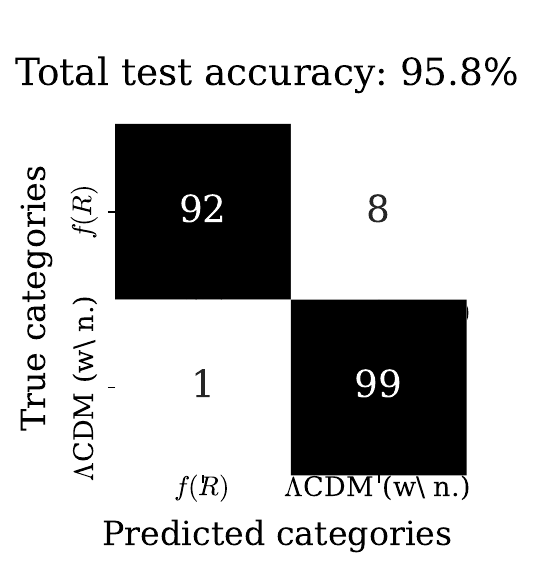}
    \includegraphics[width=0.48\columnwidth]{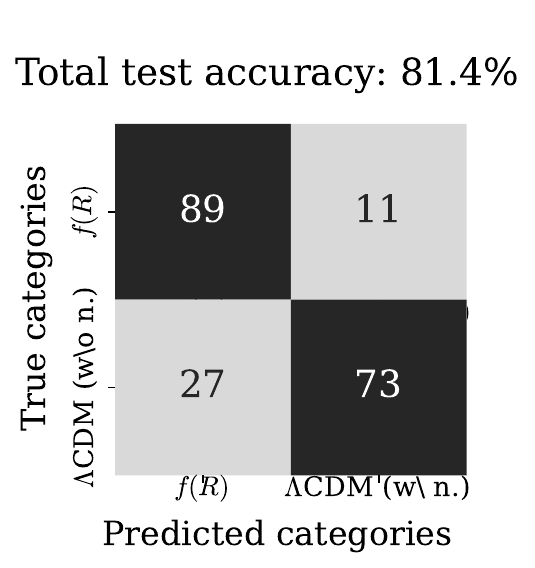}
    \caption{\textbf{Influence of massive neutrinos:} confusion matrices of two-class models trained on $f(R)$ spectra without massive neutrinos and \lcdm{} data with (right) or without (left) massive neutrinos (with a 5\% theoretical error component). The test data is the same in both cases, it contains $f(R)$ without and \lcdm{} data with massive neutrinos.}
    \label{fig:cm-fr-nu}
\end{figure}

We have done a more specific test to understand how the network is able to distinguish between  $f(R)$ and massive neutrinos, given that both of these can have strong scale-dependent effects. For that, we've trained a two-class model only with \lcdm{} and $f(R)$ data to see if the effect of massive neutrinos in a \lcdm{} cosmology can get confused with $f(R)$. 
\autoref{fig:cm-fr-nu} shows two confusion matrices produced using two distinctly trained models. Both models have been trained and tested with $f(R)$ spectra without massive neutrinos but the \lcdm{} in training varies. It either included or didn't include massive neutrinos. The test is then performed with massive neutrinos in the \lcdm{} testing data.
If we don't account for massive neutrinos in training, their effect gets mistaken as $f(R)$. But if we train with neutrinos, there is near to no degeneracy. This means that they have a broadly similar effect on the shape of the power spectrum, so they can get mistaken for one another in the lack of more class-specific information. However, when trained with spectra including both effects, then the specific imprints of $f(R)$ and massive neutrinos in the power spectrum are distinctive enough to confidently separate them. 

\autoref{fig:mg_params} shows that the changes that $f(R)$ and massive neutrinos induce in the power spectrum have different shapes on nonlinear scales. So the classification ability of the network most likely originates from including these highly nonlinear scales in the training data.
To understand more about the specific patterns that the network is looking for, we would need methods that visualise the decision making process of the network, like Class activation maps. We will implement explainable AI methods like this in our next work. Additionally, this test emphasises how important it is to accurately model all of the physics on nonlinear scales as this might otherwise lead to a false detection of modified gravity, for example.  

\subsection{Power Spectrum Prescription}
\label{sec:powerSpectrumPrescription}

We will make use of the known systematic difference between the EE2 and \hmcode{} matter power spectrum,  shown in \autoref{fig:ratio_EEHM_noise}, to test the effectiveness of our theoretical error model. This difference is found to be largely independent of the cosmological parameters. We add this fixed EE2-\hmcode{} ratio to EE2 test data sets, scaling the ratio to 1\% and 3\% (the actual ratio shows roughly a  $\sim 4$\% difference). The expectation is that testing the EE2 baseline model on these sets should exhibit an overall accuracy trend that tends towards the \hmcode{} test set.

\begin{figure}
    \centering
        \includegraphics[width=\columnwidth]{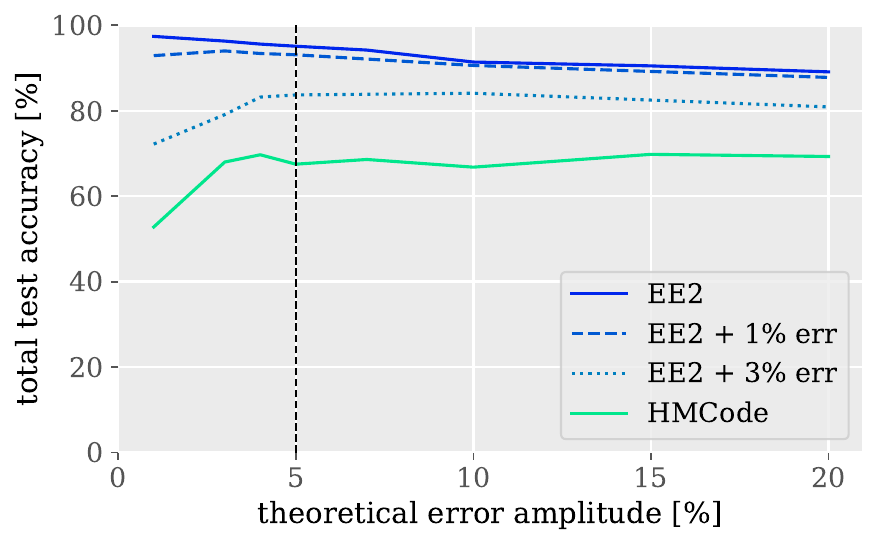}
    \caption {\textbf{Influence of power spectrum prescription and theoretical error:} Total test accuracy of the EE2 baseline model depending on the scaling of the theory error curves added to the main EE2 training data (20,000 training spectra per class). We use EE2 and \hmcode{} test sets as well as the EE2 test set with an added and scaled fixed EE2-\hmcode{} difference (as in \autoref{fig:ratio_EEHM_noise}). 5\% is marked as the theoretical error amplitude where the ideal theoretical error model would show convergence of the various test set accuracies. }
    \label{fig:sys-acc-testdata}
\end{figure}

In \autoref{fig:sys-acc-testdata} we show the test results of the EE2 baseline model for EE2 test sets with these specific systematic errors added in. As expected, smaller systematic differences between the training and test spectra lead to a higher accuracy. Increasing the amplitude of the theory error curves in training reduces the accuracy difference to EE2 test data but still does not fully converge in the case of the 3\% EE2-\hmcode{} systematic added to the test set.
This indicates that our theory error curves have the desired effect but they are not able to account for the full systematic difference between these two power spectra prescriptions, and so are likely not general enough to be used on real Stage-IV data without incurring some bias in the classification. This lack of convergence can most likely be attributed to the `short-wavelength' features that are visible in the systematic EE2-\hmcode{} difference seen in the coloured curves of \autoref{fig:ratio_EEHM_noise}.

On the other hand, if we are training and testing a network with the same code, we get consistently good classification results. \autoref{fig:cm-overview-tikz} shows confusion matrices for a test of a \halofit{} network with \halofit{} and EE2 test data. The \halofit{} test gives a total classification accuracy of $\sim$ 96\% which is nearly identical to the results in \cite{Mancarella:2020jyu}. Hence, our changes to the pipeline have not reduced the classification power of the BNN. In contrast, a \halofit{} network classifies all of the EE2 test spectra as random. This is not surprising as the random spectra are based on EE2 \lcdm{} spectra but it shows that the effect of the power spectrum prescription is dominant over the characteristics of the different cosmological classes. 

\begin{figure}
    \centering
        \includegraphics[width=\columnwidth]{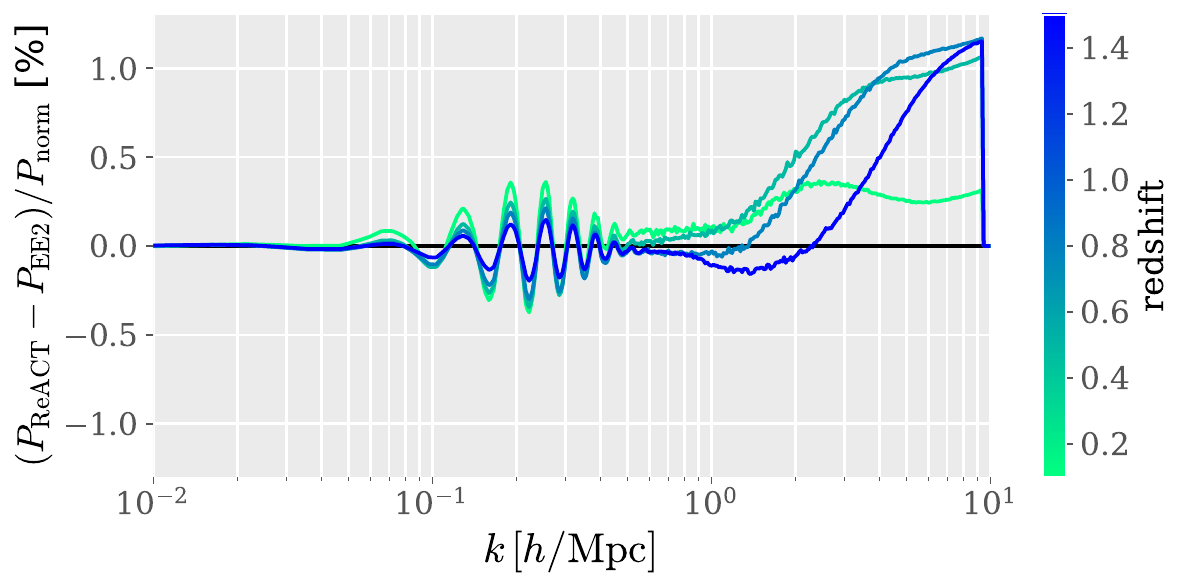}
    \caption{\textbf{Effects of the matter power spectrum prescription:} Normalised difference of $w$CDM matter power spectra produced with our standard \react{} pipeline and directly with the EE2 emulator for the same cosmological parameters. We show all the 4 redshifts considered in this work. The deviations from \lcdm{} are $\{w_0,w_a\} \approx \{-0.877, -0.298\}$ for this example and no massive neutrinos or baryons are included. }
    \label{fig:diff-wcdm-ee2-react}
\end{figure}

\begin{figure}
    \centering
        \includegraphics[width=\columnwidth]{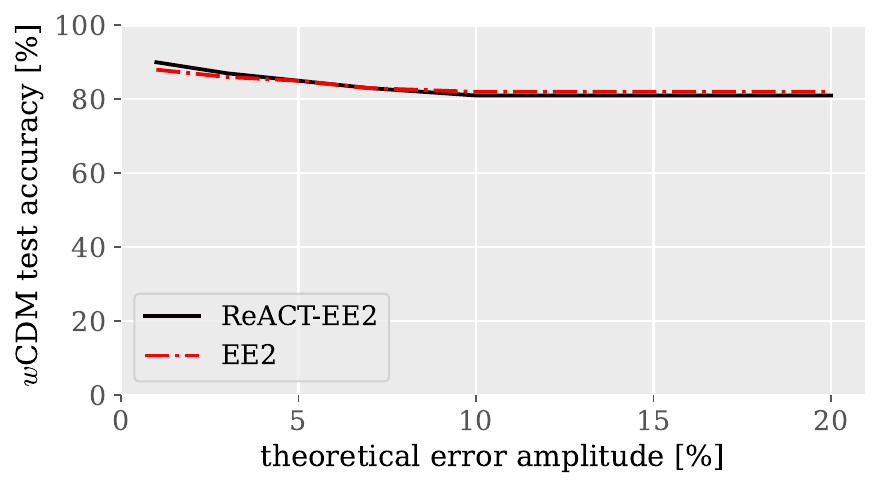}
    \caption{\textbf{Effects of the matter power spectrum prescription:} Test accuracies for correct class classifications when switching the $w$CDM test data with directly emulated $w$CDM spectra.
    The models were trained on our \react{}-EE2 test data set with various amplitudes of theory curves (20,000 training spectra per class, without massive neutrinos or baryonic feedback, $k_\mathrm{max} = 2.5 \;h/\mathrm{Mpc}$). The $w$CDM test data is either produced with the same pipeline as the training data (\react{}-EE2) or directly emulated with EE2. The test and training data have the same cosmological parameter distributions and contain no massive neutrinos or baryonic feedback effects. }
    \label{fig:class-wcdm}
\end{figure}

Motivated by these findings, we put the network to another test: comparing data produced with the \react{} pipeline compared to spectra computed directly with an emulator. This is a robustness test to see if the more accurate emulator power spectra can be correctly classified by our trained network.  Specifically, we swap the $w$CDM spectra in our test set without baryons and massive neutrinos with 1000 $w$CDM spectra produced by EE2 (this should not be confused with our EE2 labelled test data set before, as this is based on EE2 \lcdm{} spectra but did still require corrections from \react{} for the different cosmological models). The normalised difference between an EE2 $w$CDM spectrum and an \react{}-EE2 $w$CDM spectrum for the same cosmological parameter values is shown in \autoref{fig:diff-wcdm-ee2-react}. Even for the selected example with a strong deviation from \lcdm{}, the difference between the two matter power spectrum prescriptions barely reaches 1\%. 
We will now check if our test results change when we swap the $w$CDM class in our test data for the more accurate emulated spectra. \autoref{fig:class-wcdm} shows the class test accuracies for the \react{}-EE2 and EE2 $w$CDM spectra. The network is coping well. Already for the model trained with a 1\% theory error the emulated  test data has only a 2\% drop in accuracy. This vanished when adding larger theory errors during the training of the network.
Even though the absolute difference between these $w$CDM test sets is small as seen in \autoref{fig:diff-wcdm-ee2-react}, it is reassuring that the network can generalise to more accurate prescriptions and the small theory error gets fully compensated by adding theory error curves during training.

\subsection{Scale Cuts}

\begin{figure}
    \centering
        \includegraphics[width=\columnwidth]{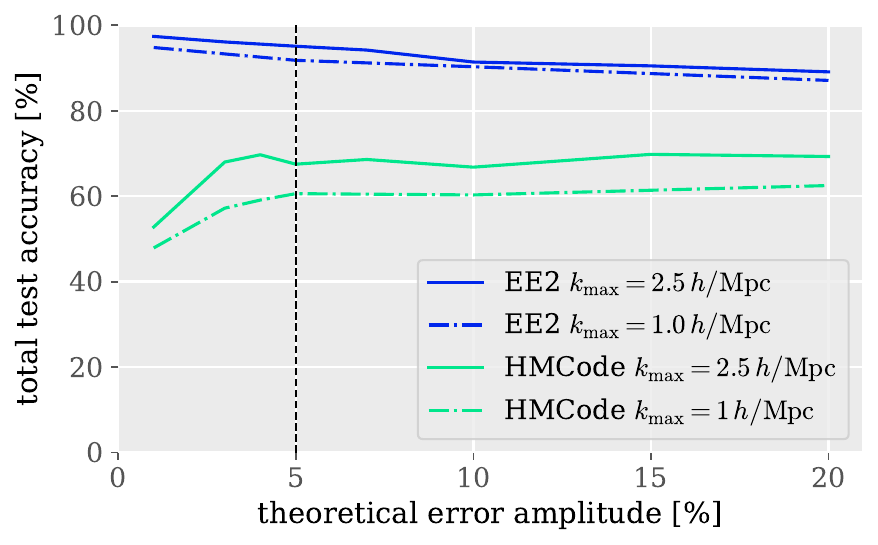}
    \caption{\textbf{Influence of scale cuts:} Total test accuracy of two EE2 models trained with the main EE2 training data set (20,000 training spectra per class) using scales up to $k_\mathrm{max} = 2.5 \;h/\mathrm{Mpc}$ or $k_\mathrm{max} = 1.0 \; h/\mathrm{Mpc}$. We show the dependence on the scaling of the theory error curves added as a noise component to the training data. All test sets have the same cosmological parameter distributions. 5\% is marked as the theoretical error amplitude where the ideal theoretical error model would show convergence based on \autoref{fig:ratio_EEHM_noise}.}
    \label{fig:sys-acc-kmax}
\end{figure}

We train two EE2 models with scale cuts at $k_\mathrm{max} = 2.5 \;h/\mathrm{Mpc}$ and $k_\mathrm{max} = 1.0 \; h/\mathrm{Mpc}$. \autoref{fig:cm-overview-tikz} displays the confusion matrix for both models when trained with 5\% theoretical error, \autoref{fig:sys-acc-kmax} shows the total accuracy of these models for various strengths of the theoretical error.

The exclusion of smaller scales leads, in all cases, to a reduction of the total accuracy. For the EE2 test data this is of the order of $\sim$ 2-3\%. This means that the network is able to extract information from the scales between $1.0 \; h/\mathrm{Mpc}$ and $2.5 \; h/\mathrm{Mpc}$ and that it is worthwhile modelling nonlinear scales. Even when we consider the lower error amplitude for a scale cut at $k_\mathrm{max} = 1.0 \; h/\mathrm{Mpc}$ and compare the 4\% model with the 5\% of $k_\mathrm{max} = 2.5 \;h/\mathrm{Mpc}$, we achieve a better accuracy going to smaller scales. The comparison of \hmcode{} tests in \autoref{fig:sys-acc-kmax} shows that test data with a generally lower accuracy is affected even stronger by the loss of the nonlinear scales.

\subsection{Prior effects and Out-Of-Distribution samples} \label{sec:priors}

To test the effect of priors on the classification we train an EE2 network with $\Omega_{\rm b}$ values from a very narrow distribution with a standard deviation of 0.001 instead of 0.016. The results are shown in \autoref{fig:cm-overview-tikz}.  If we test this model with data based on the wider $\Omega_{\rm b}$ prior then the total test accuracy gets reduced by 11\%. This is a indication that the network cannot extrapolate far beyond the learned parameter range. Many spectra get instead classified as random because the random spectra are based on \lcdm{} spectra with the full $\Omega_{\rm b}$ parameter range. For very strong deviations from Planck's best fit parameters it might be justified to flag them as new physics in the random class that is representing other unconsidered models. However in general, we can conclude that the influence of cosmological parameter values outside of the prior of the training data have a significant influence on the classification.

\begin{figure*}
        \centering
        \includegraphics[width=0.99\linewidth]{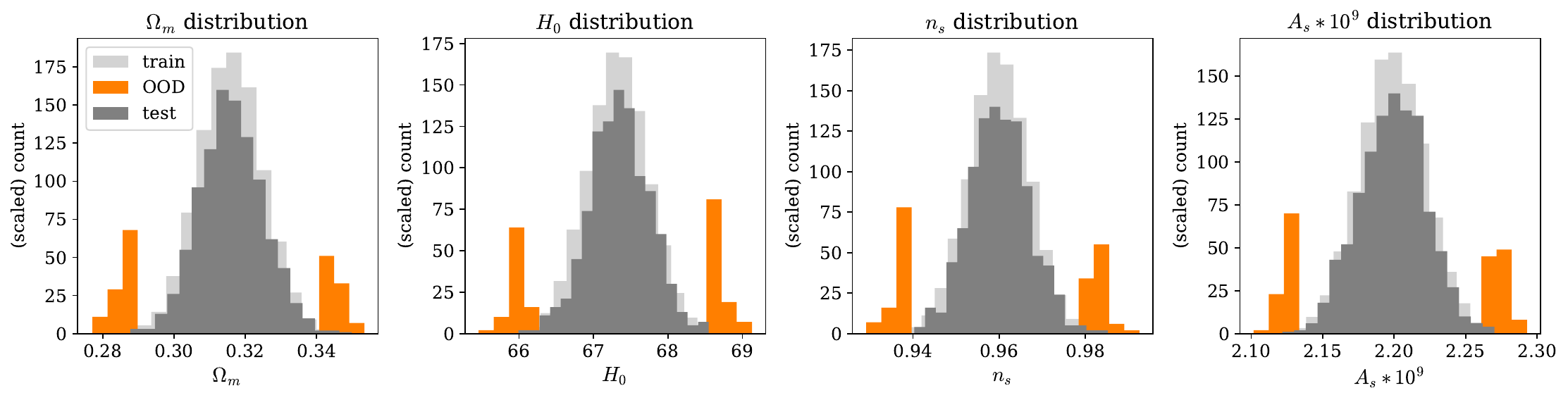}
        \includegraphics[width=0.24\linewidth]{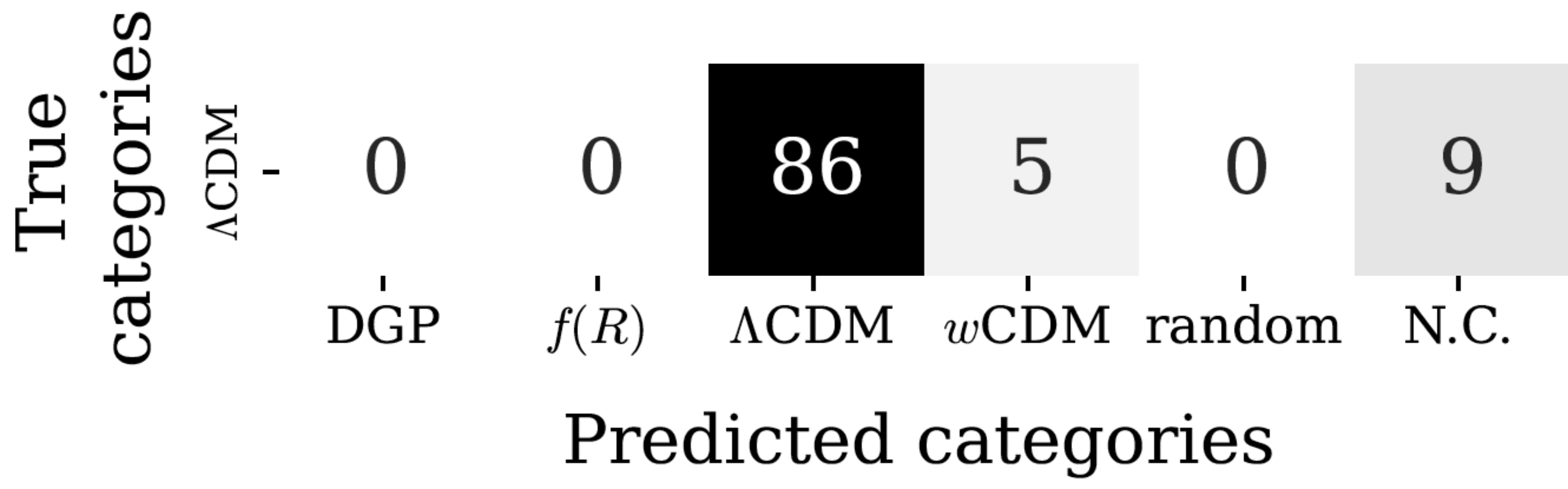} 
        \includegraphics[width=0.24\linewidth]{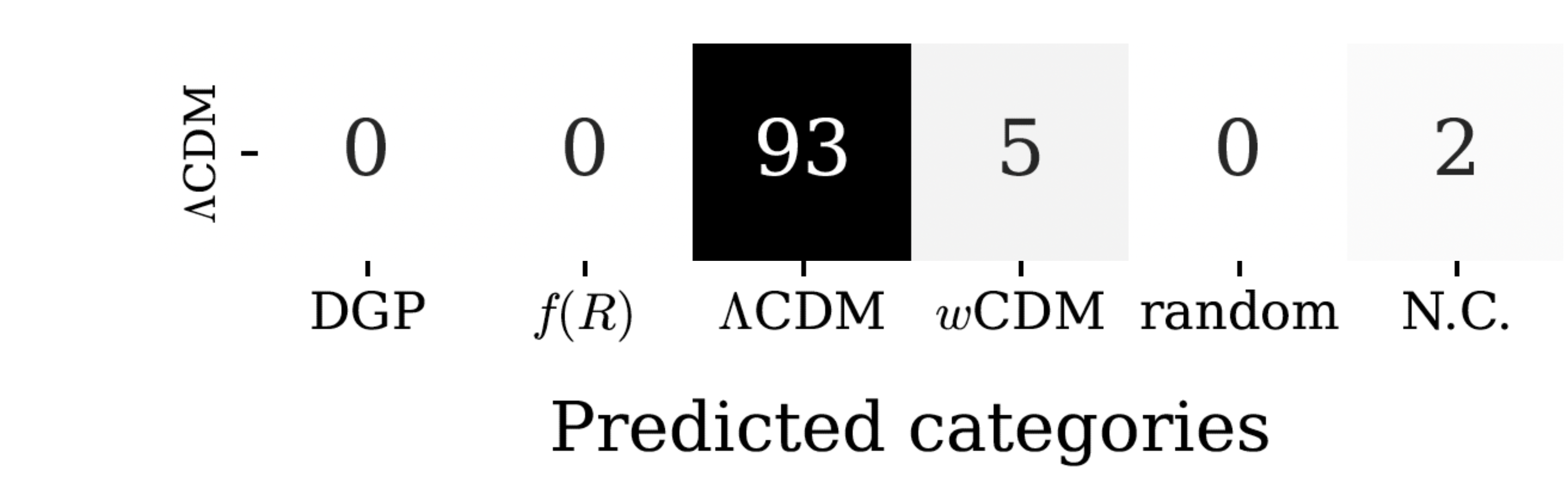} 
        \includegraphics[width=0.24\linewidth]{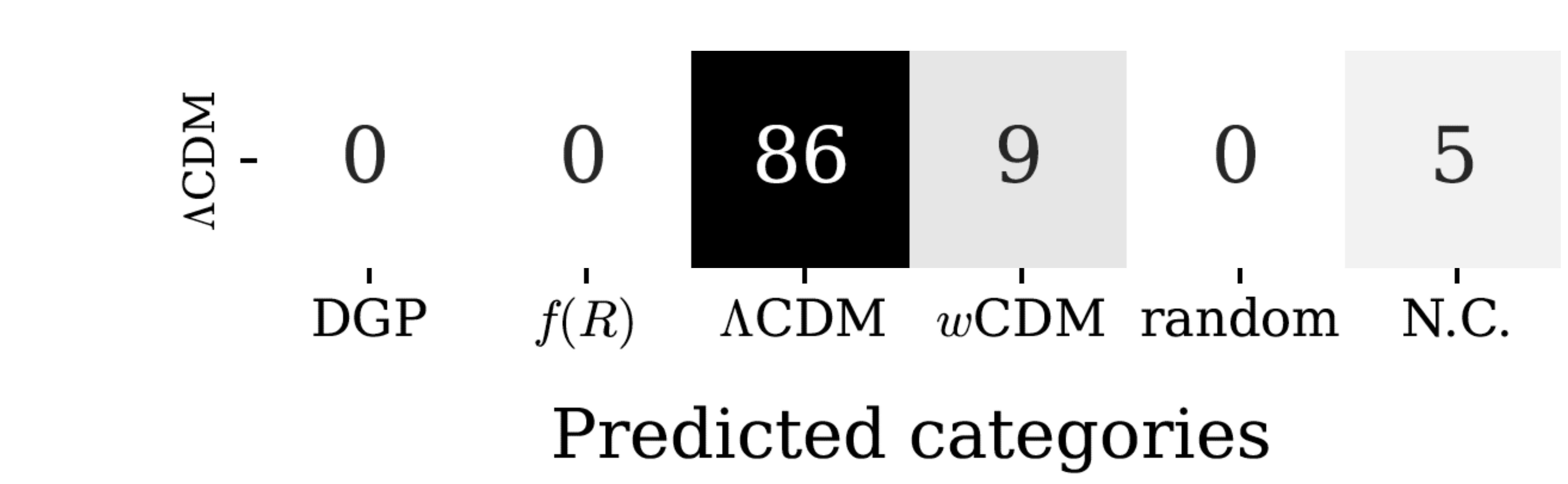} 
        \includegraphics[width=0.24\linewidth]{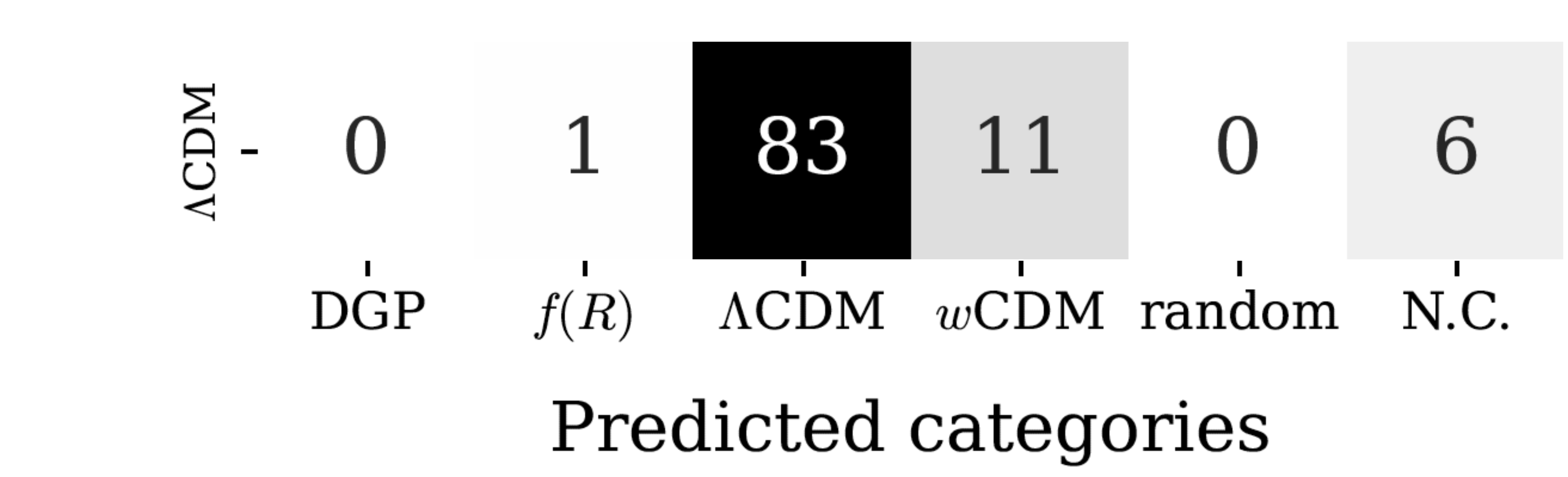} 
        \includegraphics[width=0.24\linewidth]{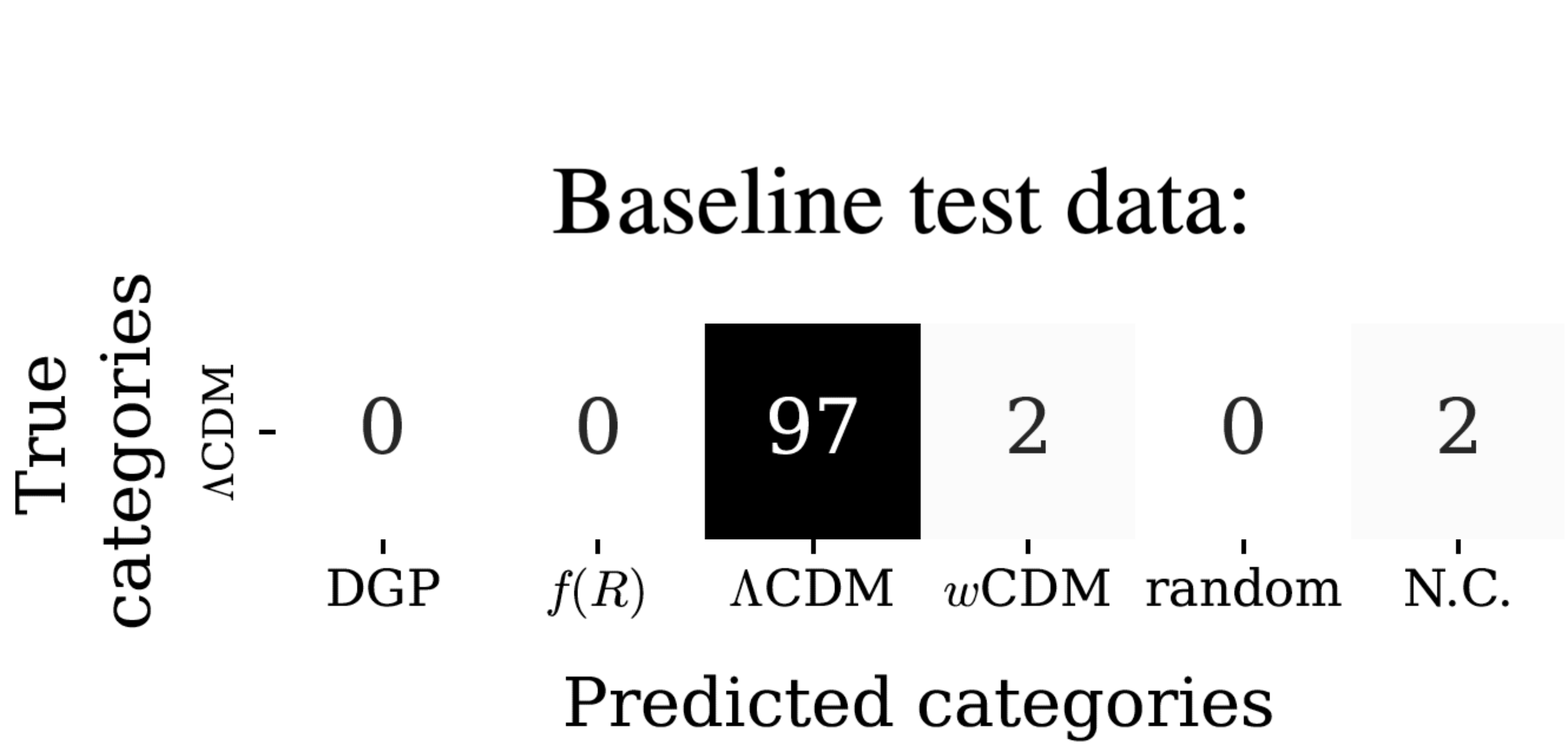} 
  \caption{\textbf{Classification of Out-Of-Distribution samples:} The histograms show the distribution of values for cosmological parameters in the \lcdm{} data of the baseline EE2 training data (dark gray) and test data (gray) as well as a set with Out-Of-Distribution (OOD) \lcdm{} data (orange). The bar height represents the full count for test and OOD data, and has been scaled for the 20,000 training samples. The second row shows the \lcdm{}-classification accuracies when the baseline EE2-based model is tested with \lcdm{} spectra that are generated from these OOD parameters. The last row features the \lcdm{}-classification accuracies of the In-Distribution baseline test data as shown in \autoref{fig:cm-EE2-5perc-kmax25}.}
  \label{fig:hist-ood}
\end{figure*}

To further probe the extent of this effect for our baseline model we have run a number of tests with Out-Of-Distribution (OOD) data. For that we have produced 200 outliers of the parameter distributions in $\Omega_{\rm m}$, $H_0$, $n_s$ and $A_s$. They are sampled to be $3 \sigma$ outliers from the \lcdm{} training set used for our baseline EE2 model. When computing the \lcdm{} power spectra for these samples we only use the OOD value for one of the parameters at a time while sampling the other parameters as normal. 

The parameter distributions used in the training data and testing data of the baseline model are shown in \autoref{fig:hist-ood}. We rescaled the height of the bars for the training data to account for the larger number of training samples. The parameter values of the OOD test set are shown in orange. The corresponding confusion matrix for the \lcdm{} class is shown below the respective histogram. Even though all the \lcdm{} spectra are produced from $3\sigma$ outliers, the \lcdm{} true-positive accuracy does not drop by more than 15\% compared to the standard test data (see \autoref{fig:cm-EE2-5perc-kmax25}).
This is a high performance given that all test samples have one parameter that is a significant outlier of the training parameter distribution. It shows that the trained network is fairly robust to these type of OOD samples and our chosen parameter priors. 

It should be noted, that none of the OOD samples is classified as random. Hence, the network is able to distinguish between cosmological parameter induced patterns in the spectrum and completely different behaviours expected from a different class of cosmological model.
However, we can see that existing degeneracies are worsened. Especially for the $n_s$ and $A_s$ outliers we see an increased mis-classification of \lcdm{} as $w$CDM. $w$CDM affects the amplitude of the spectrum as well as its scale-dependency and so this degeneracy is expected.

Despite the relative robustness to OOD samples, this exercise shows that priors should be chosen carefully and a wider choice is advantageous. Such a behaviour is not surprising and was expected both from a machine learning and cosmological perspective. 
In our case, we have chosen Planck and Stage-IV galaxy survey based priors for the training data that assume LSS and the CMB are consistent. This approach should encompass the majority of currently considered  cosmological models.

%%%%%%%%%%%%%%%%%%%%%% Conclusions ###################

\section{Conclusions}
\label{sec:conclusions}

Forthcoming Stage-IV cosmological surveys will perform measurements with unprecedented accuracies, in particular at nonlinear scales. We have improved a novel machine learning approach to detect beyond-standard-model physics in the data using a Bayesian Neural Network. The purpose of this machine-learning based method is not the replacement of a MCMC analysis, but the development of a tool that can reduce the vast model space of \lcdm{} modifications with high time efficiency. This includes signs for the preference of \lcdm{} or hints at new physics beyond the considered cosmological classes. Once preferred models are identified, a MCMC analysis can be used to obtain parameter constraints from the selected subset. This approach reduces the data analysis time in the search for new physics without the need for biased assumptions on the model or parameter space.

Based on the halo model reaction framework, we create nonlinear dark matter power spectra for a variety of modified gravity and dark energy theories.  The data shape of a matter power spectrum is chosen to roughly represent Stage IV-like survey data with four z-bins $z \in \{ 1.5,0.785,0.478, 0.1 \}$ and 100 $k$-bins up to $k \leq 2.5~h/{\rm Mpc}$ \citep{2011arXiv1110.3193L,Euclid:2019clj}. Our spectra include both baryonic and massive neutrino effects to increase the accuracy on nonlinear scales and enhance our ability to detect deviations from \lcdm{} in an unbiased way.

Our Bayesian Cosmological Network (\bacon{}) contains convolutional layers that perform an image analysis of the matter power spectrum. It uses the information in the shape of the power spectrum to classify it into one of five classes: 
\lcdm{}, $f(R)$, $w$CDM, DGP gravity and a `random' class accounting for unknown physics. Our baseline model is able to distinguish between these classes without any major degeneracies and has a total classification accuracy of $\sim$ 95\%. 
It is well calibrated with a Static Calibration Error of 1.1\%. We have also shown that we can use the distribution of output probabilities of the BNN to estimate the uncertainty of the classification result. This is an advantage over deterministic networks for the analysis of observable data, as we will be able to calculate the confidence for the test result of a single spectrum.

The trained classification network shows a strong dependency on the choice of code for the computation of the \lcdm{} spectrum. We have used direct comparisons of \hmcode{}- and EE2-based data to investigate differences in the theoretical error and test which error model can account for this most effectively. The theory error curves developed in this work have been presented in \autoref{sec:theoryerr}. They are scaled to 5\% of the normalisation spectrum to represent the error in our EE2-based data sets that has been estimated in comparison to $N$-body simulations. Additionally, we add  Stage IV-like cosmic variance to the matter power spectrum to represent the noise in Stage-IV survey measurements.

We find that the network reaches higher classification accuracies if we include smaller scales ($k_\mathrm{max} = 2.5 \;h/\mathrm{Mpc}$ instead of $k_\mathrm{max} = 1.0 \; h/\mathrm{Mpc}$.), showing it is able to gain information from nonlinear scales. 
The baryonic feedback and massive neutrinos have a strong effect on the shape of the power spectrum, especially on nonlinear scales, but their inclusion does not reduce the classification accuracy significantly. 

There are some limitations to the ability of the network to generalise: 
 It is able to extrapolate beyond the priors of the training data to a certain extent but is still dependent on the choices for the parameter ranges. This effect was to be expected and might be of use to flag strong deviations from Planck's best fit parameters. However in general one has to be aware that the selection of priors matters. Furthermore, the influence of baryonic effects, massive neutrinos and power spectrum-specific characteristics is stronger than features generated by modified gravity and dynamic dark energy theories. This can lead to a wrong classification as `random' if one of the effects is missing in the training data. Hence, a consistent construction of the random class based on accurate matter power spectrum is important. However, as the random class represents other feasible models beyond the selected ones this classification shows that \bacon{} can successfully flag the detection of new physics.

There is a range of further improvements and new applications that we are planning for \bacon{}. While the new theoretical error model is able to alleviate some of the systematic differences between power spectra prescriptions, it is not accounting for all of this yet. This is one indicator that it is not general enough to account for the `real' systematic difference between our EE2-based prescription and the real Universe. Further refinement of the theory error curves could achieve this. They can then be used to produce random curves that do not have detectable features of the specific power spectrum prescription. 

Moreover, we have seen that the network fails to classify new physics when it has not been trained with the effects of baryonic feedback. This raises the question of how different baryonic feedback models will effect the result. We also want to explore the option to constrain parameters by training a network with classes based on binning specific parameters. Lastly, with respect to the forthcoming Stage-IV surveys, it will be necessary to move to the actual weak lensing and galaxy clustering observables for our data input, i.e., cosmic shear and redshift space galaxy spectra. 

On the machine learning side, we are working on implementing explainable AI for \bacon{}. These are methods that can explain the reasoning for a classification decision of a network to a certain extent. For example Class Activation Maps (CAMs) could find the characteristic features in a spectrum that lead to its classification as a specific modified gravity theory. 

In summary, the inclusion of effects from massive neutrinos and baryonic feedback do not lead to significantly new degeneracies between the considered cosmological models.
We have shown in this work that it is possible to account for large parts of the theoretical error while maintaining a high classification accuracy of $\sim$ 95\% in our 5 class model.
This is a further improvement for the potential of BNNs to indicate the presence of new physics in cosmological datasets with an higher independence from theoretical limitations of the mock data production.

%%%%%%%%%%%%%%%%%%%%%%%%%%%%%%%%%%%%%%%%%%%%%%%%%

\section*{Acknowledgements}
B.B. was supported by a UK Research and Innovation Stephen Hawking Fellowship (EP/W005654/2). A.P. is a UK Research and Innovation Future Leaders Fellow [grant MR/X005399/1].
L.T. is funded by the Science and Technology Facilities Council (STFC) and thanks the German Academic Scholarship Foundation for their support.
L.L.~acknowledges the support by a Swiss National Science Foundation
(SNSF) Professorship grant (No.~202671). 
We are grateful to Joe Kennedy for his assistance and the useful discussions.
We thank Michele Mancarella for the development of the BaCoN architecture.
The authors thank Francesco Pace for allowing use of his modified CLASS code. 
We thank Ben Moews for his public python package \textit{smurves}.
For the purpose of open access, the author has applied a Creative Commons Attribution (CC BY) licence to any Author Accepted Manuscript version arising.
%

%%%%%%%%%%%%%%%%%%%%%%%%%%%%%%%%%%%%%%%%%%%%%%%%%%
\section*{Data Availability}

Alongside this paper we publish the pre-trained  Bayesian Cosmological Network (\bacon{}) which can be accessed at the github repository \url{https://github.com/cosmicLinux/BaCoN-II} with the training and test data available at \url{https://zenodo.org/records/10688282}.

%%%%%%%%%%%%%%%%%%%% REFERENCES %%%%%%%%%%%%%%%%%%

% The best way to enter references is to use BibTeX:

\bibliographystyle{mnras}
\bibliography{bib-bacon}

\begin{thebibliography}{}
\makeatletter
\relax
\def\mn@urlcharsother{\let\do\@makeother \do\$\do\&\do\#\do\^\do\_\do\%\do\~}
\def\mn@doi{\begingroup\mn@urlcharsother \@ifnextchar [ {\mn@doi@} {\mn@doi@[]}}
\def\mn@doi@[#1]#2{\def\@tempa{#1}\ifx\@tempa\@empty \href {http://dx.doi.org/#2} {doi:#2}\else \href {http://dx.doi.org/#2} {#1}\fi \endgroup}
\def\mn@eprint#1#2{\mn@eprint@#1:#2::\@nil}
\def\mn@eprint@arXiv#1{\href {http://arxiv.org/abs/#1} {{\tt arXiv:#1}}}
\def\mn@eprint@dblp#1{\href {http://dblp.uni-trier.de/rec/bibtex/#1.xml} {dblp:#1}}
\def\mn@eprint@#1:#2:#3:#4\@nil{\def\@tempa {#1}\def\@tempb {#2}\def\@tempc {#3}\ifx \@tempc \@empty \let \@tempc \@tempb \let \@tempb \@tempa \fi \ifx \@tempb \@empty \def\@tempb {arXiv}\fi \@ifundefined {mn@eprint@\@tempb}{\@tempb:\@tempc}{\expandafter \expandafter \csname mn@eprint@\@tempb\endcsname \expandafter{\@tempc}}}

\bibitem[\protect\citeauthoryear{Abadi et~al.}{Abadi et~al.}{2015}]{tensorflow2015-whitepaper}
Abadi M.,  et~al., 2015, {TensorFlow}: Large-Scale Machine Learning on Heterogeneous Systems, \url {https://www.tensorflow.org/}

\bibitem[\protect\citeauthoryear{Abbott et~al.}{Abbott et~al.}{2017}]{Monitor:2017mdv}
Abbott B.~P.,  et~al., 2017, {Gravitational Waves and Gamma-rays from a Binary Neutron Star Merger: GW170817 and GRB 170817A} (\mn@eprint {arXiv} {1710.05834}), \mn@doi{10.3847/2041-8213/aa920c}

\bibitem[\protect\citeauthoryear{Abbott et~al.}{Abbott et~al.}{2020}]{Abbott:2020knk}
Abbott T. M.~C.,  et~al., 2020, {Dark Energy Survey Year 1 Results: Cosmological constraints from cluster abundances and weak lensing} (\mn@eprint {arXiv} {2002.11124}), \mn@doi{10.1103/PhysRevD.102.023509}

\bibitem[\protect\citeauthoryear{Aghanim et~al.}{Aghanim et~al.}{2020a}]{Aghanim:2018eyx}
Aghanim N.,  et~al., 2020a, {Planck 2018 results. VI. Cosmological parameters} (\mn@eprint {arXiv} {1807.06209}), \mn@doi{10.1051/0004-6361/201833910}

\bibitem[\protect\citeauthoryear{Aghanim et~al.}{Aghanim et~al.}{2020b}]{Planck:2018vyg}
Aghanim N.,  et~al., 2020b, \mn@doi [A&A] {10.1051/0004-6361/201833910}, 641, A6

\bibitem[\protect\citeauthoryear{{Akeson} et~al.,}{{Akeson} et~al.}{2019}]{2019arXiv190205569A}
{Akeson} R.,  et~al., 2019, {The Wide Field Infrared Survey Telescope: 100 Hubbles for the 2020s} (\mn@eprint {arXiv} {1902.05569})

\bibitem[\protect\citeauthoryear{{Allen}, {Evrard}  \& {Mantz}}{{Allen} et~al.}{2011}]{2011ARA&A..49..409A}
{Allen} S.~W.,  {Evrard} A.~E.,   {Mantz} A.~B.,  2011, {Cosmological Parameters from Observations of Galaxy Clusters} (\mn@eprint {arXiv} {1103.4829}), \mn@doi{10.1146/annurev-astro-081710-102514}

\bibitem[\protect\citeauthoryear{Amendola et~al.}{Amendola et~al.}{2018}]{Amendola:2016saw}
Amendola L.,  et~al., 2018, {Cosmology and fundamental physics with the Euclid satellite} (\mn@eprint {arXiv} {1606.00180}), \mn@doi{10.1007/s41114-017-0010-3}

\bibitem[\protect\citeauthoryear{Anderson et~al.}{Anderson et~al.}{2013}]{Anderson:2012sa}
Anderson L.,  et~al., 2013, \mn@doi [MNRAS] {10.1111/j.1365-2966.2012.22066.x}, 427, 3435

\bibitem[\protect\citeauthoryear{Angulo, Zennaro, Contreras, Aric\`o, Pellejero-Iba\~nez  \& St\"ucker}{Angulo et~al.}{2020}]{Angulo:2020vky}
Angulo R.~E.,  Zennaro M.,  Contreras S.,  Aric\`o G.,  Pellejero-Iba\~nez M.,   St\"ucker J.,  2020, {The BACCO Simulation Project: Exploiting the full power of large-scale structure for cosmology} (\mn@eprint {arXiv} {2004.06245})

\bibitem[\protect\citeauthoryear{Angulo, Zennaro, Contreras, Arico, Pellejero-Iba{\~n}ez  \& St{\"u}cker}{Angulo et~al.}{2021}]{2020arXiv200406245A}
Angulo R.~E.,  Zennaro M.,  Contreras S.,  Arico G.,  Pellejero-Iba{\~n}ez M.,   St{\"u}cker J.,  2021, The BACCO simulation project: exploiting the full power of large-scale structure for cosmology

\bibitem[\protect\citeauthoryear{Aric\`o, Angulo, Contreras, Ondaro-Mallea, Pellejero-Iba\~nez  \& Zennaro}{Aric\`o et~al.}{2021}]{Arico:2020lhq}
Aric\`o G.,  Angulo R.~E.,  Contreras S.,  Ondaro-Mallea L.,  Pellejero-Iba\~nez M.,   Zennaro M.,  2021, {The BACCO simulation project: a baryonification emulator with neural networks} (\mn@eprint {arXiv} {2011.15018}), \mn@doi{10.1093/mnras/stab1911}

\bibitem[\protect\citeauthoryear{Arjona \& Nesseris}{Arjona \& Nesseris}{2020}]{Arjona-2020-WhatCanMachine}
Arjona R.,  Nesseris S.,  2020, \mn@doi [Physical Review D] {10.1103/PhysRevD.101.123525}, 101, 123525

\bibitem[\protect\citeauthoryear{Arnold \& Li}{Arnold \& Li}{2019}]{Arnold:2019zup}
Arnold C.,  Li B.,  2019, \mn@doi [MNRAS] {10.1093/mnras/stz2690}, 490, 2507

\bibitem[\protect\citeauthoryear{Arnold, Li, Giblin, Harnois-D\'eraps  \& Cai}{Arnold et~al.}{2021}]{Arnold:2021xtm}
Arnold C.,  Li B.,  Giblin B.,  Harnois-D\'eraps J.,   Cai Y.-C.,  2021, {FORGE -- the f(R) gravity cosmic emulator project I: Introduction and matter power spectrum emulator} (\mn@eprint {arXiv} {2109.04984})

\bibitem[\protect\citeauthoryear{Atayde, Frusciante, Bose, Casas, Barreira  \& Li}{Atayde et~al.}{2024}]{luis2024}
Atayde L.,  Frusciante N.,  Bose B.,  Casas S.,  Barreira A.,   Li B.,  2024, {In prep.}

\bibitem[\protect\citeauthoryear{Baker, Bellini, Ferreira, Lagos, Noller  \& Sawicki}{Baker et~al.}{2017}]{Baker:2017hug}
Baker T.,  Bellini E.,  Ferreira P.~G.,  Lagos M.,  Noller J.,   Sawicki I.,  2017, {Strong constraints on cosmological gravity from GW170817 and GRB 170817A} (\mn@eprint {arXiv} {1710.06394}), \mn@doi{10.1103/PhysRevLett.119.251301}

\bibitem[\protect\citeauthoryear{Battye, Pace  \& Trinh}{Battye et~al.}{2018}]{Battye:2018ssx}
Battye R.~A.,  Pace F.,   Trinh D.,  2018, {Gravitational wave constraints on dark sector models} (\mn@eprint {arXiv} {1802.09447}), \mn@doi{10.1103/PhysRevD.98.023504}

\bibitem[\protect\citeauthoryear{Bernardo, Bose, Franzmann, Hagstotz, He, Litsa  \& Niedermann}{Bernardo et~al.}{2023}]{Bernardo:2022cck}
Bernardo H.,  Bose B.,  Franzmann G.,  Hagstotz S.,  He Y.,  Litsa A.,   Niedermann F.,  2023, \mn@doi [Universe] {10.3390/universe9020063}, 9, 63

\bibitem[\protect\citeauthoryear{Beutler et~al.}{Beutler et~al.}{2017}]{Beutler:2016arn}
Beutler F.,  et~al., 2017, \mn@doi [MNRAS] {10.1093/mnras/stw3298}, 466, 2242

\bibitem[\protect\citeauthoryear{{Bird}, {Viel}  \& {Haehnelt}}{{Bird} et~al.}{2012}]{2012MNRAS.420.2551B}
{Bird} S.,  {Viel} M.,   {Haehnelt} M.~G.,  2012, {Massive neutrinos and the non-linear matter power spectrum} (\mn@eprint {arXiv} {1109.4416}), \mn@doi{10.1111/j.1365-2966.2011.20222.x}

\bibitem[\protect\citeauthoryear{{Bird}, {Ali-Ha{\"\i}moud}, {Feng}  \& {Liu}}{{Bird} et~al.}{2018}]{2018MNRAS.481.1486B}
{Bird} S.,  {Ali-Ha{\"\i}moud} Y.,  {Feng} Y.,   {Liu} J.,  2018, {An efficient and accurate hybrid method for simulating non-linear neutrino structure} (\mn@eprint {arXiv} {1803.09854}), \mn@doi{10.1093/mnras/sty2376}

\bibitem[\protect\citeauthoryear{Blanchard et~al.}{Blanchard et~al.}{2020a}]{Blanchard:2019oqi}
Blanchard A.,  et~al., 2020a, {Euclid preparation: VII. Forecast validation for Euclid cosmological probes} (\mn@eprint {arXiv} {1910.09273}), \mn@doi{10.1051/0004-6361/202038071}

\bibitem[\protect\citeauthoryear{Blanchard et~al.}{Blanchard et~al.}{2020b}]{Euclid:2019clj}
Blanchard A.,  et~al., 2020b, \mn@doi [A&A] {10.1051/0004-6361/202038071}, 642, A191

\bibitem[\protect\citeauthoryear{Blas, Lesgourgues  \& Tram}{Blas et~al.}{2011}]{Blas:2011rf}
Blas D.,  Lesgourgues J.,   Tram T.,  2011, \mn@doi [JCAP] {10.1088/1475-7516/2011/07/034}, 07, 034

\bibitem[\protect\citeauthoryear{{Blas}, {Garny}, {Konstandin}  \& {Lesgourgues}}{{Blas} et~al.}{2014}]{2014JCAP...11..039B}
{Blas} D.,  {Garny} M.,  {Konstandin} T.,   {Lesgourgues} J.,  2014, {Structure formation with massive neutrinos: going beyond linear theory} (\mn@eprint {arXiv} {1408.2995}), \mn@doi{10.1088/1475-7516/2014/11/039}

\bibitem[\protect\citeauthoryear{Blundell, Cornebise, Kavukcuoglu  \& Wierstra}{Blundell et~al.}{2015}]{Blundell2015}
Blundell C.,  Cornebise J.,  Kavukcuoglu K.,   Wierstra D.,  2015, in International conference on machine learning. pp 1613--1622

\bibitem[\protect\citeauthoryear{Bose, Cataneo, Tr\"oster, Xia, Heymans  \& Lombriser}{Bose et~al.}{2020}]{Bose:2020wch}
Bose B.,  Cataneo M.,  Tr\"oster T.,  Xia Q.,  Heymans C.,   Lombriser L.,  2020, {On the road to per cent accuracy IV: ReACT \textendash{} computing the non-linear power spectrum beyond \ensuremath{\Lambda}CDM} (\mn@eprint {arXiv} {2005.12184}), \mn@doi{10.1093/mnras/staa2696}

\bibitem[\protect\citeauthoryear{Bose et~al.,}{Bose et~al.}{2021}]{Bose:2021mkz}
Bose B.,  et~al., 2021, {On the road to per cent accuracy \textendash{} V. The non-linear power spectrum beyond \ensuremath{\Lambda}CDM with massive neutrinos and baryonic feedback} (\mn@eprint {arXiv} {2105.12114}), \mn@doi{10.1093/mnras/stab2731}

\bibitem[\protect\citeauthoryear{Bose, Tsedrik, Kennedy, Lombriser, Pourtsidou  \& Taylor}{Bose et~al.}{2022}]{Bose:2022vwi}
Bose B.,  Tsedrik M.,  Kennedy J.,  Lombriser L.,  Pourtsidou A.,   Taylor A.,  2022, {Fast and accurate predictions of the nonlinear matter power spectrum for general models of Dark Energy and Modified Gravity} (\mn@eprint {arXiv} {2210.01094}), \mn@doi{10.1093/mnras/stac3783}

\bibitem[\protect\citeauthoryear{Brax \& Valageas}{Brax \& Valageas}{2017}]{Brax:2016kin}
Brax P.,  Valageas P.,  2017, {Goldstone models of modified gravity} (\mn@eprint {arXiv} {1611.08279}), \mn@doi{10.1103/PhysRevD.95.043515}

\bibitem[\protect\citeauthoryear{Brax, Casas, Desmond  \& Elder}{Brax et~al.}{2021}]{Brax:2021wcv}
Brax P.,  Casas S.,  Desmond H.,   Elder B.,  2021, \mn@doi [Universe] {10.3390/universe8010011}, 8, 11

\bibitem[\protect\citeauthoryear{Burgess}{Burgess}{2004}]{Burgess:2004ib}
Burgess C.~P.,  2004, {Towards a natural theory of dark energy: Supersymmetric large extra dimensions} (\mn@eprint {arXiv} {hep-th/0411140}), \mn@doi{10.1063/1.1848343}

\bibitem[\protect\citeauthoryear{Burrage}{Burrage}{2019}]{Burrage:2019yle}
Burrage C.,  2019, {What laboratory experiments can teach us about cosmology: A chameleon example} (\mn@eprint {arXiv} {1901.01784}), \mn@doi{10.1051/epjconf/201921905001}

\bibitem[\protect\citeauthoryear{Carrion, Carrilho, Spurio~Mancini, Pourtsidou  \& Hidalgo}{Carrion et~al.}{2024}]{Carrion:2024itc}
Carrion K.,  Carrilho P.,  Spurio~Mancini A.,  Pourtsidou A.,   Hidalgo J.~C.,  2024, {Dark Scattering: accelerated constraints from KiDS-1000 with $\tt{ReACT}$ and $\tt{CosmoPower}$} (\mn@eprint {arXiv} {2402.18562})

\bibitem[\protect\citeauthoryear{Cataneo, Lombriser, Heymans, Mead, Barreira, Bose  \& Li}{Cataneo et~al.}{2019}]{Cataneo:2018cic}
Cataneo M.,  Lombriser L.,  Heymans C.,  Mead A.,  Barreira A.,  Bose S.,   Li B.,  2019, {On the road to percent accuracy: non-linear reaction of the matter power spectrum to dark energy and modified gravity} (\mn@eprint {arXiv} {1812.05594}), \mn@doi{10.1093/mnras/stz1836}

\bibitem[\protect\citeauthoryear{Cataneo, Emberson, Inman, Harnois-Deraps  \& Heymans}{Cataneo et~al.}{2020}]{Cataneo:2019fjp}
Cataneo M.,  Emberson J.~D.,  Inman D.,  Harnois-Deraps J.,   Heymans C.,  2020, {On the road to per cent accuracy \textendash{} III. Non-linear reaction of the matter power spectrum to massive neutrinos} (\mn@eprint {arXiv} {1909.02561}), \mn@doi{10.1093/mnras/stz3189}

\bibitem[\protect\citeauthoryear{Charnock, {Perreault-Levasseur}  \& Lanusse}{Charnock et~al.}{2020}]{Charnock-2020-BayesianNeuralNetworks}
Charnock T.,  {Perreault-Levasseur} L.,   Lanusse F.,  2020, Bayesian {{Neural Networks}} (\mn@eprint {arXiv} {2006.01490}), \mn@doi{10.48550/arXiv.2006.01490}

\bibitem[\protect\citeauthoryear{Charnock, Perreault-Levasseur  \& Lanusse}{Charnock et~al.}{2022}]{Charnock2020}
Charnock T.,  Perreault-Levasseur L.,   Lanusse F.,  2022, in , Artificial Intelligence for High Energy Physics.
World Scientific, pp 663--713

\bibitem[\protect\citeauthoryear{Chevallier \& Polarski}{Chevallier \& Polarski}{2001}]{Chevallier:2000qy}
Chevallier M.,  Polarski D.,  2001, \mn@doi [Int. J. Mod. Phys. D] {10.1142/S0218271801000822}, 10, 213

\bibitem[\protect\citeauthoryear{Chisari et~al.}{Chisari et~al.}{2019}]{Chisari:2019tus}
Chisari N.~E.,  et~al., 2019, {Modelling baryonic feedback for survey cosmology} (\mn@eprint {arXiv} {1905.06082}), \mn@doi{10.21105/astro.1905.06082}

\bibitem[\protect\citeauthoryear{Clifton, Ferreira, Padilla  \& Skordis}{Clifton et~al.}{2012}]{cliftonModifiedGravityCosmology2012}
Clifton T.,  Ferreira P.~G.,  Padilla A.,   Skordis C.,  2012, \mn@doi [Physics Reports] {10.1016/j.physrep.2012.01.001}, 513, 1

\bibitem[\protect\citeauthoryear{Clowe, Bradac, Gonzalez, Markevitch, Randall, Jones  \& Zaritsky}{Clowe et~al.}{2006}]{Clowe:2006eq}
Clowe D.,  Bradac M.,  Gonzalez A.~H.,  Markevitch M.,  Randall S.~W.,  Jones C.,   Zaritsky D.,  2006, {A direct empirical proof of the existence of dark matter} (\mn@eprint {arXiv} {astro-ph/0608407}), \mn@doi{10.1086/508162}

\bibitem[\protect\citeauthoryear{Copeland, Sami  \& Tsujikawa}{Copeland et~al.}{2006}]{Copeland:2006wr}
Copeland E.~J.,  Sami M.,   Tsujikawa S.,  2006, \mn@doi [Int. J. Mod. Phys. D] {10.1142/S021827180600942X}, 15, 1753

\bibitem[\protect\citeauthoryear{Corbelli \& Salucci}{Corbelli \& Salucci}{2000}]{Corbelli:1999af}
Corbelli E.,  Salucci P.,  2000, {The Extended Rotation Curve and the Dark Matter Halo of M33} (\mn@eprint {arXiv} {astro-ph/9909252}), \mn@doi{10.1046/j.1365-8711.2000.03075.x}

\bibitem[\protect\citeauthoryear{Creminelli \& Vernizzi}{Creminelli \& Vernizzi}{2017}]{Creminelli:2017sry}
Creminelli P.,  Vernizzi F.,  2017, {Dark Energy after GW170817 and GRB170817A} (\mn@eprint {arXiv} {1710.05877}), \mn@doi{10.1103/PhysRevLett.119.251302}

\bibitem[\protect\citeauthoryear{Creminelli, Lewandowski, Tambalo  \& Vernizzi}{Creminelli et~al.}{2018}]{Creminelli:2018xsv}
Creminelli P.,  Lewandowski M.,  Tambalo G.,   Vernizzi F.,  2018, {Gravitational Wave Decay into Dark Energy} (\mn@eprint {arXiv} {1809.03484}), \mn@doi{10.1088/1475-7516/2018/12/025}

\bibitem[\protect\citeauthoryear{{Dillon} et~al.,}{{Dillon} et~al.}{2017}]{dillon2017tensorflow}
{Dillon} J.~V.,  et~al., 2017, arXiv e-prints, \href {https://ui.adsabs.harvard.edu/abs/2017arXiv171110604D} {}

\bibitem[\protect\citeauthoryear{Donald-McCann, Beutler, Koyama  \& Karamanis}{Donald-McCann et~al.}{2021}]{Donald-McCann:2021nxc}
Donald-McCann J.,  Beutler F.,  Koyama K.,   Karamanis M.,  2021, {$\texttt{matryoshka}$: Halo Model Emulator for the Galaxy Power Spectrum} (\mn@eprint {arXiv} {2109.15236})

\bibitem[\protect\citeauthoryear{Dvali, Gabadadze  \& Porrati}{Dvali et~al.}{2000}]{Dvali:2000hr}
Dvali G.~R.,  Gabadadze G.,   Porrati M.,  2000, {4-D gravity on a brane in 5-D Minkowski space} (\mn@eprint {arXiv} {hep-th/0005016}), \mn@doi{10.1016/S0370-2693(00)00669-9}

\bibitem[\protect\citeauthoryear{{Escamilla-Rivera}, Quintero  \& Capozziello}{{Escamilla-Rivera} et~al.}{2020}]{Escamilla-Rivera-2020-DeepLearningApproach}
{Escamilla-Rivera} C.,  Quintero M. A.~C.,   Capozziello S.,  2020, \mn@doi [Journal of Cosmology and Astroparticle Physics] {10.1088/1475-7516/2020/03/008}, 2020, 008

\bibitem[\protect\citeauthoryear{Esteban, Gonzalez-Garcia, Maltoni, Schwetz  \& Zhou}{Esteban et~al.}{2020}]{Esteban:2020cvm}
Esteban I.,  Gonzalez-Garcia M.~C.,  Maltoni M.,  Schwetz T.,   Zhou A.,  2020, \mn@doi [JHEP] {10.1007/JHEP09(2020)178}, 09, 178

\bibitem[\protect\citeauthoryear{Ezquiaga \& Zumalacárregui}{Ezquiaga \& Zumalacárregui}{2017}]{Ezquiaga:2017ekz}
Ezquiaga J.~M.,  Zumalacárregui M.,  2017, {Dark Energy After GW170817: Dead Ends and the Road Ahead} (\mn@eprint {arXiv} {1710.05901}), \mn@doi{10.1103/PhysRevLett.119.251304}

\bibitem[\protect\citeauthoryear{Frusciante et~al.}{Frusciante et~al.}{2023}]{Euclid:2023rjj}
Frusciante N.,  et~al., 2023, {Euclid: Constraining linearly scale-independent modifications of gravity with the spectroscopic and photometric primary probes} (\mn@eprint {arXiv} {2306.12368})

\bibitem[\protect\citeauthoryear{{Gal} \& {Ghahramani}}{{Gal} \& {Ghahramani}}{2016}]{gal2016dropout}
{Gal} Y.,  {Ghahramani} Z.,  2016, in international conference on machine learning. pp 1050--1059 (\mn@eprint {arXiv} {1506.02142})

\bibitem[\protect\citeauthoryear{Gangopadhyay, Sami  \& Sharma}{Gangopadhyay et~al.}{2023}]{Gangopadhyay-2023-PhantomDarkEnergy}
Gangopadhyay M.~R.,  Sami M.,   Sharma M.~K.,  2023, Phantom Dark Energy as a Natural Selection of Evolutionary Processes \${\textbackslash}hat\{{\textbackslash}rm A\}\$ \${\textbackslash}textit\{la\}\$ \${\textbackslash}textit\{genetic Algorithm\}\$ and Cosmological Tensions (\mn@eprint {arXiv} {2303.07301}), \mn@doi{10.48550/arXiv.2303.07301}

\bibitem[\protect\citeauthoryear{{Garc{\'i}a-Farieta}, Hort{\'u}a  \& Kitaura}{{Garc{\'i}a-Farieta} et~al.}{2024}]{Garcia-Farieta-2024-BayesianDeepLearning}
{Garc{\'i}a-Farieta} J.~E.,  Hort{\'u}a H.~J.,   Kitaura F.-S.,  2024, \mn@doi [Astronomy \& Astrophysics] {10.1051/0004-6361/202347929}, 684, A100

\bibitem[\protect\citeauthoryear{Gatti et~al.}{Gatti et~al.}{2021}]{DES:2021lsy}
Gatti M.,  et~al., 2021, {Dark Energy Survey Year 3 results: cosmology with moments of weak lensing mass maps} (\mn@eprint {arXiv} {2110.10141})

\bibitem[\protect\citeauthoryear{Giri \& Schneider}{Giri \& Schneider}{2021}]{Giri:2021qin}
Giri S.~K.,  Schneider A.,  2021, \mn@doi [JCAP] {10.1088/1475-7516/2021/12/046}, 12, 046

\bibitem[\protect\citeauthoryear{{G{\'o}mez-Vargas}, Briones~Andrade  \& V{\'a}zquez}{{G{\'o}mez-Vargas} et~al.}{2023}]{Gomez-Vargas-2023-NeuralNetworksOptimized}
{G{\'o}mez-Vargas} I.,  Briones~Andrade J.,   V{\'a}zquez J.~A.,  2023, \mn@doi [Physical Review D] {10.1103/PhysRevD.107.043509}, 107, 043509

\bibitem[\protect\citeauthoryear{Graves}{Graves}{2011}]{NIPS2011_4329}
Graves A.,  2011, in Shawe-Taylor J.,  Zemel R.~S.,  Bartlett P.~L.,  Pereira F.,   Weinberger K.~Q.,  eds, , Advances in Neural Information Processing Systems 24.
Curran Associates, Inc., pp 2348--2356, \url {http://papers.nips.cc/paper/4329-practical-variational-inference-for-neural-networks.pdf}

\bibitem[\protect\citeauthoryear{Gubitosi, Piazza  \& Vernizzi}{Gubitosi et~al.}{2013}]{Gubitosi:2012hu}
Gubitosi G.,  Piazza F.,   Vernizzi F.,  2013, {The Effective Field Theory of Dark Energy} (\mn@eprint {arXiv} {1210.0201}), \mn@doi{10.1088/1475-7516/2013/02/032}

\bibitem[\protect\citeauthoryear{Guo, Pleiss, Sun  \& Weinberger}{Guo et~al.}{2017}]{pmlr-v70-guo17a}
Guo C.,  Pleiss G.,  Sun Y.,   Weinberger K.~Q.,  2017, in Precup D.,  Teh Y.~W.,  eds,  Proceedings of Machine Learning Research Vol. 70, Proceedings of the 34th International Conference on Machine Learning. PMLR, pp 1321--1330, \url {https://proceedings.mlr.press/v70/guo17a.html}

\bibitem[\protect\citeauthoryear{{Heymans} et~al.,}{{Heymans} et~al.}{2021}]{2021A&A...646A.140H}
{Heymans} C.,  et~al., 2021, {KiDS-1000 Cosmology: Multi-probe weak gravitational lensing and spectroscopic galaxy clustering constraints} (\mn@eprint {arXiv} {2007.15632}), \mn@doi{10.1051/0004-6361/202039063}

\bibitem[\protect\citeauthoryear{Hildebrandt et~al.}{Hildebrandt et~al.}{2017}]{Hildebrandt:2016iqg}
Hildebrandt H.,  et~al., 2017, \mn@doi [MNRAS] {10.1093/mnras/stw2805}, 465, 1454

\bibitem[\protect\citeauthoryear{Hollemans}{Hollemans}{2020}]{Hollemans-2024-Reliabilitydiagrams}
Hollemans M.,  2020, Reliability-Diagrams, \url {https://github.com/hollance/reliability-diagrams}

\bibitem[\protect\citeauthoryear{Hort{\'u}a, Volpi, Marinelli  \& Malag{\`o}}{Hort{\'u}a et~al.}{2020}]{Hortua-2020-ParameterEstimationCosmic}
Hort{\'u}a H.~J.,  Volpi R.,  Marinelli D.,   Malag{\`o} L.,  2020, \mn@doi [Physical Review D] {10.1103/PhysRevD.102.103509}, 102, 103509

\bibitem[\protect\citeauthoryear{Hort{\'u}a, Garc{\'i}a  \& Casta{\~n}eda~C.}{Hort{\'u}a et~al.}{2023}]{Hortua-2023-ConstrainingCosmologicalParameters}
Hort{\'u}a H.~J.,  Garc{\'i}a L.~{\'A}.,   Casta{\~n}eda~C. L.,  2023, \mn@doi [Frontiers in Astronomy and Space Sciences] {10.3389/fspas.2023.1139120}, 10

\bibitem[\protect\citeauthoryear{Hu \& Sawicki}{Hu \& Sawicki}{2007}]{Hu:2007nk}
Hu W.,  Sawicki I.,  2007, \mn@doi [Phys. Rev. D] {10.1103/PhysRevD.76.064004}, 76, 064004

\bibitem[\protect\citeauthoryear{Huterer \& Shafer}{Huterer \& Shafer}{2018}]{Huterer:2017buf}
Huterer D.,  Shafer D.~L.,  2018, {Dark energy two decades after: Observables, probes, consistency tests} (\mn@eprint {arXiv} {1709.01091}), \mn@doi{10.1088/1361-6633/aa997e}

\bibitem[\protect\citeauthoryear{Islam, Jia  \& Bruce}{Islam et~al.}{2019}]{islamHOWMUCHPOSITION2020}
Islam M.~A.,  Jia S.,   Bruce N.~D.,  2019, in International Conference on Learning Representations.

\bibitem[\protect\citeauthoryear{Jordan, Ghahramani, Jaakkola  \& Saul}{Jordan et~al.}{1999}]{Jordan1999}
Jordan M.~I.,  Ghahramani Z.,  Jaakkola T.~S.,   Saul L.~K.,  1999, \mn@doi [Machine Learning] {10.1023/A:1007665907178}, 37, 183

\bibitem[\protect\citeauthoryear{Khoury \& Weltman}{Khoury \& Weltman}{2004}]{Khoury:2003rn}
Khoury J.,  Weltman A.,  2004, \mn@doi [Phys. Rev. D] {10.1103/PhysRevD.69.044026}, 69, 044026

\bibitem[\protect\citeauthoryear{Kingma \& Ba}{Kingma \& Ba}{2014}]{kingma2017adam}
Kingma D.~P.,  Ba J.,  2014, Adam: A Method for Stochastic Optimization (\mn@eprint {arXiv} {1412.6980})

\bibitem[\protect\citeauthoryear{{Kingma} \& {Welling}}{{Kingma} \& {Welling}}{2013}]{Kingma2013}
{Kingma} D.~P.,  {Welling} M.,  2013, arXiv e-prints, \href {https://ui.adsabs.harvard.edu/abs/2013arXiv1312.6114K} {}

\bibitem[\protect\citeauthoryear{{Kingma}, {Salimans}  \& {Welling}}{{Kingma} et~al.}{2015}]{Kingma2015}
{Kingma} D.~P.,  {Salimans} T.,   {Welling} M.,  2015, arXiv e-prints, \href {https://ui.adsabs.harvard.edu/abs/2015arXiv150602557K} {}

\bibitem[\protect\citeauthoryear{Kiureghian \& Ditlevsen}{Kiureghian \& Ditlevsen}{2009}]{Kiureghian-2009-AleatoryEpistemicDoes}
Kiureghian A.~D.,  Ditlevsen O.,  2009, \mn@doi [Structural Safety] {10.1016/j.strusafe.2008.06.020}, 31, 105

\bibitem[\protect\citeauthoryear{Knabenhans et~al.}{Knabenhans et~al.}{2021}]{Euclid:2020rfv}
Knabenhans M.,  et~al., 2021, {Euclid preparation: IX. EuclidEmulator2 \textendash{} power spectrum emulation with massive neutrinos and self-consistent dark energy perturbations} (\mn@eprint {arXiv} {2010.11288}), \mn@doi{10.1093/mnras/stab1366}

\bibitem[\protect\citeauthoryear{Kobayashi, Nishimichi, Takada, Takahashi  \& Osato}{Kobayashi et~al.}{2020}]{Kobayashi:2020zsw}
Kobayashi Y.,  Nishimichi T.,  Takada M.,  Takahashi R.,   Osato K.,  2020, {Accurate emulator for the redshift-space power spectrum of dark matter halos and its application to galaxy power spectrum} (\mn@eprint {arXiv} {2005.06122}), \mn@doi{10.1103/PhysRevD.102.063504}

\bibitem[\protect\citeauthoryear{Koyama}{Koyama}{2016}]{Koyama:2015vza}
Koyama K.,  2016, {Cosmological Tests of Modified Gravity} (\mn@eprint {arXiv} {1504.04623}), \mn@doi{10.1088/0034-4885/79/4/046902}

\bibitem[\protect\citeauthoryear{Kullback \& Leibler}{Kullback \& Leibler}{1951}]{kullback1951}
Kullback S.,  Leibler R.~A.,  1951, \mn@doi [Ann. Math. Statist.] {10.1214/aoms/1177729694}, 22, 79

\bibitem[\protect\citeauthoryear{Kwon, Won, Kim  \& Paik}{Kwon et~al.}{2020}]{uncertainty-cov}
Kwon Y.,  Won J.-H.,  Kim B.~J.,   Paik M.~C.,  2020, \mn@doi [Computational Statistics \& Data Analysis] {https://doi.org/10.1016/j.csda.2019.106816}, 142, 106816

\bibitem[\protect\citeauthoryear{{LSST Dark Energy Science Collaboration}}{{LSST Dark Energy Science Collaboration}}{2012}]{Abate:2012za}
{LSST Dark Energy Science Collaboration} 2012, {Large Synoptic Survey Telescope: Dark Energy Science Collaboration} (\mn@eprint {arXiv} {1211.0310})

\bibitem[\protect\citeauthoryear{{Laureijs} et~al.,}{{Laureijs} et~al.}{2011}]{2011arXiv1110.3193L}
{Laureijs} R.,  et~al., 2011, \mn@doi [arXiv e-prints] {10.48550/arXiv.1110.3193}, \href {https://ui.adsabs.harvard.edu/abs/2011arXiv1110.3193L} {p. arXiv:1110.3193}

\bibitem[\protect\citeauthoryear{{Lawrence} et~al.,}{{Lawrence} et~al.}{2017}]{2017ApJ...847...50L}
{Lawrence} E.,  et~al., 2017, {The Mira-Titan Universe. II. Matter Power Spectrum Emulation} (\mn@eprint {arXiv} {1705.03388}), \mn@doi{10.3847/1538-4357/aa86a9}

\bibitem[\protect\citeauthoryear{LeCun, Bengio  \& Hinton}{LeCun et~al.}{2015}]{LeCun2015}
LeCun Y.,  Bengio Y.,   Hinton G.,  2015, Deep learning, \mn@doi{10.1038/nature14539}, \url {https://doi.org/10.1038/nature14539}

\bibitem[\protect\citeauthoryear{Lesgourgues}{Lesgourgues}{2011}]{Lesgourgues:2011re}
Lesgourgues J.,  2011, {The Cosmic Linear Anisotropy Solving System (CLASS) I: Overview} (\mn@eprint {arXiv} {1104.2932})

\bibitem[\protect\citeauthoryear{{Levi} et~al.,}{{Levi} et~al.}{2019}]{Levi:2019ggs}
{Levi} M.,  et~al., 2019, in Bulletin of the American Astronomical Society. p.~57 (\mn@eprint {arXiv} {1907.10688})

\bibitem[\protect\citeauthoryear{Lin, Tegmark  \& Rolnick}{Lin et~al.}{2017}]{Lin2017}
Lin H.~W.,  Tegmark M.,   Rolnick D.,  2017, Why Does Deep and Cheap Learning Work So Well?, \mn@doi{10.1007/s10955-017-1836-5}, \url {https://doi.org/10.1007/s10955-017-1836-5}

\bibitem[\protect\citeauthoryear{Linder}{Linder}{2003}]{Linder:2002et}
Linder E.~V.,  2003, \mn@doi [Phys. Rev. Lett.] {10.1103/PhysRevLett.90.091301}, 90, 091301

\bibitem[\protect\citeauthoryear{Lombriser \& Lima}{Lombriser \& Lima}{2017}]{Lombriser:2016yzn}
Lombriser L.,  Lima N.~A.,  2017, {Challenges to Self-Acceleration in Modified Gravity from Gravitational Waves and Large-Scale Structure} (\mn@eprint {arXiv} {1602.07670}), \mn@doi{10.1016/j.physletb.2016.12.048}

\bibitem[\protect\citeauthoryear{Lombriser \& Taylor}{Lombriser \& Taylor}{2016}]{Lombriser:2015sxa}
Lombriser L.,  Taylor A.,  2016, {Breaking a Dark Degeneracy with Gravitational Waves} (\mn@eprint {arXiv} {1509.08458}), \mn@doi{10.1088/1475-7516/2016/03/031}

\bibitem[\protect\citeauthoryear{MacKay}{MacKay}{1992}]{MacKay}
MacKay D. J.~C.,  1992, Neural Computation, 4, 448

\bibitem[\protect\citeauthoryear{Mancarella, Kennedy, Bose  \& Lombriser}{Mancarella et~al.}{2022}]{Mancarella:2020jyu}
Mancarella M.,  Kennedy J.,  Bose B.,   Lombriser L.,  2022, \mn@doi [Physical Review D] {10.1103/PhysRevD.105.023531}, 105, 023531

\bibitem[\protect\citeauthoryear{Martinelli et~al.}{Martinelli et~al.}{2021}]{Euclid:2020tff}
Martinelli M.,  et~al., 2021, \mn@doi [A&A] {10.1051/0004-6361/202039835}, 649, A100

\bibitem[\protect\citeauthoryear{Massara, Villaescusa-Navarro  \& Viel}{Massara et~al.}{2014}]{Massara:2014kba}
Massara E.,  Villaescusa-Navarro F.,   Viel M.,  2014, {The halo model in a massive neutrino cosmology} (\mn@eprint {arXiv} {1410.6813}), \mn@doi{10.1088/1475-7516/2014/12/053}

\bibitem[\protect\citeauthoryear{McCarthy, Schaye, Bird  \& Le~Brun}{McCarthy et~al.}{2017}]{McCarthy:2016mry}
McCarthy I.~G.,  Schaye J.,  Bird S.,   Le~Brun A. M.~C.,  2017, \mn@doi [MNRAS] {10.1093/mnras/stw2792}, 465, 2936

\bibitem[\protect\citeauthoryear{Mead, Peacock, Heymans, Joudaki  \& Heavens}{Mead et~al.}{2015}]{Mead:2015yca}
Mead A.,  Peacock J.,  Heymans C.,  Joudaki S.,   Heavens A.,  2015, \mn@doi [MNRAS] {10.1093/mnras/stv2036}, 454, 1958

\bibitem[\protect\citeauthoryear{{Mead}, {Heymans}, {Lombriser}, {Peacock}, {Steele}  \& {Winther}}{{Mead} et~al.}{2016a}]{2016MNRAS.459.1468M}
{Mead} A.~J.,  {Heymans} C.,  {Lombriser} L.,  {Peacock} J.~A.,  {Steele} O.~I.,   {Winther} H.~A.,  2016a, {Accurate halo-model matter power spectra with dark energy, massive neutrinos and modified gravitational forces} (\mn@eprint {arXiv} {1602.02154}), \mn@doi{10.1093/mnras/stw681}

\bibitem[\protect\citeauthoryear{Mead, Heymans, Lombriser, Peacock, Steele  \& Winther}{Mead et~al.}{2016b}]{Mead:2016zqy}
Mead A.,  Heymans C.,  Lombriser L.,  Peacock J.,  Steele O.,   Winther H.,  2016b, {Accurate halo-model matter power spectra with dark energy, massive neutrinos and modified gravitational forces} (\mn@eprint {arXiv} {1602.02154}), \mn@doi{10.1093/mnras/stw681}

\bibitem[\protect\citeauthoryear{Mead, Brieden, Tr\"oster  \& Heymans}{Mead et~al.}{2020}]{Mead:2020vgs}
Mead A.,  Brieden S.,  Tr\"oster T.,   Heymans C.,  2020, {HMcode-2020: Improved modelling of non-linear cosmological power spectra with baryonic feedback} (\mn@eprint {arXiv} {2009.01858}), \mn@doi{10.1093/mnras/stab082}

\bibitem[\protect\citeauthoryear{{Mehta}, {Bukov}, {Wang}, {Day}, {Richardson}, {Fisher}  \& {Schwab}}{{Mehta} et~al.}{2019}]{2019PhR...810....1M}
{Mehta} P.,  {Bukov} M.,  {Wang} C.-H.,  {Day} A. r. G.~R.,  {Richardson} C.,  {Fisher} C.~K.,   {Schwab} D.~J.,  2019, {A high-bias, low-variance introduction to Machine Learning for physicists} (\mn@eprint {arXiv} {1803.08823}), \mn@doi{10.1016/j.physrep.2019.03.001}

\bibitem[\protect\citeauthoryear{Moews et~al.,}{Moews et~al.}{2019}]{moews2019stress}
Moews B.,  et~al., 2019, Physical Review D, 99, 123529

\bibitem[\protect\citeauthoryear{{Mummery}, {McCarthy}, {Bird}  \& {Schaye}}{{Mummery} et~al.}{2017}]{Mummery2017}
{Mummery} B.~O.,  {McCarthy} I.~G.,  {Bird} S.,   {Schaye} J.,  2017, {The separate and combined effects of baryon physics and neutrino free streaming on large-scale structure} (\mn@eprint {arXiv} {1702.02064}), \mn@doi{10.1093/mnras/stx1469}

\bibitem[\protect\citeauthoryear{Neal}{Neal}{1996}]{Neal}
Neal R.~M.,  1996, Bayesian Learning for Neural Networks.
Springer-Verlag, Berlin, Heidelberg

\bibitem[\protect\citeauthoryear{Nishimichi et~al.}{Nishimichi et~al.}{2019}]{Nishimichi:2018etk}
Nishimichi T.,  et~al., 2019, {Dark Quest. I. Fast and Accurate Emulation of Halo Clustering Statistics and Its Application to Galaxy Clustering} (\mn@eprint {arXiv} {1811.09504}), \mn@doi{10.3847/1538-4357/ab3719}

\bibitem[\protect\citeauthoryear{Nixon, Dusenberry, Zhang, Jerfel  \& Tran}{Nixon et~al.}{2019}]{nixon2019measuring}
Nixon J.,  Dusenberry M.~W.,  Zhang L.,  Jerfel G.,   Tran D.,  2019, in CVPR workshops.

\bibitem[\protect\citeauthoryear{Noller}{Noller}{2020}]{Noller:2020afd}
Noller J.,  2020, {Cosmological constraints on dark energy in light of gravitational wave bounds} (\mn@eprint {arXiv} {2001.05469}), \mn@doi{10.1103/PhysRevD.101.063524}

\bibitem[\protect\citeauthoryear{Ocampo, {Ca{\~n}as-Herrera}  \& Nesseris}{Ocampo et~al.}{2024a}]{Ocampo-2024-NeuralNetworksCosmological}
Ocampo I.,  {Ca{\~n}as-Herrera} G.,   Nesseris S.,  2024a, Neural {{Networks}} for Cosmological Model Selection and Feature Importance Using {{Cosmic Microwave Background}} Data

\bibitem[\protect\citeauthoryear{Ocampo, Alestas, Nesseris  \& Sapone}{Ocampo et~al.}{2024b}]{Ocampo-2024-EnhancingCosmologicalModel}
Ocampo I.,  Alestas G.,  Nesseris S.,   Sapone D.,  2024b, Enhancing {{Cosmological Model Selection}} with {{Interpretable Machine Learning}} (\mn@eprint {arXiv} {2406.08351})

\bibitem[\protect\citeauthoryear{Parimbelli, Carbone, Bel, Bose, Calabrese, Carella  \& Zennaro}{Parimbelli et~al.}{2022}]{Parimbelli:2022pmr}
Parimbelli G.,  Carbone C.,  Bel J.,  Bose B.,  Calabrese M.,  Carella E.,   Zennaro M.,  2022, \mn@doi [JCAP] {10.1088/1475-7516/2022/11/041}, 11, 041

\bibitem[\protect\citeauthoryear{Peel, Lalande, Starck, Pettorino, Merten, Giocoli, Meneghetti  \& Baldi}{Peel et~al.}{2019}]{Peel-2019-DistinguishingStandardModified}
Peel A.,  Lalande F.,  Starck J.-L.,  Pettorino V.,  Merten J.,  Giocoli C.,  Meneghetti M.,   Baldi M.,  2019, \mn@doi [Physical Review D] {10.1103/PhysRevD.100.023508}, 100, 023508

\bibitem[\protect\citeauthoryear{Perenon, Bel, Maartens  \& de~la Cruz-Dombriz}{Perenon et~al.}{2019}]{Perenon:2019dpc}
Perenon L.,  Bel J.,  Maartens R.,   de~la Cruz-Dombriz A.,  2019, {Optimising growth of structure constraints on modified gravity} (\mn@eprint {arXiv} {1901.11063}), \mn@doi{10.1088/1475-7516/2019/06/020}

\bibitem[\protect\citeauthoryear{Ramachandra, Valogiannis, Ishak  \& Heitmann}{Ramachandra et~al.}{2021}]{Ramachandra:2020lue}
Ramachandra N.,  Valogiannis G.,  Ishak M.,   Heitmann K.,  2021, {Matter Power Spectrum Emulator for f(R) Modified Gravity Cosmologies} (\mn@eprint {arXiv} {2010.00596}), \mn@doi{10.1103/PhysRevD.103.123525}

\bibitem[\protect\citeauthoryear{Sakstein \& Jain}{Sakstein \& Jain}{2017}]{Sakstein:2017xjx}
Sakstein J.,  Jain B.,  2017, {Implications of the Neutron Star Merger GW170817 for Cosmological Scalar-Tensor Theories} (\mn@eprint {arXiv} {1710.05893}), \mn@doi{10.1103/PhysRevLett.119.251303}

\bibitem[\protect\citeauthoryear{Salvatelli, Piazza  \& Marinoni}{Salvatelli et~al.}{2016}]{Salvatelli:2016mgy}
Salvatelli V.,  Piazza F.,   Marinoni C.,  2016, {Constraints on modified gravity from Planck 2015: when the health of your theory makes the difference} (\mn@eprint {arXiv} {1602.08283}), \mn@doi{10.1088/1475-7516/2016/09/027}

\bibitem[\protect\citeauthoryear{Schneider, Teyssier, Stadel, Chisari, Le~Brun, Amara  \& Refregier}{Schneider et~al.}{2019}]{Schneider:2018pfw}
Schneider A.,  Teyssier R.,  Stadel J.,  Chisari N.~E.,  Le~Brun A. M.~C.,  Amara A.,   Refregier A.,  2019, {Quantifying baryon effects on the matter power spectrum and the weak lensing shear correlation} (\mn@eprint {arXiv} {1810.08629}), \mn@doi{10.1088/1475-7516/2019/03/020}

\bibitem[\protect\citeauthoryear{Schneider, Stoira, Refregier, Weiss, Knabenhans, Stadel  \& Teyssier}{Schneider et~al.}{2020a}]{Schneider:2019snl}
Schneider A.,  Stoira N.,  Refregier A.,  Weiss A.~J.,  Knabenhans M.,  Stadel J.,   Teyssier R.,  2020a, {Baryonic effects for weak lensing. Part I. Power spectrum and covariance matrix} (\mn@eprint {arXiv} {1910.11357}), \mn@doi{10.1088/1475-7516/2020/04/019}

\bibitem[\protect\citeauthoryear{Schneider et~al.,}{Schneider et~al.}{2020b}]{Schneider:2019xpf}
Schneider A.,  et~al., 2020b, {Baryonic effects for weak lensing. Part II. Combination with X-ray data and extended cosmologies} (\mn@eprint {arXiv} {1911.08494}), \mn@doi{10.1088/1475-7516/2020/04/020}

\bibitem[\protect\citeauthoryear{Semboloni, Hoekstra, Schaye, van Daalen  \& McCarthy}{Semboloni et~al.}{2011}]{Semboloni:2011fe}
Semboloni E.,  Hoekstra H.,  Schaye J.,  van Daalen M.~P.,   McCarthy I.~J.,  2011, \mn@doi [MNRAS] {10.1111/j.1365-2966.2011.19385.x}, 417, 2020

\bibitem[\protect\citeauthoryear{Song et~al.,}{Song et~al.}{2015}]{Song:2015oza}
Song Y.-S.,  et~al., 2015, \mn@doi [Phys. Rev.] {10.1103/PhysRevD.92.043522}, D92, 043522

\bibitem[\protect\citeauthoryear{{Springel} et~al.,}{{Springel} et~al.}{2018}]{Springel2018}
{Springel} V.,  et~al., 2018, {First results from the IllustrisTNG simulations: matter and galaxy clustering} (\mn@eprint {arXiv} {1707.03397}), \mn@doi{10.1093/mnras/stx3304}

\bibitem[\protect\citeauthoryear{Spurio~Mancini \& Bose}{Spurio~Mancini \& Bose}{2023}]{SpurioMancini:2023mpt}
Spurio~Mancini A.,  Bose B.,  2023, {On the degeneracies between baryons, massive neutrinos and f(R) gravity in Stage IV cosmic shear analyses} (\mn@eprint {arXiv} {2305.06350}), \mn@doi{10.21105/astro.2305.06350}

\bibitem[\protect\citeauthoryear{Takahashi, Sato, Nishimichi, Taruya  \& Oguri}{Takahashi et~al.}{2012}]{Takahashi:2012em}
Takahashi R.,  Sato M.,  Nishimichi T.,  Taruya A.,   Oguri M.,  2012, \mn@doi [Astrophys. J.] {10.1088/0004-637X/761/2/152}, 761, 152

\bibitem[\protect\citeauthoryear{Tram, Brandbyge, Dakin  \& Hannestad}{Tram et~al.}{2019}]{Tram:2018znz}
Tram T.,  Brandbyge J.,  Dakin J.,   Hannestad S.,  2019, {Fully relativistic treatment of light neutrinos in $N$-body simulations} (\mn@eprint {arXiv} {1811.00904}), \mn@doi{10.1088/1475-7516/2019/03/022}

\bibitem[\protect\citeauthoryear{Tr\"oster et~al.}{Tr\"oster et~al.}{2021}]{KiDS:2020ghu}
Tr\"oster T.,  et~al., 2021, \mn@doi [A&A] {10.1051/0004-6361/202039805}, 649, A88

\bibitem[\protect\citeauthoryear{Tsedrik, Bose, Carrilho, Pourtsidou, Pamuk, Casas  \& Lesgourges}{Tsedrik et~al.}{2024}]{maria2024}
Tsedrik M.,  Bose B.,  Carrilho P.,  Pourtsidou A.,  Pamuk S.,  Casas S.,   Lesgourges J.,  2024, {In prep.}

\bibitem[\protect\citeauthoryear{Vainshtein}{Vainshtein}{1972}]{Vainshtein:1972sx}
Vainshtein A.~I.,  1972, \mn@doi [Phys. Lett. B] {10.1016/0370-2693(72)90147-5}, 39, 393

\bibitem[\protect\citeauthoryear{{Valentin Jospin}, {Buntine}, {Boussaid}, {Laga}  \& {Bennamoun}}{{Valentin Jospin} et~al.}{2020}]{jospin2020handson}
{Valentin Jospin} L.,  {Buntine} W.,  {Boussaid} F.,  {Laga} H.,   {Bennamoun} M.,  2020, arXiv e-prints, \href {https://ui.adsabs.harvard.edu/abs/2020arXiv200706823V} {}

\bibitem[\protect\citeauthoryear{Wen, Vicol, Ba, Tran  \& Grosse}{Wen et~al.}{2018}]{wen2018flipout}
Wen Y.,  Vicol P.,  Ba J.,  Tran D.,   Grosse R.,  2018, in International Conference on Learning Representations. \url {https://openreview.net/forum?id=rJNpifWAb}

\bibitem[\protect\citeauthoryear{Will}{Will}{2014}]{Will:2014kxa}
Will C.~M.,  2014, {The Confrontation between General Relativity and Experiment} (\mn@eprint {arXiv} {1403.7377}), \mn@doi{10.12942/lrr-2014-4}

\bibitem[\protect\citeauthoryear{Winther, Casas, Baldi, Koyama, Li, Lombriser  \& Zhao}{Winther et~al.}{2019}]{Winther:2019mus}
Winther H.,  Casas S.,  Baldi M.,  Koyama K.,  Li B.,  Lombriser L.,   Zhao G.-B.,  2019, {Emulators for the nonlinear matter power spectrum beyond $\Lambda$CDM} (\mn@eprint {arXiv} {1903.08798}), \mn@doi{10.1103/PhysRevD.100.123540}

\bibitem[\protect\citeauthoryear{Wright, Winther  \& Koyama}{Wright et~al.}{2017}]{Wright:2017dkw}
Wright B.~S.,  Winther H.~A.,   Koyama K.,  2017, {COLA with massive neutrinos} (\mn@eprint {arXiv} {1705.08165}), \mn@doi{10.1088/1475-7516/2017/10/054}

\bibitem[\protect\citeauthoryear{Wright, Koyama, Winther  \& Zhao}{Wright et~al.}{2019}]{Wright:2019qhf}
Wright B.~S.,  Koyama K.,  Winther H.~A.,   Zhao G.-B.,  2019, {Investigating the degeneracy between modified gravity and massive neutrinos with redshift-space distortions} (\mn@eprint {arXiv} {1902.10692}), \mn@doi{10.1088/1475-7516/2019/06/040}

\bibitem[\protect\citeauthoryear{Zhong, Gallagher, Liu, Kailkhura, Hiszpanski  \& Han}{Zhong et~al.}{2022}]{zhongExplainableMachineLearning2022}
Zhong X.,  Gallagher B.,  Liu S.,  Kailkhura B.,  Hiszpanski A.,   Han T. Y.-J.,  2022, npj Computational Materials, 8, 204

\bibitem[\protect\citeauthoryear{Zhou, Khosla, Lapedriza, Oliva  \& Torralba}{Zhou et~al.}{2016}]{zhouLearningDeepFeatures2016}
Zhou B.,  Khosla A.,  Lapedriza A.,  Oliva A.,   Torralba A.,  2016, in Proceedings of the IEEE conference on computer vision and pattern recognition. pp 2921--2929

\bibitem[\protect\citeauthoryear{de Rham \& Melville}{de~Rham \& Melville}{2018}]{deRham:2018red}
de Rham C.,  Melville S.,  2018, {Gravitational Rainbows: LIGO and Dark Energy at its Cutoff} (\mn@eprint {arXiv} {1806.09417}), \mn@doi{10.1103/PhysRevLett.121.221101}

\bibitem[\protect\citeauthoryear{{van Daalen}, {Schaye}, {Booth}  \& {Dalla Vecchia}}{{van Daalen} et~al.}{2011}]{vanDaalen2011}
{van Daalen} M.~P.,  {Schaye} J.,  {Booth} C.~M.,   {Dalla Vecchia} C.,  2011, {The effects of galaxy formation on the matter power spectrum: a challenge for precision cosmology} (\mn@eprint {arXiv} {1104.1174}), \mn@doi{10.1111/j.1365-2966.2011.18981.x}

\bibitem[\protect\citeauthoryear{van Daalen, McCarthy  \& Schaye}{van Daalen et~al.}{2020}]{vanDaalen:2019pst}
van Daalen M.~P.,  McCarthy I.~G.,   Schaye J.,  2020, {Exploring the effects of galaxy formation on matter clustering through a library of simulation power spectra} (\mn@eprint {arXiv} {1906.00968}), \mn@doi{10.1093/mnras/stz3199}

\makeatother
\end{thebibliography}

% Alternatively you could enter them by hand, like this:
% This method is tedious and prone to error if you have lots of references
%\begin{thebibliography}{99}
%\bibitem[\protect\citeauthoryear{Author}{2012}]{Author2012}
%Author A.~N., 2013, Journal of Improbable Astronomy, 1, 1
%\bibitem[\protect\citeauthoryear{Others}{2013}]{Others2013}
%Others S., 2012, Journal of Interesting Stuff, 17, 198
%\end{thebibliography}

%%%%%%%%%%%%%%%%%%%%%%%%%%%%%%%%%%%%%%%%%%%%%%%%%%

%%%%%%%%%%%%%%%%% APPENDICES %%%%%%%%%%%%%%%%%%%%%

\appendix

\section{Random Spectra}\label{app:randoms}
\begin{figure}
    \centering
        \includegraphics[width=\columnwidth]{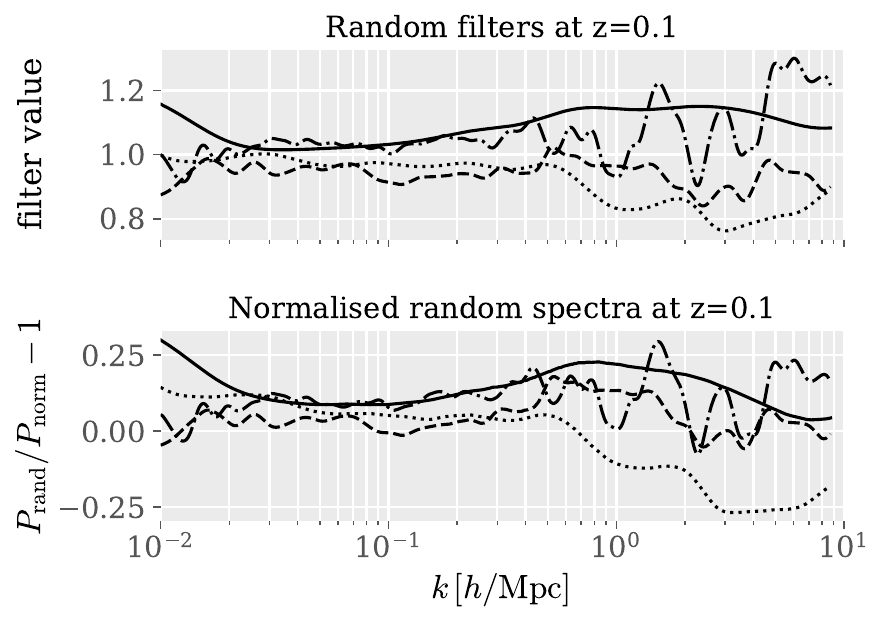}
    \caption{Four different random filters are shown for $z = 0.1$ in the upper panel. They are then each multiplied with a randomly drawn \lcdm{} spectrum to produce a random spectrum. These are shown in the lower panel after being normalised with the Planck \lcdm{} spectrum.}
    \label{fig:random_spectra}
\end{figure}

The random class represents models that are currently unknown. It was added to identify a cosmology which is not amongst our selected classes instead of  wrongly classifying it as one of them. It is not actually completely random but instead contains features that other feasible models might introduce in the power spectrum.
We create the random class by combining a \lcdm{} matter power spectrum with random filters.
The random filters should contain signatures that are $k$ and $z$ correlated to mimic real physical models and an algorithm to produce these was presented in \cite{Mancarella:2020jyu}. We have updated these filters since our last work to account for more realistic and subtle deviations from currently known theories. This was done by including wider variances in the correlation length in both $k$ and $z$ as well as reducing the fluctuations allowed at large scales. A set of four filters with randomly generated features that are correlated in $k$ and $z$ is shown at the top of \autoref{fig:random_spectra}. The bottom panel shows spectra of yet unknown physics after the filters have been multiplied with randomly selected \lcdm{} spectra. 
We have seen in  \autoref{sec:results} that it is important to build the random class from spectra that include effects of known physics like baryonic feedback and then use the random filters to account for unknown physics on top of that. For this reason we base them on \lcdm{} spectra from the most accurate EE2 training and test data set with massive neutrinos and baryonic feedback. The Mathematica notebook including the algorithm is available \href{https://zenodo.org/records/10688282}{here}. 

\section{Parameter Effects on the Power Spectrum}\label{app:priors}

\begin{figure*}
    \centering
        \includegraphics[width=\textwidth]{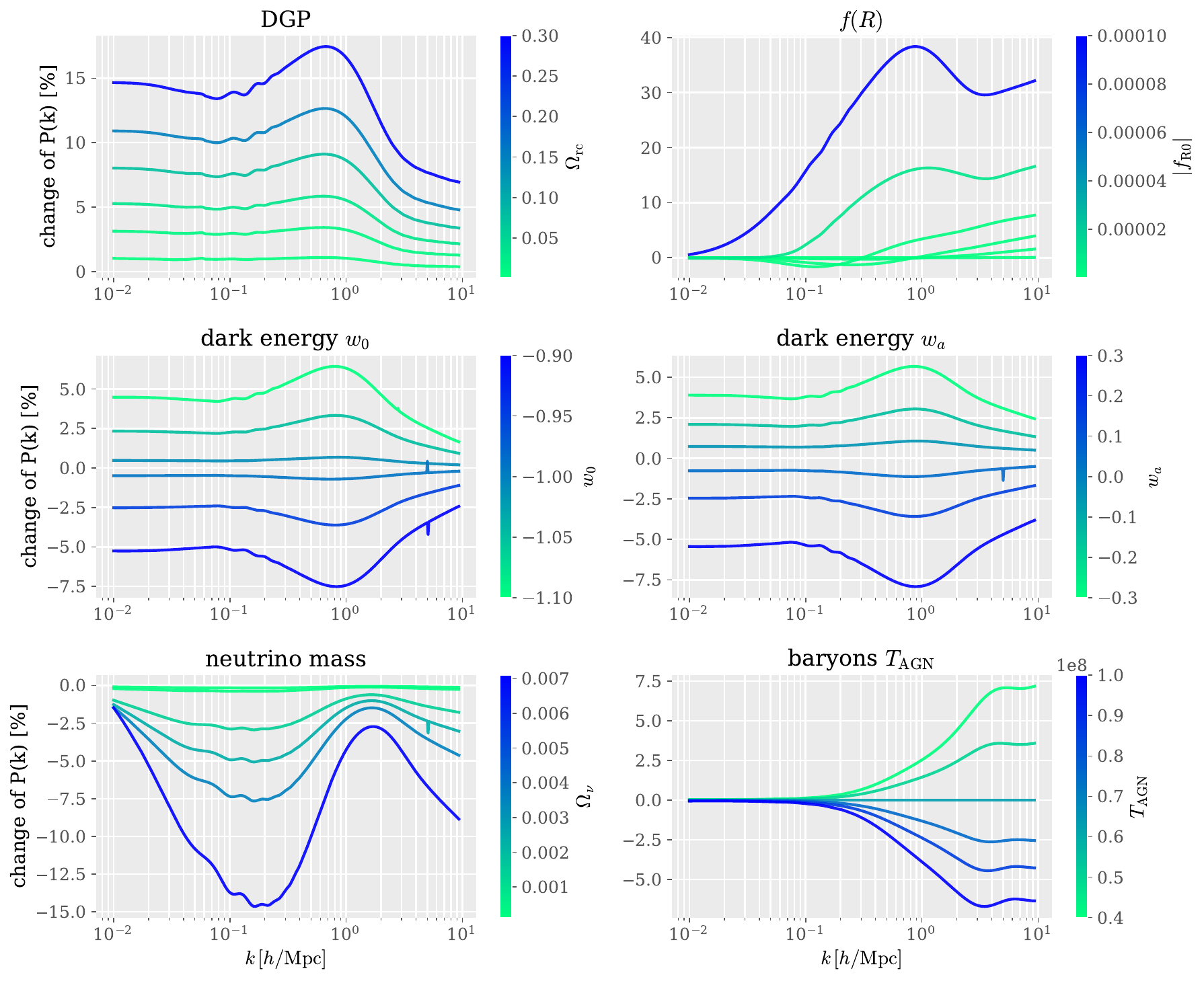}
    \caption{Influence of cosmological model parameters on the power spectrum. Other cosmological parameters are fixed to the mean values given in \autoref{tab:bac2data}. We normalise the spectra by the Planck \lcdm{} spectrum without massive neutrinos but with baryonic feedback assuming $T_\mathrm{AGN} = 6.3 \times 10^7 \, \mathrm{K}$. All values are within the prior ranges of the training data.}
    \label{fig:mg_params}
\end{figure*}

This section visualises the effects of model parameters for dynamical dark energy and modified gravity as well as the sum of neutrino masses and the strength of baryonic feedback on the matter power spectrum. \autoref{fig:mg_params} shows the characteristic changes when one parameter is varied while the other cosmological parameters are fixed to the Planck best-fit values (see \autoref{tab:bac2data}). The plotted values are representative of the prior ranges described in  \autoref{sec:data}. 
The selected classes show different characteristic influences on the power spectrum. Stronger modifications of DGP lead to increased amplitudes of the power spectrum on large scales while $f(R)$ introduces a strong scale dependence on small scales. 
Baryonic feedback has a very strong influence on nonlinear scales while the neutrino mass also affects BAO-scales. These specific features allow the network to distinguish the different cosmological classes even when the cosmological parameters are varied.
Some of the curves show a small spike which is an artefact from the \react{} computation. These numerical glitches are above the scales we consider so they do not affect our results. 

%%%%%%%%%%%%%%%%%%%%%%%%%%%%%%%%%%%%%%%%%%%%%%%%%%

% Don't change these lines
\bsp	% typesetting comment
\label{lastpage}
\end{document}